\documentclass[natbib,doc,12pt,floatsintext]{apa6}
\usepackage{amsfonts, amsmath, amsthm, amssymb}
\usepackage{relsize}
\usepackage{epstopdf}
\usepackage{bm}
\usepackage{todonotes}
\usepackage{epigraph} 
\usepackage{graphicx,psfrag,epsf} 
\usepackage{subcaption}
\usepackage{nicefrac}
\usepackage{color}

\theoremstyle{plain}                          

\newcommand{\given}{\, | \,}

\newcommand{\Hc}{\mathcal{H}}

\newcommand{\Mc}{\mathcal{M}}
\newcommand{\Nc}{\mathcal{N}}

\newcommand{\BF}{\text{BF}}
\newcommand{\der}{\mathrm{d}}

\newcommand{\bibOrder}[1]{}

\newcommand{\SR}[1]{} 

\newcommand{\rev}[1]{\textcolor{black}{#1}}

\title{History and Nature of the Jeffreys-Lindley Paradox}

\shorttitle{The Jeffreys-Lindley Paradox}

\twoauthors{Eric-Jan~Wagenmakers}{Alexander~Ly}

\twoaffiliations{ORCID: 0000-0003-1596-1034\\University of Amsterdam}{ORCID: 0000-0003-3925-3833\\University of Amsterdam\\Centrum Wiskunde \& Informatica}

\authornote{This research was supported by a Netherlands Organisation for Scientific Research (NWO) grant to EJW (016.Vici.170.083), as well as an Advanced ERC grant to EJW (743086 UNIFY). Centrum Wiskunde \& Informatica (CWI) is the national research institute for mathematics and computer science in the Netherlands. Correspondence should be sent to Eric-Jan Wagenmakers, University of Amsterdam, Nieuwe Achtergracht 129 B, 1018 WT Amsterdam, The Netherlands. E-mail may be sent to EJ.Wagenmakers@gmail.com.}

\abstract{The Jeffreys-Lindley paradox exposes a rift between Bayesian and frequentist hypothesis testing that strikes at the heart of statistical inference. Contrary to what most current literature suggests, the paradox was central to the Bayesian testing methodology developed by Sir Harold Jeffreys in the late 1930s. Jeffreys showed that the evidence \rev{for} a point-null hypothesis $\mathcal{H}_0$ scales with $\sqrt{n}$ and repeatedly argued that it would therefore be mistaken to set a threshold for rejecting $\mathcal{H}_0$ at a constant multiple of the standard error. Here we summarize Jeffreys's early work on the paradox and clarify his reasons for including the $\sqrt{n}$ term. The prior distribution is seen to play a crucial role; by implicitly correcting for selection, small parameter values are identified as relatively surprising under $\mathcal{H}_1$. We highlight the general nature of the paradox by presenting both a fully frequentist and a fully Bayesian version. We also demonstrate that the paradox does not depend on assigning prior mass to a point hypothesis, as is commonly believed.}


\begin{document}
	
\maketitle
\pagebreak

The Jeffreys-Lindley paradox (e.g., \citealp{Jeffreys1935,Lindley1957,Bartlett1957}) refers to the fact that, as sample size increases indefinitely and the $p$-value remains constant at any non-zero value (e.g., $p=.005$), we inevitably arrive at a conflict between $p$-values and Bayes factors, in the sense that the $p$-value suggests that the point-null hypothesis $\mathcal{H}_0$ should be rejected, whereas the Bayes factor indicates that $\mathcal{H}_0$ decisively outpredicts the alternative hypothesis $\mathcal{H}_1$. This conflict will arise regardless of the $p$-value under consideration and regardless of the prior distribution on the test-relevant parameter in $\mathcal{H}_1$ (under regularity conditions). Thus, a frequentist statistician may specify any non-zero $\alpha$-level whatever, a Bayesian statistician may specify any continuous prior distribution on the test-relevant parameter under $\mathcal{H}_1$, and a third party could then infallibly construct data sets for which the point-null hypothesis $\mathcal{H}_0$ would be simultaneously rejected by the frequentist and accepted by the Bayesian.\footnote{Note that the Jeffreys-Lindley paradox is a veridical paradox: it is an \emph{apparent} contradiction (e.g., \citealp[p. 310]{Jeffreys1938Continuous}). A sufficiently knowledgeable and confident statistician may therefore rightly proclaim that the Jeffreys-Lindley paradox is not at all paradoxical (to them). Veridical paradoxes are in the eye of the beholder. See also \citet{Cousins2017} and \citet{Pericchi2011}.}

Although the paradox is often associated with \citet{Lindley1957}, and sometimes with \citet{Bartlett1957}, it was already derived, demonstrated, explained, and emphasized by Sir Harold Jeffreys in his articles and books on Bayesian hypothesis testing from the second half of the 1930s (i.e., \citealp[p. 205]{Jeffreys1935}; \citealp[p. 345 and p. 353]{Jeffreys1936}; \citealp[p. 417]{Jeffreys1936Further}; \citealp[p. 494]{Jeffreys1937Tests}; \citealp[pp. 250-251 and p. 259]{Jeffreys1937SI}; \citealp[p. 1004]{Jeffreys1937Nature}; \citealp[pp. 377-381]{Jeffreys1938Comparison2}; \citealp[p. 161]{Jeffreys1938}; \citealp[p. 148]{Jeffreys1938MLE}; \rev{\citealp[p. 114]{Jeffreys1938aftershock};} \citealp[p. 310]{Jeffreys1938Continuous}; \citealp[pp. 194-195 and pp. 359-360]{Jeffreys1939} -- see p. 248 and pp. 435-436 in \citealp{Jeffreys1961}). The paradox has remained a source of inspiration for statisticians and philosophers alike (e.g., \citealp{Bernardo1980,Bernardo2011,berrar2017jeffreys,Colquhoun2019,Cornfield1966,Cousins2017,EdwardsEtAl1963,Good1980,Jefferys1990,Leamer1978,nasir2020operational,OrmerodEtAl2017,Robert2014,Royall1986,Senn2001,Shafer1982,spanos2013should,sprenger2013testing,VillaWalker2017,yin2020demystify,Wagenmakers2007}; \citealp[Chapter 10]{Zellner19711996}), but we believe that the neglect of Jeffreys's original work on the paradox has led to considerable confusion. Indeed, the paradox has caused statisticians to question the usefulness of Bayesian statistics as a whole (e.g., \citealp{Shafer1982,spanos2013should}), to reject Bayes factor hypothesis testing in favor of Bayesian parameter estimation (e.g., \citealp{Bernardo1980}), and to develop alternative forms of Bayesian hypothesis testing (e.g., \citealp{Aitkin1991,Andrews1994,KamaryEtAl2014,PereiraEtAl2008,VehtariEtAl2017}). We do not wish to disparage this work but we do believe the original arguments by Jeffreys have been underappreciated if not entirely forgotten (for a notable exception see \citealp{Cousins2017}). 

The goal of this paper is therefore threefold. First we aim to demonstrate the extent to which the paradox had already been treated by Jeffreys prior to the 1957 articles by Lindley and by Bartlett. The appendix lists Jeffreys's discussions of the paradox after 1957. Contrary to popular belief, our analysis reveals that the paradox played a central role in Jeffreys's system of Bayes factor hypothesis tests, and did so from the outset. Although Jeffreys often downplayed the practical ramification of the paradox for moderate sample sizes, he also repeatedly stressed that his Bayesian hypothesis test depended not just on how many standard errors the maximum likelihood estimate is away from zero (as in the classical method) but also involved a $\sqrt{n}$ term. Crucially, this means that the criterion for ``significance'' in Jeffreys's tests is not given by a constant multiple of the standard error. Jeffreys presented almost every Bayes factor he proposed in the same form, with a $\sqrt{n}$ factor outside of an exponential term and a multiple-of-the-standard-error factor inside the exponential term; these expressions leave no doubt about the large-$n$ conflict between Jeffreys's Bayes factors and $p$-values. Moreover, throughout his published work Jeffreys highlighted the effect of sample size on his tests by means of tables; he discussed the reasons for the appearance of the $\sqrt{n}$ term, and he explicitly stated that including this term was both desirable and dictated by the application of Bayesian probability theory to the problem of hypothesis testing. The common notion that Jeffreys mentioned the paradox only in passing is therefore seriously incorrect.

The second goal of this paper is to revive Jeffreys's original line of argumentation, which was that the paradox, instead of being ``certainly embarrassing to the Bayesian'' \citep[p. 17]{SzaboVaart2019}, or ``difficult to accept'' \citep[p. 174]{Bernardo2009} was rather the inevitable consequence of any reasonable definition of evidence. In other words, Jeffreys felt that no sensible measure of evidence can be based on a constant multiple of the standard error, independent of sample size. 

The third goal of this manuscript is to highlight the general nature of the paradox. Specifically, we demonstrate that the paradox can be given both a fully frequentist interpretation and a fully Bayesian interpretation. Moreover, and in contrast to popular belief, we show that the essence of the paradox does not depend on the fact that the model comparison involves a sharp null hypothesis $\mathcal{H}_0$ with a point-mass at zero. Instead, the paradox will manifest itself for any Bayes factor where the prior distribution for effect size under the sceptic's $\mathcal{H}_0$ is more heavily concentrated around zero than the prior distribution for effect size under the proponent's $\mathcal{H}_1$, a condition so mild as to be almost tautological.  


\section*{Statistical Background}
In the early 20th century Sir Ronald Fisher promoted the idea of null hypothesis significance testing (NHST) using \( p \)-values. Informally, the \( p \)-value is the chance under the null hypothesis of finding a test statistic at least as extreme as the one obtained (e.g., \citealp{wasserstein2016asa}). The idea of NHST is loosely similar to that of a proof by contradiction: to show that there exists an effect, one assumes the opposite (i.e., the null model $\mathcal{H}_0$) and demonstrates that the data make this assumption unlikely \citep{WagenmakersEtAlScrutinyBook2017}. In NHST, the data are believed to cast doubt on $\mathcal{H}_0$ when the obtained \( p \)-value is sufficiently small. Fisher deemed a $p$-value of $.05$ or lower sufficient grounds to reject the null hypothesis. In Chapter~{3} of \emph{Statistical Methods for Research Workers}, Fisher discusses the normal distribution and notes that %

\begin{quote}
``The value for which P = .05, or 1 in 20, is 1.96 or nearly 2 ; it is convenient to take this point as a limit in judging whether a deviation is to be considered significant or not. Deviations exceeding twice the standard deviation are thus formally regarded as significant. Using this criterion, we should be led to follow up a false indication only once in 22 trials, even if the statistics were the only guide available. Small effects will still escape notice if the data are insufficiently numerous to bring them out, but no lowering of the standard of significance would meet this difficulty.''
(p. 45, \citealp{fisher1934statistical5})
\end{quote}

Despite constant criticism from within the statistical community, Fisher's rule has since been institutionalised in academic practice. Researchers routinely conclude that results constitute a significant deviation from the null model whenever \( p < .05 \), that is, whenever the observed value of the test statistic falls more than two standard errors away from the value postulated by the null model.\footnote{\rev{As noted by \citet[pp. 18-19]{Cornfield1966}, the $\alpha$-level (i.e., the critical value below which the $p$-value is deemed to indicate a statistically significant deviation from the null model) is often viewed as a ``universal yardstick'' and the underlying intuition is that ``All hypotheses rejected at the same critical level have equal amounts of evidence against them.'' (i.e., Cornfield's ``$\alpha$-postulate'', which he sought to undercut).}} 



As an alternative to \( p \)-value significance testing, Sir Harold Jeffreys developed and advocated a series of Bayesian hypothesis tests whose key outcome is now known as the \emph{Bayes factor} (e.g., \citealp{KassRaftery1995}). The philosophical foundation of the Bayes factor goes back to Jeffreys's work with Dorothy Wrinch in the early 1920s \citep{WrinchJeffreys1919,WrinchJeffreys1921,WrinchJeffreys1923}, but the concrete statistical development was initiated and largely completed by Jeffreys in the second half of the 1930s (e.g., \citealp{Jeffreys1935,Jeffreys1939}; for a modern appreciation see \citealp{etz2017haldane,Howie2002,LyEtAl2016,LyEtAl2016Rejoinder,LyEtAl2020Alternative}, and \citealp{RobertEtAl2009}). To learn from data Jeffreys proposed to assign prior model probabilities \( P(\Mc_{0}) \) and \( P(\Mc_{1}) \) to the null hypothesis and the alternative hypothesis, respectively. In light of data \( y \) these probabilities can then be updated to posterior model probabilities using Bayes' rule. The ratio of the posterior models probabilities then leads to 
\begin{align}
\label{defBf10}
\underbrace{\frac{P(\Mc_{1} \, | \, y)}{P(\Mc_{0} \, | \, y)}}_{\text{posterior model odds}} = \,\,\,\, \underbrace{\frac{p(y \, | \, \Mc_{1})}{p(y \, | \, \Mc_{0})}}_{\BF_{10} (y)} \,\,\,\, \times \underbrace{\frac{ P(\Mc_{1})}{P(\Mc_{0})}}_{\text{prior model odds}},
\end{align}
where \( p(y \, | \, \Mc_{k}) \) is known as the marginal likelihood, that is, the likelihood function of the free parameters \( \theta_{k} \) under \( \Mc_{k} \) integrated out with respect to a prior distribution \( \pi( \theta \, | \, \Mc_{k}) \):
\begin{align}
p(y \, | \, \Mc_{k}) := \int_{\Theta_{k}} f ( y \, | \, \theta_{k}, \Mc_{k}) \pi( \theta_{k} \, | \, \Mc_{k}) \,\der \theta_{k} . 
\end{align}
The purpose of the Bayes factor $\text{BF}_{10}(y)$ is to ``grade the decisiveness of the evidence'' \citep[p. 432]{Jeffreys1961}. In contrast to the \( p \)-value, this pertains to both $\mathcal{M}_0$ and $\mathcal{M}_1$. Specifically, a $\text{BF}_{10}(y)$ much larger than 1 indicates evidence for $\mathcal{M}_1$ over $\mathcal{M}_0$; a $\text{BF}_{10}(y)$ near 0 indicates evidence for $\mathcal{M}_0$ over $\mathcal{M}_1$ (i.e., ``evidence of absence''); and a $\text{BF}_{10}(y)$ near 1 indicates that the data are insufficiently diagnostic (``absence of evidence''; \citealp{KeysersEtAl2020}). Note that for the construction of a Bayes factor a \emph{pair of priors} needs to be selected, one for each model. Jeffreys did so with great care for various statistical models and documented the results in his magnum opus \emph{Theory of Probability} \citep{Jeffreys1939,Jeffreys1948,Jeffreys1961}.\footnote{\rev{For his Bayes factor tests, Jeffreys proposed prior distributions that did not reflect strong advance knowledge and that obeyed several logical desiderata (e.g., \citealp{BayarriEtAl2012,ConsonniEtAl2018,LyEtAl2016}). Note that the paradox manifests itself regardless of how the prior distribution is defined, under regularity conditions.}} As will become apparent below, one of the defining features of the Bayes factor is that it does not depend on a constant multiple of the standard error. The additional involvement of sample size is what generates the paradox.

To set the stage we start by discussing the 1957 article from Dennis Lindley\nocite{Lindley1957}. We then list Jeffreys's work on the paradox as expressed in a series of \rev{16} articles and two books from 1935 to 1957. In order to drive home the point that the paradox was central to Jeffreys's tests, our treatment aims to be comprehensive. The included quotations are unusual both in their number and in their length, but we believe this is necessary in order to (1) irrevocably refute the common misconception that Jeffreys had ignored or neglected the paradox; (2) support the claim that the paradox in fact presents a defining feature of the Bayes factor hypothesis test; (3) demonstrate the different ways in which Jeffreys explained why a measure of evidence cannot depend on a constant multiple of the standard error.



\section*{The 1957 Contribution by Lindley}
Dennis \citet{Lindley1957} started his famous article \emph{A statistical paradox} as follows:
\begin{quotation}
``An example is produced to show that, if $H$ is a simple hypothesis and $x$ the result of an experiment, the following two phenomena can occur simultaneously:

(i) a significance test for $H$ reveals that $x$ is significant at, say, the 5\% level;

(ii) the posterior probability of $H$, given $x$, is, for quite small prior probabilities of $H$, as high as 95\%.

Clearly the common-sense interpretations of (i) and (ii) are in direct conflict. The phenomenon is fairly general with significance tests and casts doubts on the meaning of a significance level in some circumstances.'' (p. 187)
\end{quotation}

Later in the article, Lindley elaborates:
\begin{quotation}
``Now in our example we have taken situations in which the significance level is fixed because, as explained above, we wish to see whether its interpretation as a measure of lack of conviction about the null hypothesis does mean the same in different circumstances. The Bayesian probability is all right, by the arguments above; and since we now see that it varies strikingly with \( n \) for fixed significance level, in an extreme case producing a result in direct conflict with the significance level, the degree of conviction is not even approximately the same in two situations with equal significance levels. \emph{5\% in to-day's small sample does not mean the same as 5\% in to-morrow's large one.}'' (\citealp[p. 189]{Lindley1957}, italics added for emphasis)
\end{quotation}

\noindent Lindley explicitly acknowledges the fact that Jeffreys noted the paradox earlier:
\begin{quotation}
``The paradox is not, in essentials, new, although few statisticians are aware of it. The difference between the two approaches has been noted before by Jeffreys (see, in particular, 1948, Appendix)\nocite{Jeffreys1948}, who is the originator of significance tests based on Bayes's theorem and a concentration of prior probability on the null value. But Jeffreys is concerned to emphasize the similarity between his tests and those due to Fisher and the discrepancies are not emphasized.'' (p. 190)
\end{quotation}

\noindent We believe that Lindley's assessment requires revision. Below we demonstrate that Jeffreys repeatedly emphasized the theoretical difference between the two approaches throughout \rev{16} articles and two books published from 1935 to 1957.

\section*{The Contributions by Jeffreys from 1935 to 1957}

In order to follow the quotations from the works cited below, note that Jeffreys uses $K$ to refer to the Bayes factor for $\mathcal{H}_0$ over $\mathcal{H}_1$, that is, $K \equiv \text{BF}_{01}$. In addition, Jeffreys denotes $\mathcal{H}_0$ by $q$ and $\mathcal{H}_1$ by $\sim  \! q$ or $q'$. For a modern-day reader, it may be confusing that Jeffreys used Greek letters for observed data -- in particular, he often used $\theta$ to denote observed data rather than an unobserved parameter. Finally, Jeffreys often conditioned all probability statements on background knowledge, which he denoted by \emph{h} or \emph{H} (`history') -- not to be mistaken for the modern-day use of \emph{H} for `hypothesis'. A complete translation of Jeffreys's notation can be found in Table D.4 of \citet{LyEtAl2016}.  

\subsection{1. The 1934 letter to Fisher}
The first hint that Jeffreys is interested in developing a Bayesian significance test is found in a 1934 letter to Fisher:
\begin{quotation}
``The sort of thing that bothers me is this. In seismology we get times of transmission to various distances, and fit a polynomial of degree 3, say, to them. The significance of the last term really involves the prior probability that such a term will be present. The usual thing is to keep it if it is some arbitrary multiple of its standard error, but I think it ought to be possible to frame a rule with \emph{some} sort of argument behind it...''\\
Sir Harold Jeffreys, in a letter to Sir Ronald Fisher, 1934 \citep[p. 156]{Bennett1990}\footnote{Curiously, the letter as given in Bennett is incomplete. The original, complete letter can be found at \url{https://digital.library.adelaide.edu.au/dspace/bitstream/2440/67780/109/1934-03-21.pdf}.} 
\end{quotation}
The Bayes factor rule that Jeffreys later derived turned out to be different from ``the usual thing'': the strength of the Bayes factor is not proportional to a constant multiple of the standard error, but also involves sample size. This is the paradox. Thus, the 1934 letter to Fisher shows that the seeds of the paradox were sown even before Jeffreys had started to develop his tests.

\subsection{2. The 1935 article \emph{Some tests of significance, treated by the theory of probability}}
This was the first article in which Jeffreys developed a series of concrete Bayes factor hypothesis tests. The introductory paragraph immediately sets up the key issue, in similar fashion to the 1934 letter to Fisher:
\begin{quotation}
``It often happens that when two sets of data obtained by observation give slightly different estimates of the true value we wish to know whether the difference is significant. The usual procedure is to say that it is significant if it exceeds a certain rather arbitrary multiple of the standard error; but this is not very satisfactory, and it seems worth while to see whether any precise criterion can be obtained by a thorough application of the theory of probability.'' \citep[p. 203]{Jeffreys1935}
\end{quotation}

First Jeffreys turns to a comparison of two proportions:
\begin{quotation}
``Suppose that two different large, but not infinite, populations have been sampled in respect of a certain property. One gives $x$ specimens with the property, $y$ without; the other gives $x'$ and $y'$ respectively. The question is, whether the difference between $x/y$ and $x'/y'$ gives any ground for inferring a difference between the corresponding ratios in the complete population.'' \citep[p. 203]{Jeffreys1935}
\end{quotation}

Jeffreys (p. 204, Eq. 11) then shows that the posterior odds for $q$ over $\sim \! q$ is given by
\begin{equation*}
    \frac{P(q \mid \theta, h)}{P(\sim \! q \mid \theta, h)} = \frac{(x+x')! \, (y+y')! \, (x+y+1)! \, (x'+y'+1)!}{x!\,y!\,x'!\,y'!\, (x+x'+y+y'+1)!},
\end{equation*}
where $\theta$ denotes the observed data and $h$ (`history') denotes background knowledge. For large samples, Jeffreys obtains the following approximation (p. 205, Eq. 15):
\begin{multline*}
    \frac{P(q \mid \theta, h)}{P(\sim \! q \mid \theta, h)} \sim \left\{\frac{(x+x'+y+y')(x+y)(x'+y')}{2\pi(x+x')(y+y')}\right\}^{\frac{1}{2}}\\ 
    \exp{\left\{ -\tfrac{1}{2} \frac{(x+x'+y+y')(xy'-x'y)^2}{(x+x')(y+y')(x+y)(x'+y')} \right\}}.
\end{multline*}
Jeffreys then continues and identifies the phenomenon that lies at the heart of the paradox:
\begin{quotation}
``The theory therefore shows that a small difference between the sampling ratios may establish a high probability that the ratios in the main populations are equal, while a large one may show that they are different. This is in accordance with ordinary practice, but has not, so far as I know, been related to the general theory before. \emph{In one respect, however, there is a departure from ordinary practice}. It would be natural to define a standard error of $xy' - x'y$ in terms of the coefficient of its square in the exponential; but the range of values of the exponent that make the ratio of the posterior probabilities greater than 1 is not a constant, since it depends on the outside factor, which increases with the sizes of the samples. This variability is of course connected directly with the fact that agreement between the two populations becomes more probable if the samples are large and the difference of the sampling ratios are small; when the ratio is large at $xy'-x'y=0$, a larger value of the exponent is obviously needed to reduce the product to unity.

Some numerical values are given by way of illustration. In each case $x=y$, $x'+y'=x+y$, but in general $x' \neq y'$. The table gives $x+y$, the maximum value of the ratio of the posterior probabilities, and that of $x'-y'$ needed to make the ratio equal to unity.

\begin{table}[h]
	\centering
	\caption{Table reproduced from \citealp[p. 205]{Jeffreys1935}.}
	\label{tableJeffreys35TwoProportions}
	{
		\begin{tabular}{|r|c|c|c|c}
			\toprule
			\( x+ y \) & \( P(q) / P(\sim \! q)  \) & \( x' - y' \) & \( (x' - y')/(x+y)^{\frac{1}{2}} \) \\
			\cmidrule[0.4pt]{1-4}
			\( 40 \) & \( 3.57 \) & \( 14.3 \) & \( 2.26 \)  \\
			\( 100 \) & \( 5.65 \) & \( 26.4 \) & \( 2.64 \)  \\
			\( 200 \) & \( 7.97 \) & \( 40.8 \) & \( 2.89 \) \\
			\( 400 \) & \( 11.3 \)  \, & \( 61.5 \) & \( 3.07 \) \\
			\( 1,000 \) & \( 17.8 \)  \, & \( 107.3 \) & 3.39 \\
			\( 10,000 \) & \( 56.4 \)  \, & \( 401 \) \, \,  & \( 4.01 \) \\
			\( 100,000 \) & \( 178 \) \,  \, & \( 1440 \) \, \, & \( 4.57 \) \\
			\bottomrule
		\end{tabular}
	}
\end{table}

\noindent The ratio of the critical value of $x'-y'$ to $(x+y)^{\frac{1}{2}}$ is given in a further column to show how little it varies when the sizes of the samples change by a factor of 2500.'' (\citealp[pp. 205-206]{Jeffreys1935}; italics added for emphasis)
\end{quotation}

Later on Jeffreys draws the same conclusion for a test between two means with the standard error known:
\begin{quotation}
``It is therefore not correct to say that a systematic difference becomes significant when it reaches any constant multiple of its standard error'' \citep[p. 207]{Jeffreys1935}
\end{quotation}

\noindent Jeffreys returns to this theme several times throughout the article, for different tests (e.g., correlation, periodicity). The overall impression is that in the 1935 article Jeffreys emphasized the theoretical aspect of the paradox but at the same time downplayed its practical ramifications. 



\subsection{3. The 1936 article \emph{On some criticisms of the theory of probability}}
One year later Jeffreys again raises the key issue:
\begin{quotation}
``The results show that the probability that such a term is needed is increased or decreased according as the coefficient is more or less than a certain multiple of its standard error; \emph{the multiple needed, however, increases with the number of observations.}'' (\citealp[p. 345]{Jeffreys1936}; italics added for emphasis) 
\end{quotation}

\noindent Jeffreys elaborates and discusses the problem of a least-squares fit to a regression equation: 
\begin{quotation}
``When one unknown is determined at a time by least squares the criterion\footnote{For this result Jeffreys includes a footnote to \citet{Jeffreys1936Further} (relevant pages: 432-440) which was \emph{in press} at the time.} that the last determined shall be supported by the observations is that
\begin{equation*}
    \frac{b^2}{\sigma_b^2} > \log_e \frac{2n}{\pi},
\end{equation*}
where $n$ is the number of observations.'' \citep[p. 352]{Jeffreys1936}
\end{quotation}
As $b$ is the least-squares parameter point estimate, and $\sigma_b$ is the standard error, the equation shows that for support to remain constant as $n$ increases, the multiple of the standard error will need to increase as well. To underscore this point Jeffreys provides a table, reproduced here as Table~\ref{tableJeffreys36}, which ``gives the critical ratios that an unknown found by least squares from $n$ observations shall be supported by the observations.'' \citep[p. 352]{Jeffreys1936}. For instance, when $n=10$ we have $b/\sigma_b = \sqrt{\log_e 20/\pi} \approx 1.36$ and for $n=100$ we have $b/\sigma_b = \sqrt{\log_e 200/\pi} \approx 2.04$.

\begin{table}[h]
	\centering
	\caption{Table reproduced from \citealp[p. 352]{Jeffreys1936}. Here \( b/ \sigma_{b} \) indicates the ratio of a least squares point estimate $b$ to its standard error $\sigma_b$ that results in a Bayes factor of 1. This critical ratio increases with $n$.}
	\label{tableJeffreys36}
	{
		\begin{tabular}{rrrrrr}
			\toprule
			\( n. \) & \( b/ \sigma_{b}.  \) & \( n. \) & \( b/ \sigma_{b}.  \) & \( n. \) & \( b/ \sigma_{b}.  \) \\
			\cmidrule[0.4pt]{1-6}
			\( 5 \) & \( 1.07 \) & \( 200 \) & \( 2.20 \) & \( 10,000 \) & \( 2.96 \) \\
			\( 10 \) & \( 1.36 \) & \( 500 \) & \( 2.40 \) & \( 20,000 \) &  \( 3.07 \) \\
			\( 20 \) & \( 1.59 \) & \( 1000 \) & \( 2.54 \) & \( 50,000 \) & \( 3.22 \) \\
			\( 50 \) & \( 1.86 \) & \( 2000 \) & \( 2.67 \) & \( 100,000 \) & \( 3.33 \) \\
			\( 100 \) & \( 2.04 \) & \( 5000 \) & \( 2.84 \) &  &  \\
			\bottomrule
		\end{tabular}
	}
\end{table}

Jeffreys then explains the consequences of this sample-size induced increase of the critical ratio, and explicitly discusses the paradox: 
\begin{quotation}
``The usual practice has been to regard a departure from a simple law as genuine if it amounts to some constant multiple of the standard error, usually 2 or 3 times. \emph{The ratio given above is not constant, but depends on the number of observations.} If a ratio of 2 or 3 is really needed when the number is small, it expresses a prior belief in the simple law to the extent of saying that the odds in its favour are 6 to 1 or 90 to 1, or else a criterion of convenience that we must not complicate future computations except for specially strong reasons. In either case corresponding, but smaller, increases would be needed throughout the table. \emph{When the number of observations is large the critical ratio exceeds the arbitrary standard, which will thus for $100,000$ observations lead to coefficients between 2 and $3.33$ times their standard errors being accepted as genuine, when in fact the observations render them less probable than before.} Thus there will be mistakes in all cases where there is no real departure and yet the computed departure is between 2 and $3.33$ times its standard error. Fortunately the latter event does not occur very often; nevertheless it has arisen.'' (\citealp[pp. 353-354]{Jeffreys1936}; italics added for emphasis)
\end{quotation}

Jeffreys concludes the article by demonstrating and explaining the paradox in the field of astronomy with a concrete example.\footnote{This example also features in later papers, discussed below.} In a regression model for the motion of the node of Venus, there were $12,319$ observations. The Bayes factor is about 6 in favor of $\mathcal{H}_0$. However,
\begin{quotation}
``On the usual theory the probability of an accidental variation exceeding $3.5$ times its standard error is $4 \times 10^{-4}$, and the anomaly would have to be taken as real. Such a value will in any case be exceptional, but with the actual number and accuracy of the observations it is more exceptional on the hypothesis that it is real than on the hypothesis that it is due to accidental error.'' \citep[p. 445]{Jeffreys1936}
\end{quotation}

\subsection{4. The 1936 article \emph{Further significance tests}}
In the same year Jeffreys again stresses the same issue:
\begin{quotation}
``The results are usually of the form $\alpha n^{\frac{1}{2}}\exp(-\frac{1}{2}x^2/\sigma^2)$, where $n$ is the number of observations and $x$ is the difference found statistically, which may be a difference of two sampling ratios or measurements, a correlation or a harmonic coefficient. $\sigma$ is the standard error of $x$ as found from the usual statistical theories. $\alpha$ is usually a moderate coefficient. The form of the results can be explained simply in general terms. Suppose that the difference which we are trying to find might have had any value within a range $m$. It is actually found to be within a certain small range of length $\tau$ about $x$. Then, on the hypothesis that there is a real difference, the probability that the results would be in this range is $\tau/m$. But on the hypothesis that there is no real difference the corresponding probability is $\tau(2\pi\sigma^2)^{-\frac{1}{2}} \exp(-\frac{1}{2}x^2/\sigma^2)$. Hence by the theorem of inverse probability the probabilities of no real difference and of a real difference are in the ratio $(m/\sigma)(2\pi)^{-\frac{1}{2}}\exp(-\frac{1}{2}x^2/\sigma^2)$. But if the accuracy of the observations remains constant the standard error of the mean decreases like $n^{-\frac{1}{2}}$; hence the outside factor is of order $n^{\frac{1}{2}}$. (...) 

To put the matter in other words, if an observed difference is found to be of order $\sigma$, then on the hypothesis that there is no real difference this is what would be expected; but if there was a real difference that might have been anywhere within a range $m$ it is a remarkable coincidence that it should have happened to be in just this particular stretch near zero. On the other hand if the observed difference is several times its standard error it is very unlikely to have occurred if there was no real difference, but it is as likely as ever to have occurred if there was a real difference. In this case beyond a certain value of $x$ the more remarkable coincidence is for the hypothesis of no real difference, and as we have to decide from the facts as presented we shall accept the difference. The theory merely develops these elementary considerations quantitatively and evaluates the factor $\alpha$. If $P(q\given \theta h)> \frac{1}{2}$, we shall expect the difference found to persist with more and more accurate observations; if it is less than $\frac{1}{2}$ we shall expect the estimated difference to diminish.

\emph{The usual statistical method is to evaluate the observed difference and its standard error, and to say that it is not significant if it is less than a certain constant multiple of this error. No explanation of this rule is given, the probability of the observations being found only on the hypothesis that there is no difference, and not compared with that on the alternative hypothesis. The present method provides an explanation; but the multiple found is not constant, depending on the number of observations} and on the ratio of the standard error of one observation to the whole difference possible, but since it involves these numbers only through the square roots of their logarithms the variation in actual cases is not very large.
'' (\citealp[p. 417]{Jeffreys1936Further}; italics added for emphasis)
\end{quotation}
These quotations show that in 1935 and 1936, Jeffreys had already discovered, understood, published, emphasized, explained, and illustrated the paradox.  

\subsection{5. The 1937 article \emph{The tests for sampling differences and contingency}}
In this article Jeffreys's final paragraph again describes the phenomenon:
\begin{quotation}
``Attention is called to the fact that in my tests the ratio of the critical value of a difference to the standard error of the latter varies a little with the number of observations. A difference of twice the standard error may be just significant [in the sense of Jeffreys's Bayes factor test -- EWAL] when it rests on five observations, but not when it rests on 100. For application of the tests it is therefore necessary to know the number of observations, and in many cases this is not given explicitly in published work and can be disentangled with great difficulty, if at all. In other words a difference of $1.0 \pm 0.5$ units may be worth considering further if it rests on five observations each with a standard error of $1.2$ units; if it rests on 100 observations each with a standard error of 5 units it is not. This comes from pure probability theory and does not allow for the possibility of systematic error of observation, which might be considered at a later stage and would accentuate the effect.'' \citep[p. 494]{Jeffreys1937Tests}
\end{quotation}

\subsection{6. The 1937 addenda to the first edition of \emph{Scientific inference}}
Jeffreys's book \emph{Scientific Inference} first appeared in 1931, before Jeffreys had started to work on Bayes factors in earnest. A 1937 reissue \emph{Scientific Inference}, however, contains addenda that describe the Bayes factor hypothesis test and a description of the reasoning that underpins the paradox (cf. \citealp{Jeffreys1936Further} above):
\begin{quotation}
``Suppose we consider as a serious possibility that a quantity $x$ may be zero; denote this proposition by $q$, with prior probability $\frac{1}{2}$. The proposition that $x$ is not zero is denoted by $\sim \! q$, also with prior probability $\frac{1}{2}$; but if $x$ is not zero it may be anywhere in a range of length $m$. An actual determination from data $\theta$ suggests a value of $x_0 \pm \sigma$. Now, if $x$ is really 0, the probability of finding a mean in a range $dx_0$ about $dx_0$ is $\frac{1}{\sqrt{(2\pi)} \sigma} \exp\left(-\frac{x_0^2}{2\sigma^2} \right)dx_0$. But if $x$ is not 0, the probability that it would be in such a range is $dx_0/m$. Given then that $x_0$ has actually been found in such a range, the posterior probabilities of $q$ and $\sim \! q$ are in the ratio of these two expressions, namely
\begin{equation*}
        \frac{P(q \mid \theta h)}{P(\sim \! q \mid \theta h)} = \frac{m}{\sqrt{(2\pi)}\sigma} \exp\left(-\frac{x_0^2}{2\sigma^2}\right).
\end{equation*}
When $x_0$ is large compared with $\sigma$, this is small, $q$ has a small posterior probability, and we can assert with confidence that $x$ is different from zero. But $\sigma$, the standard error of the mean, is proportional to $n^{-\frac{1}{2}}$, where $n$ is the number of observations; hence if $n$ is large the first factor is large of order $\sqrt{n}$, and the ratio will be large if $x_0$ is less than $\sigma$. Thus a discrepancy less than a certain amount increases the probability that the parameter sought is zero; one more than this amount decreases it and indicates that the parameter is needed. In the cases examined the critical value, with ordinary numbers of observations, ranges from about $1.5$ to 3 times the standard error, increasing with the number of observations. The larger the number of observations the stronger the support for the simple law $x=0$ if the empirical value turns out to be within its standard error. To put the argument in words, if $x_0$ is of order $\sigma$, this is what we should expect if $x$ is zero, but if $x$ might be anywhere in a range $m$ it is a remarkable coincidence that it should be in just this one. On the other hand, if $x_0$ is substantially more than $\sigma$, we should not expect it if $x$ is zero, but we should expect it if $x$ is not zero; in both cases we adopt the less remarkable coincidence.'' \citep[pp. 250-251]{Jeffreys1937SI}
\end{quotation}

\noindent A few pages later, Jeffreys provides a concrete example of the paradox (cf. \citealp[p. 445]{Jeffreys1936} above):
\begin{quotation}
``In current statistical practice the word ``significance'' appears to be used in several different senses, corresponding to different questions, but it is apparently often supposed that they will have the same answers. I have used it in the case where we want to know whether the observations support a new parameter; this is one that regularly occurs, for instance, in astronomy. The multiple of the standard error used to indicate a statistical difference is about the same as my theory gives for ordinary numbers of observations, but it is taken constant. \emph{My fuller theory shows that it should increase somewhat with the number of observations.} I have only once come upon a case where the difference between the criteria would affect the decision, namely the excess motion of the node of Venus, which, if genuine, is inconsistent with Einstein's law of gravitation. It is $3.5$ times the standard error, and by the usual rules would have to be taken as real. But the number of observations used is so large that by my rule it is even more likely to be a random error. In fact Sir Arthur Eddington, who does not accept the theory of probability, adopted the decision it gives and not that given by his own theory.'' (\citealp[p. 256]{Jeffreys1937SI}, italics added for emphasis)
\end{quotation}

\noindent Jeffreys then draws the explicit comparison to $p$-values:
\begin{quotation}
``A constant significance limit, in relation to the standard error, would however be equivalent to saying that the prior probability of a zero value varies with the number of observations, which is absurd; or, alternatively, that the chance of a real difference exceeding the standard error is the same no matter how small the standard error is made by increasing the number of observations. Actually, however, my significance limit varies very slowly with the number of observations and with ordinary numbers does not differ much from Fisher's limits based on the arbitrary 5 per cent. and 1 per cent.; in the great majority of actual cases the decisions will be the same. Accordingly it appears that Fisher's practice does not follow from his postulates, but it, or something very like it, follows from mine.'' \citep[p. 259]{Jeffreys1937SI}
\end{quotation}
It is noteworthy that the two ``absurdities'' that Jeffreys identifies in this fragment (i.e., as $n$ increases, either lower the probability of $\mathcal{H}_0$ or narrow the prior parameter distribution under $\mathcal{H}_1$) would later be proposed by \citet{Robert1993} (see also \citealp{BurnhamAnderson2004}) and \citet{Bartlett1957}, respectively. 

\subsection{7. The 1937 correspondence with Fisher}
The sample-size induced discrepancy between Bayes factor and $p$-values was also noted explicitly in a 1937 letter that Jeffreys wrote to Fisher (note that this example was also presented in \citealp[p. 445]{Jeffreys1936} and in \citealp[p. 256]{Jeffreys1937SI}, as discussed above):
\begin{quotation}
``A question has just arisen about the excess motion of the node of Venus. It is $3.5$ times the standard error, the probability of a random deviation exceeding which is $0.00041$. Eddington says that as it is one of 15 it can be accepted as normal. The p$.$ that one of 15 would exceed $3.5\sigma$ is $0.006$. What I should like to know from you is whether there is another case on record where a statistician has accepted at sight a deviation beyond your 1\% limit as random? (The other 14 give a $\chi^2$ of 15).

By my test the thing is probably random on account of the large number of observations combined, but there's not much to spare, and the situation would be altered if some \emph{specific} systematic error was before the House.''

Sir Harold Jeffreys, in a letter to Sir Ronald Fisher, 1937 (\citealp[p. 161]{Bennett1990}; italics in original)
\end{quotation}

\noindent Later that year, Fisher replied as follows:

\begin{quotation}
``I should be inclined, naturally, to accept Eddington's judgement on an astronomical point, especially as your own test seems to confirm it. On the other hand, \emph{prima facie}, i.e. on an assumption ordinarily made, the probability $0.006$ is amply small enough to claim significance, and would be used for this purpose with complete confidence, I have no doubt, if anyone had a theory which required such a deviation.''

Sir Ronald Fisher, in a letter to Sir Harold Jeffreys, 1937 (\citealp[p. 162]{Bennett1990}; italics in original)
\end{quotation}

Fisher's answer is somewhat ambiguous, but it does appear as if he believed a $p$-value of $.006$ to be sufficiently compelling for declaring a deviation significant, regardless of sample size. Instead of pushing Fisher on the issue, Jeffreys's response strikes a conciliatory tone:
\begin{quotation}
``Your letter confirms my previous impression that it would only be once in a blue moon that we would disagree about the inference to be drawn in any particular case, and that in the exceptional cases we would both be a bit doubtful. (...)

I am writing this because there is a tendency about to attribute what I believe to be an entirely exaggerated idea of our disagreement to us, for which we are both possibly partly responsible, and I think an occasional mention of cases where we agree would be for the good of the subject.''

Sir Harold Jeffreys, in a letter to Sir Ronald Fisher, 1937 (\citealp[pp. 162-163]{Bennett1990}; italics in original)\footnote{In a 1983 interview with Dennis Lindley, Jeffreys referred to this exchange as follows: ``[the correspondence with Fisher] was after I'd said that on most things we should agree and when we disagreed we would both be doubtful. After that, Fisher and I were great friends.'' (``Transcription of a Conversation between Sir Harold Jeffreys and Professor D.V. Lindley,'' Exhibit A25, St John's College Library, Papers of Sir Harold Jeffreys).}
\end{quotation}

\subsection{8. The 1937 article \emph{Modern Aristotelianism: Contribution to Discussion}}

In this one-page discussion on the role of induction in science, Jeffreys mentions the common elements in the statistical frameworks advocated by Karl Pearson and Ronald Fisher, and then states:
\begin{quotation}
``I should expect the decisions by my methods to lead to the correct decisions most rapidly, because the method contains more explicit provision for allowing for the whole of the data; but many rules given by Fisher, and others accepted by him, are of exactly the same form as mine [EWAL: point estimates] and would in practice be used in the same way, while in other cases where there are differences [EWAL: Bayes factors vs. $p$-values] the actual limits recommended are such that it would be extremely rarely that the decisions would differ in any specific application, and then we should both be doubtful.'' \citep[p. 1004]{Jeffreys1937Nature}
\end{quotation}

As in the 1935 article, Jeffreys downplays the practical ramifications of the paradox -- a theme that will recur in the appendix of Jeffreys's book \emph{Theory of Probability}. In later sections we speculate about Jeffreys's reasons for doing so.  

\subsection{9. The 1938 article \emph{The comparison of series of measures on different hypotheses concerning the standard errors}}
In this article \citet[p. 378]{Jeffreys1938Comparison2} gives the Bayes factor in the case of a $t$-test:
\begin{equation*}
\begin{split}
    K =& \left(\frac{2n}{\pi}\right)^{\frac{1}{2}} \, \left(1+\frac{\bar{x}^2}{\sigma^2}\right)^{-\frac{1}{2}(n-3)}\\
    =& \left(\frac{2n}{\pi}\right)^{\frac{1}{2}} \, \left(1+\frac{t^2}{n-1}\right)^{-\frac{1}{2}(n-3)},
\end{split}
\end{equation*}
as $t^2 = (n-1)\bar{x}^2/\sigma^2$. This is followed by a table that shows the values of $K$ associated with Fisher's 5\% values of $t$ for various sample sizes $n$, reproduced here as Table~\ref{tableJeffreys38}. 
\begin{table}[h]
	\centering
	\caption{Table reproduced from \citealp[p. 379]{Jeffreys1938Comparison2}.}
	\label{tableJeffreys38}
	{
		\begin{tabular}{ccccc}
			\( n \) (Fisher's \( n+1 \)) & \( K  \) & \( n \) (Fisher's \( n+1 \)) & \( K  \) \\
			\( 5 \) & \( 0.610 \) & \( 9 \) & \( 0.519 \)  \\
			\( 6 \) & \( 0.551 \) & \( 10 \) & \( 0.522 \)  \\
			\( 7 \) & \( 0.529 \) & \( 20 \) & \( 0.612 \) \\
			\( 8 \) & \( 0.520 \)  & \( 30 \) & \( 0.719 \) \\
		\end{tabular}
	}
\end{table}

Jeffreys then mentions the paradox:
\begin{quotation}
``\noindent For the first few entries my formula may be appreciably inaccurate, but for $n=8$ and more it should be fairly good. It appears therefore that the 5\% point of the $t$ distribution never corresponds to a value of $K$ less than about $0.5$, or to 2 to 1 odds on the need for the new parameter. If we are entitled to interpret this as indicating at what value of $K$ we may consider a new parameter as worth introducing, the value should be about $0.5$; but there will then be just about as much confidence in the need for it as in a statement that an estimate of a parameter, whose relevance is not in doubt, is right within its standard error.

The inequality is reversed at large numbers of observations; thus for $K=1$ and large $n$ we have the approximation
\begin{equation*}
    t^2 = \log_e \, 2n/\pi,
\end{equation*}
whereas the 5\% point of the $t$ distribution tends to $t=1.96$. The properties of the logarithm make the rise very slow; when $n=100,000$, $t$ is still only $3.32$. \emph{But if the 5\% rule was used habitually there would be cases, with large numbers of observations, when a new parameter is asserted on evidence that is actually against it.} Users of the rule usually advocate it with considerable caution, which would agree with the indications of the present theory up to about 30 observations, but at large numbers it is definitely too lax.'' (\citealp[pp. 379]{Jeffreys1938Comparison2}; italics added for emphasis)
\end{quotation}

\noindent Jeffreys then explains,
\begin{quotation}
``It may be worth while to call attention again to the reason for the increase of $t$ and its analogues in other tests when the number of observations is very large. If we start with the minimum of information about the new parameter, which is quite likely to be zero but might account for most of the outstanding variation until we have actually analysed the data, then as we increase the number of observations the standard error of the estimate steadily falls. If the parameter is not zero, however, it is independent of the number of observations, and will ultimately become several times the standard error of its estimate and asserted to be genuine. If the estimate persists within the order of magnitude of its standard error, our confidence that this is because the parameter is really zero will naturally increase, on the ground that with a large number of observations it is increasingly unlikely that we should have failed to find it if it was there. This of course is a well-known phenomenon in physics, where an estimated difference, always in doubt, strengthens that doubt by diminishing every time the number of observations is increased or the experimental technique improved; and it is represented in the present theory by the increase of the outside factor in $K$. When the number of observations is small, this factor is not much more than 1, and it is impossible to obtain strong support for $q$ however well the observations may agree with it; and in sampling problems in similar conditions it is also impossible to obtain strong support for $\sim \! q$. It may be recalled that in the problem of sampling to test an even chance it took an $80:80$ sample to give 10 to 1 support for $q$ and a $7:0$ one to give 10 to 1 support for $\sim \! q$. It is in such cases that we say that there is not enough evidence to make a decision, and any definite rule will make a considerable number of mistakes of one kind or the other. Mathematically, the ratio of the estimate to its standard error must increase with the number of observations because it has to counteract this factor to reduce $K$ to any fixed value. In general terms, it must increase because the number of cases where $q$ is still acceptable remains the same, but those where it is untrue and its falsehood still undetected become fewer. (I am not here considering cases where selection of an extreme value, or previous knowledge indicating a restriction on the possible values of a new parameter, needs to be taken into account; they only complicate the matter without altering the general principle.)''
\citep[pp. 379-380]{Jeffreys1938Comparison2}
\end{quotation}

\subsection{10. The 1938 article \emph{Significance tests when several degrees of freedom arise simultaneously}}
Here Jeffreys first describes the Bayes factor and immediately points out its dependence on sample size:
\begin{quotation}
``If a set of observations are analysed for a new parameter $a$, which is initially as likely as not to be zero, and the possible range of whose values is $s$ if it is not zero, we can denote the proposition that it is 0 by $q$, and the proposition that it is not 0 by $\sim \! q$. [EWAL: Here Jeffreys inserts the following footnote: ``My $q$ is always what Fisher (1935) calls a ``null hypothesis''.''\nocite{Fisher1935b}] Then the prior probabilities of $q$ and $\sim \! q$ are given by 
\begin{equation} \tag{1}
    P(q \mid h) = P(\sim \! q \mid h) = \tfrac{1}{2},
\end{equation}
and the posterior probabilities on data $\theta$ are shown, by an approximate argument (Jeffreys 1937b, p. 250 [EWAL: This refers to the fragment from \emph{Scientific Inference} provided earlier]), to be given by
\begin{equation} \tag{2}
    K = \frac{P(q \mid \theta h)}{P(\sim \! q \mid \theta h)} \Big/ \frac{P(q \mid h)}{P(\sim \! q \mid h)} = \frac{s}{\sqrt(2\pi)\sigma_\alpha} \exp\left(-\frac{\alpha^2}{2\sigma_\alpha^2}\right),
\end{equation}
where $\alpha$ is the maximum likelihood solution for $a$ and $\sigma_\alpha$ its standard error. Since $s$ is initially fixed and $\sigma_\alpha$ decreases like $n^{-\frac{1}{2}}$ when $n$, the number of observations, increases, the outside factor is proportional to $\sqrt n$. If $K$ is less than 1, the observations support the introduction of the new parameter; if $K$ is more than 1 they do not. In the cases so far examined the critical value of $\alpha/\sigma_\alpha$ ranges from about $1.8$ to 3 as the number of observations rises from 5 to 5000.'' 
\citep[p. 161]{Jeffreys1938}
\end{quotation}

Later in the same article, the dependence of the Bayes factor on sample size (as $\sqrt{n}$) plays a crucial role. For instance, on p. 164 Jeffreys remarks that ``the outside factor in the support for $q$ is of order $n^{\frac{1}{2}}$; this factor would be the support provided if the estimates happened to agree exactly with the predictions made by $q$.'' (see also p. 172). However, in this article Jeffreys does not engage in an explicit comparison between Bayes factors and $p$-values.

\subsection{11. The 1938 article \emph{Maximum likelihood, inverse probability and the method of moments}}
In this article Jeffreys hints at the paradox but underplays its practical importance:
\begin{quotation}
``(...) a moderate fraction of the prior probability of $a$ [i.e., a parameter] is concentrated in a particular value $a_0$. This is the case where a possible value of $a$ is already assigned and the observations are to be used to test whether this value is correct. (...) The result, which I had hardly expected to find, was that if $\alpha - a_0$ is less than a certain multiple of $\sigma_a$ (varying somewhat with $n$ and the type of problem), the observations increase the probability that $a$ is equal to $a_0$. This connects up significance tests with the principle of inverse probability, but the results do not differ greatly from those that statisticians have found to work well in practice. The relation to the method of maximum likelihood is that the apparently arbitrary rejection of small differences found by that method is now explained in terms of the general theory.'' \citep[p. 148]{Jeffreys1938MLE}
\end{quotation}

\subsection{12. The 1938 article \emph{Significance tests for continuous departures from suggested distributions of chance}}

This article features a more explicit comparison to $p$-values. Here Jeffreys sets out to test the null hypothesis that a set of frequencies are uniformly distributed. He arrives at the familiar $\sqrt{n}$ form of his test and then engages explicitly with the paradox:
\begin{quotation}
``Hence (...) 
\begin{equation*} \tag{15}
    K = \frac{P(q \mid \theta h)}{P(\sim \! q \mid \theta h)} = \sqrt{ \left(\frac{n}{2\pi}\right)} c \exp(-\tfrac{1}{2}na_0^2).
\end{equation*}
The term in $f(t)$ will therefore be supported if $a_0$ [the MLE -- EWAL] is such as to make this less than 1. The standard error of $a_0$, in this notation, is $n^{-\frac{1}{2}}$, so that the exponential factor has the usual form $\exp(-\tfrac{1}{2}\chi^2)$.

The following table, for various values of $n$, gives $K$ for $a_0=0$ for the two values of $c$, and the values of $\chi^2$ and $a_0 n^{\frac{1}{2}}$ that make $K = 1$. For comparison we may notice that Fisher's (1936, Table III) \nocite{Fisher1936} 5~\% and 1~\% limits, for one degree of freedom, are at $\chi^2 = 3.84$ and $6.64$; the former would agree in the first case at about 200 observations, the latter at about 4000. In the second case the agreements would come at about 100 and 1700 observations. His test, of course, does not mean quite the same thing; it says when an observed result would be surprising on hypothesis $q$, whereas mine, for the larger numbers of observations, may admit this and yet say that it would be still more surprising on $\sim \! q$. In any event cases where the observed $a_0$ would come in the disputable region would be expected to be rare if \emph{either} of the hypotheses compared was correct, and some third alternative may suggest itself.'' (\citealp[p. 310]{Jeffreys1938Continuous}; italics in original; table reproduced as Table~\ref{tableJeffreys38d})
\end{quotation}

\begin{table}[h]
	\centering
	\caption{Table reproduced from \citealp[pp. 310]{Jeffreys1938Continuous}. Increases in sample size $n$ need to be accompanied by increases in the $\chi^2$ value so that the Bayes factor $K$ remains constant at $K=1$.}
	\label{tableJeffreys38d}
	{
		\begin{tabular}{rrrrrrr}
			\multicolumn{1}{c}{\( %
			\begingroup
			\color{white} 
			\overbrace{\hspace{35pt}}^{\textstyle \textcolor{black}{n}} 
			\endgroup %
			\) %
			} & \multicolumn{2}{c}{\( \overbrace{\hspace{35pt}}^{\textstyle K} \)} & \multicolumn{2}{c}{ \( \overbrace{\hspace{35pt}}^{\textstyle a_{0} n^{\tfrac{1}{2}}} \)} & \multicolumn{2}{c}{\( \overbrace{\hspace{35pt}}^{\textstyle \chi^{2}} \)} \\
			5 & 1.03 & 1.55 & 0.25 & 0.94 & 0.06 & 0.88  \\
			10 & 1.46 & 2.19 & 0.87 & 1.25 & 0.76 & 1.57  \\
			20 & 2.06 & 3.09 & 1.20 & 1.50 & 1.45 & 2.26  \\
			50 & 3.26 & 4.89 & 1.54 & 1.78 & 2.36 & 3.17  \\
			100 & 4.61 & 6.92 & 1.75 & 1.97 & 3.06 & 3.87  \\
			200 & 6.51 & 9.76 & 1.94 & 2.13 & 3.75 & 4.56  \\
			500 & 10.31 & 15.46 & 2.16 & 2.34 & 4.67 & 5.48  \\
			1,000 & 14.6 \, & 21.9 \, & 2.32 & 2.48 & 5.36 & 6.17  \\
			2,000 & 20.6 \, & 30.9 \, & 2.46 & 2.62 & 6.05 & 6.86  \\
			5,000 & 32.6 \, & 48.9 \, & 2.46 & 2.79 & 6.97 & 7.78  \\
			10,000 & 46.1 \, & 69.2 \, &2.77 & 2.86 & 7.66 & 8.19  \\
		\end{tabular}
	}
\end{table}

\subsection{\rev{13. The 1938 article \emph{Aftershocks and periodicity in earthquakes}}}
\rev{In this article Jeffreys studies the hypothesis that earthquakes are independent events that do not excite one another. Jeffreys first elaborates on standard practice:}

\begin{quotation}
\rev{``The usual statistical procedure, recommended in particular by R. A. Fisher, is to reject the trial hypothesis if the contribution to $\chi^2$ examined is such that the probability of a larger $\chi^2$, if the hypothesis was correct, is less than $0.05$; if high confidence is required the trial hypothesis will be rejected (and correspondingly the modified one accepted) if this probability is less than $0.01$. The former criterion is somewhat mild, since it would imply the acceptance as genuine of all discrepancies more than twice their standard errors.'' \citep[p. 114]{Jeffreys1938aftershock}}
\end{quotation}

\rev{Jeffreys then describes his own significance test and mentions the paradox:}
\begin{quotation}
\rev{``The type of significance test that I have introduced depends on the general theory of probability; the observed values appear in the results only through the contributions to $\chi^2$ from the degrees of freedom actually considered, so that the tests provide an explanation of the importance of $\chi^2$, which was introduced somewhat arbitrarily by Pearson, though it has properties of symmetry under transformation that would make it commendable by themselves. For a given number of degrees of freedom, the value of $\chi^2$ that makes it more probable than not, on the data, that a new parameter or set of parameters is required, is found to vary somewhat with the number of observations$^2)$. In the case of a periodicity inferred from observed frequencies, I find that for 200 observations the periodicity is just supported if $\chi^2 = 7.1$; for 500, 8.3; for 1000, 8.9. To establish a 10 to 1 probability that the periodicity is genuine these values must be increased by 5.1. Fisher's 5 per cent. limit for two degrees of freedom is at $\chi^2 = 5.99$, his 1 per cent. one at 9.21. With these numbers of observations his 5 per cent. criterion would therefore sometimes accept a periodicity that mine would reject, though the agreement is good at somewhat smaller values.'' \citep[p. 114]{Jeffreys1938aftershock}} 
\end{quotation}

\rev{Jeffreys's footnote 2 lists several of his earlier works in which the paradox is evident:}
\begin{quotation}
\rev{``$^2)$ Scientific Inference, \textbf{1937}, 249--252 and 266--9 (General Discussion and Summary of Results). -- Proc. Camb. Philos. Soc. \textbf{31} (1935) 213--217 (Correlation). -- Proc. Camb. Philos. Soc. \textbf{32} (1936) 432--445 (Representation of a Series of Measures by Assigned Functions and Tests of Randomness). -- Proc. Camb. Philos. Soc. \textbf{33} (1937) 35--40 (Comparison of Series of Measures). -- Proc. Roy. Soc. London (A) \textbf{162} (1937) 479--495 (Contingency and tests for agreement of Sampling Ratios. Improved discussions of some problems treated in earlier papers are given.)'' \citep[p. 114]{Jeffreys1938aftershock}}  
\end{quotation}

\subsection{\rev{14}. The 1939 first edition of \emph{Theory of Probability}}
The first edition of Jeffreys's magnum opus \emph{Theory of Probability} describes a scenario similar to that covered in the addenda of the 1937 reissue of \emph{Scientific Inference}. Specifically, Jeffreys introduces the Bayesian hypothesis test by defining the null hypothesis $q$ and the alternative hypothesis $\sim \! q$. Under $\sim \! q$, there is a new parameter $\alpha$. Let $m$ denote the possible range of values for $\alpha$ about 0 within which the prior probability may be taken as uniformly distributed, and let $a$ denote the maximum likelihood estimate and $s$ its standard error. Then, if $s$ is much smaller than $m$, Jeffreys approximates the Bayes factor $K$ (i.e., $\text{BF}_{01}$) as 
\begin{equation*}
    K = \frac{P(q \mid aH)}{P(\sim \! q \mid aH)} = \frac{m}{\sqrt{(2\pi)}s} \exp\left(-\frac{a^2}{2s^2}\right),
\end{equation*}
where $H$ indicates background knowledge. Jeffreys then continues:
\begin{quotation}
``If $a$ is $s$ or less, and $s$ is much less than $m$, $K$ will be large and the observations support $q$, that is, they say that the parameter $\alpha$ is probably not needed. But if $a$ is much larger than $s$, the exponential will be very small and the observations will support the need for the new parameter. There will be a critical value of $a/s$ such that $K=1$ and no decision is reached.

In most cases $s$, being the standard error of $a$, diminishes with increasing $n$ like $n^{-\nicefrac{1}{2}}$; hence the first factor in $K$ increases like $n^{\nicefrac{1}{2}}$. Thus the larger the number of observations the stronger the support for $q$ will be if $a<s$. This is a satisfactory feature; the more thorough the investigation has been, the more ready we shall be to suppose that if we have failed to find evidence for $\alpha$ it is because $\alpha$ is really 0. But it carries with it the consequence that the critical value of $a/s$ increases with $n$ (though that of $a$ of course diminishes); the increase is very slow, since it depends on $\sqrt{(\log{n})}$, but it is appreciable. The test does not draw the line at a fixed value of $a/s$. (\citealp[p. 194]{Jeffreys1939}; echoed in \citealp[pp. 221-222]{Jeffreys1948} and \citealp[p. 248]{Jeffreys1961}) 
\end{quotation}

In Appendix I, Jeffreys again explicitly compares the Bayes factor against the $p$-value. Jeffreys concludes:
\begin{quotation}
``In spite of the difference between the nature of my tests and those based on the $P$ integrals, and the omission of the latter to give the increases of the critical values for large $n$ (dictated essentially by the fact that in testing a small departure found from a large number of observations we are selecting a value out of a long range and should allow for selection), it appears that there is not much difference in the practical recommendations. Users of these tests speak of the 5 per cent. point in much the same way as I should speak of the $K=10^{-\nicefrac{1}{2}}$ point, and of the 1 per cent. point as I should speak of the $K=10^{-1}$ point; and for moderate numbers of observations the points are not very different. At large numbers of observations there is a difference, since the tests based on the integral would sometimes assert significance at departures that would actually give $K>1$. Thus there may be opposite decisions in such cases. But they will be very rare.'' (\citealp[pp. 359-360]{Jeffreys1939}; echoed in \citealp[p. 399]{Jeffreys1948} and \citealp[p. 435]{Jeffreys1961})
\end{quotation}
\noindent Appendix I then concludes with four tables associated with different statistical scenarios. Each table shows that a constant level of Bayes factor support requires that larger sample sizes yield a higher multiple of the standard error.  

\subsection{\rev{15}. The 1940 article \emph{Note on the Behrens-Fisher formula}}
In this article Jeffreys briefly outlines his hypothesis test and adds that the threshold for accepting the alternative hypothesis is not a constant `as usually defined':
\begin{quotation}
``A definite limit is then found for $z$, such that larger values support the need for the new parameter while smaller ones support the null hypothesis, but this limit is not given by any single value of $P(t)$ as usually defined.'' \citep[p. 49]{Jeffreys1940}
\end{quotation}

\subsection{\rev{16}. The 1942 article \emph{On the significance tests for the introduction of new functions to represent measures}}
In this article Jeffreys once more emphasizes the dependence of the Bayes factor $K$ on sample size. After providing the equation for $K$ in its familiar form, Jeffreys provides a table that shows how $K$ increases with $n$ when $t$ is fixed at 0, and how $t^2$ increases with $n$ when $K$ is fixed at 1. Jeffreys remarks,
\begin{quotation}
``It is interesting that the values of $t^2$ for $K = 1$ increase steadily with $n$, just as the corresponding values of $\chi^2$ do. This of course is the level where the test is quite indecisive.'' \citep[p. 260]{Jeffreys1942Functions}
\end{quotation}

\subsection{\rev{17}. The 1948 second edition of \emph{Theory of Probability}}

Although this second edition is 31 pages longer than the 380-page first edition, the paradox-related content (i.e., pp. 221-222, p. 399) has remained mostly unchanged, except for a small change in notation and for a partly adjusted and expanded set of tables in the appendix.

\subsection{\rev{18}. The 1950 article \emph{Bertrand Russell on Probability}}
In this article Jeffreys describes his generic Bayes factor, including the $\sqrt{n}$ term that exposes the paradox:
\begin{quotation}
``But if we are at liberty to modify a law arbitrarily to any extent we can fit any set of observations exactly, and some of these possibilities would fit any further observation whatever; consequently if there is no limitation on the choice of laws no prediction from observations is possible. (...) [a solution] is given in my \emph{Theory of Probability}, Chapters 5 and 6. This is that where a suggested modification of a law involves an increase in the number of adjustable parameters, half the prior probability is concentrated in the old law; in other words, when a modification is suggested it is as likely to be needed as not. This has been shown to lead to satisfactory significance tests in the standard problems of statistics, though there is much more to be done. The results are of the approximate form
\begin{equation*}
    \frac{\text{P}(q/\theta H)}{\text{P}(q'/\theta H)} = \sqrt (An) \text{e}^{-\frac{1}{2}a^2/s_a^2}
\end{equation*}
Here if the new parameter considered is $\alpha$, it is defined so as to be zero on the old law $q$, but on the modified law $q'$ it has to be estimated from the observations; $H$ is the previous information and $\theta$ the observational evidence. $A$ is a constant usually of order 1, $n$ the number of observations, $a$ the estimate of $\alpha$ by the usual statistical methods, and $s_a$ its standard error. The expression is of order $\sqrt n$ if $a/s_a$ is less than 1, but very small if $a/s_a$ is large. Consequently observations support the old law for $a/s_a < 1$ and the new one if it is large. This choice of the prior probability is what I call the simplicity postulate.'' (\citealp[p. 316]{Jeffreys1950Russell}; italics in original)
\end{quotation}

\subsection{\rev{19}. The 1953 comment on Lindley's article \emph{Statistical inference}}

Historically, the 1953 Lindley article \emph{Statistical inference} is particularly relevant, as it can be considered the conceptual forerunner to the 1957 paradox article. Inspired by the work of Abraham Wald, Lindley studied statistical procedures that minimize a weighted sum of Type I and Type II errors.\footnote{At this time Lindley was still a frequentist, as witness statements such as ``...the use of inverse probability solutions as a \emph{general} rule can hardly be considered satisfactory, though in special circumstances they may be adequate.'' (\citealp[p. 45]{Lindley1953}; see also \citealp{Fienberg2003}).} Lindley showed that for consistency to hold regardless of the weight assigned to the errors, the critical value has to increase with sample size: ``...the critical value (...) increases with $n$, although very slowly. In this it agrees with the test proposed by Jeffreys (1948).'' \citep[p. 60]{Lindley1953}.

In a comment published alongside Lindley's original article, Jeffreys elaborates on the agreement:
\begin{quotation}
The appearance of log $n$ [in Lindley's tests -- EWAL] is interesting in relation to my significance tests. At first sight the origins of this term look quite different, since in mine \emph{it expresses an allowance for selection; we reasonably discount an exceptional result if we have looked specially hard for one.} In Mr. Lindley's it is an allowance for the cost of installing a new plant when the benefit would be small.

It is easy to see, however, that a similarity might have been expected. If the prior probability distribution for a parameter $\mu$ is $P(d\mu \given H) = f(\mu) \,d\mu$, the likelihood of a set of data $\theta$ is $L(\mu,\theta)$, and the benefits of two courses of action, depending on $\mu$, are $K_1(\mu)$, $K_2(\mu)$, the posterior probability distribution of $\mu$ is $P(d\mu \given \theta H) \propto f(\mu) L(\mu,\theta) \, d\mu$, and the expectations of benefit are $\int K_1(\mu)f(\mu)L(\mu,\theta) \,d\mu$, $\int K_2(\mu)f(\mu)L(\mu,\theta) \,d\mu$. Thus $K$ enters in combination with $f$, as Mr. Lindley finds. This might have been expected, since Bayes defined probabilities in terms of ratios of expectations of benefits, and in an economic application $K$ and $f$ will always be combined.'' (\citealp[p. 72]{Jeffreys1953}; italics added for emphasis)
\end{quotation}

Lindley then replied to Jeffreys as follows:
\begin{quotation}
``His connection between the log $n$ term in our two derivations is most interesting, and in conjunction with his statement that, in some circumstances, one should maximize the expected benefit, it makes me realize that my ideas on inference are much closer to Professor Jeffreys' than I had thought. \citep[p. 76]{Lindley1953}
\end{quotation}

It should not go unmentioned that, in a different comment, Lindley's contribution was evaluated positively by Egon Pearson himself:
\begin{quotation}
``We see at once the practical ``hunch'' to which Lindley's approach is here trying to give expression. If we keep $\alpha$ fixed as $n$ increases from 20 to 100 we have a rapidly increasing chance of establishing that a difference is significant when, say, $\mu - \mu_0 = 0.4$. Could we not well afford to sacrifice some of this additional power in order to reduce the risk of rejecting the null hypothesis when it is true, i.e., of making the decision $d_1$ wrongly? (...)

Lindley points out that the test proposed by Jeffreys has similar properties to his tests (...). The same practical objective may be attained if desired by the \emph{quite legitimate device of reducing $\alpha$ as $n$ increases}. If the exponents of usually accepted test theory had not thought of this possibility before, it only serves to illustrate the value of looking at a problem of statistical inference from several points of view and making numerical comparisons.'' (\citealp[p. 69]{Pearson1953}; italics added for emphasis)
\end{quotation}

In a later section, we will elaborate on the idea that the paradox undercuts only the Fisherian interpretation of a $p$-value as `evidence against the null hypothesis'; in the Neyman-Pearson paradigm, however, the brunt of the paradox can be avoided by adopting a lower value of $\alpha$ when power is known to be high.  

\subsection{\rev{20}. The 1955 article \emph{The present position in probability theory}}
Here Jeffreys again presents his generic Bayes factor equation including the $\sqrt{n}$ term:
\begin{quotation}
``In most cases the results are of very similar form when the number of observations, $n$, is large. If the straightforward estimate of $\alpha_m$, apart from the significance question, would be $a_m \pm s_m$, we usually get ($\theta$ standing for the data collectively)
\begin{equation*}
    K = \frac{P(q|\theta H)}{P(q^{'}|\theta H)} \doteqdot 
    \frac{{A_n}^\frac{1}{2}}{f(0)} \exp\left(-\frac{a^2_m}{2s^2_m}\right).
\end{equation*}
$A$ is a constant of order 1. We must have $f(0) > 0$, otherwise the null hypothesis would always be asserted [see also \citet[p. 251]{Jeffreys1961} -- EWAL]. If $f(0) > 0$ and $|a_m| < s_m$, $K$ is large and $q$ has a high probability. If $|a_m|$ greatly exceeds $s_m$, $K$ is small and $q^{'}$ has a high probability in comparison with $q$. In practice $s^2_m$ usually decreases with $n$ like $1/n$, and $K=1$ for a moderate value of $|a_m|/s_m$, usually 2 to 4.'' \citep[p. 282]{Jeffreys1955}
\end{quotation}

\subsection{\rev{21}. The 1957 second edition of \emph{Scientific Inference}}

In the second edition of \emph{Scientific Inference}, Jeffreys now presents the generic approximate Bayes factor in the main text (p. 72; as he did in the first and second editions of \emph{Theory of Probability}), where it was previously presented in the addenda of the 1937 reissued first edition. In contrast to that first edition, Jeffreys no longer engages in an explicit comparison between Bayes factors and $p$-values, and only hints a the paradox when he writes:
\begin{quotation}
``The main point is that the null hypothesis is in general strongly supported if the maximum likelihood estimate of the new parameter is less than its standard error; but the introduction of the new parameter is strongly supported if the estimate is much more than the standard error. \emph{With ordinary numbers of observations} (from 20 to 1000) the transition comes at about 3 times the standard error in most problems.'' (\citealp[p. 72]{Jeffreys1957SI}; italics added for emphasis) 
\end{quotation}

\subsection{\rev{22}. The 1957 article \emph{Probability theory in astronomy}}
Jeffreys again presents his approximate form:
\begin{quotation}
``The theory leads to rules of significance for the introduction of new parameters in laws. They are usually approximately of the form
\begin{equation*}
    K = \frac{P(q|\theta p)}{P(q'|\theta p)} \doteqdot 
    (An)^{1/2} \exp\left(-\frac{a^2}{2s^2_a}\right).
\end{equation*}
Here $q$ is the hypothesis that the new parameter $\alpha$ is zero, that is, that the previous law needs no alteration; $q'$ the hypothesis that $\alpha$ is needed, having a value to be estimated from the observations; $a$ and $s_a$ are the estimate of $\alpha$ and its standard error as given by the method of least squares; $n$ is the number of observations; and $A$ is a constant, usually not far from 1. If $|a| < s_a$, the factor $n^{1/2}$ makes $K>1$ and the old law is supported; but with ordinary numbers of observations, if $|a|>2 s_a$ or $3s_a$, $K<1$ and the new law is supported. To apply a test of this sort it is of course of the first importance that the number of observations shall be stated. This is in fact not often done by physicists, but thanks mainly to the work of Fisher (with whom I do not always agree) biologists usually do it, but with different rules.'' \citep[p. 349]{Jeffreys1957}
\end{quotation}

In sum, it appears that at the time of writing, Lindley was unaware of the extent to which Jeffreys had already identified, explained, and explored the paradox. The single reference to the appendix from the 1948 edition of \emph{Theory of Probability} certainly does not do justice to the central position that the paradox occupied in Jeffreys's philosophy; nor is the reference to the 1948 edition historically accurate, as Jeffreys had completed his work related to the paradox already in the second half of the 1930s. The idea that Lindley may not have been fully aware of Jeffreys's prior work on the paradox receives support from the following fragment of Lindley's obituary of Jeffreys:
\begin{quotation}
``He was one of the finest writers of scientific English, with an accurate, yet almost melodious, style. Like Joyce, he used the language sparingly, condensing many ideas into few words. A paradox that has been much discussed, and erroneously associated with my name, occupies two sentences in the Theory (p. 248).'' \citep[p. 417 ]{Lindley1989}
\end{quotation}
As outlined above, Jeffreys devoted many more than two sentences to the paradox. The fact that Lindley was only somewhat aware of the extent of Jeffreys's contributions is also consistent with the following remark: 
\begin{quotation}
``Having produced MEU [maximization of expected utility -- EWAL] as the constructive device for producing statistical methods, we tried to apply it to standard problems, finding sometimes that it agreed, as in the use of sufficient statistics, but more often finding that it did not, for example in the use of the tail area in a significance test. (Interestingly Jeffreys had pointed this out in 1939 but none of us had fully appreciated what he was saying. This is especially ridiculous in my case since I had attended Jeffreys's lectures in Cambridge in 1947; the only excuse I can offer, apart from my own stupidity, is that he was a bad lecturer. But that is not valid since his book is, at least seen through today's eyes, lucid and still worth reading.)'' \citep[p. 8]{Lindley2000ISBA}
\end{quotation}
As outlined above, Jeffreys's pointed out the paradox as early as 1935, returning to the same theme many times prior to the first edition of \emph{Theory of Probability}.

\section*{The 1957 Contribution from Bartlett}

For over two decades, Jeffreys had repeatedly pointed out the potential conflict between $p$-values and Bayes factors. However, Jeffreys's work on Bayes factors had been largely ignored. Instead, it was the 1957 article by Lindley that brought the paradox into the limelight. Although Lindley's conclusions were qualitatively correct, he did omit an important term from his equations, an oversight that was quickly corrected by \citet{Bartlett1957}:

\begin{quotation}
``I would agree that he [Lindley -- EWAL] establishes the point that one must be cautious when using a fixed significance level for testing a null hypothesis irrespective of the size of sample one is taking. However, there is a slip, in his expression for $K$ under his equation (1), that appears to me, unless corrected, to lead to an overstatement of his point. The prior distribution for $\theta$, given that $\theta \neq \theta_0$, was assumed to be uniform over an interval $I$, and hence its density function should be $1/I$ in this interval. This leads to the extra factor $1/I$ in the second term in the expression for $K$.[Here Bartlett adds a footnote: ``There is also a further dropping of a factor $1/\sigma$ in the last formula on p. 191, but this is a more trivial slip.'' -- EWAL] This expression then becomes consistent with Jeffreys's equation (10), \S5.0 in his book (second edition, 1948) [This is the equation for $K$ given above in the section on the 1939 first edition of \emph{Theory of Probability} -- EWAL].'' \citep[p. 533]{Bartlett1957}
\end{quotation}

In an editorial note following Bartlett's paper, Sir Maurice Kendall stated that ``Mr Lindley agrees and apologizes'' for omitting the $1/I$ term from his first equation. However, Kendall points out that this oversight affects neither Lindley's general argument nor his concrete examples. 

After including the $1/I$ term omitted by Lindley, Bartlett notes that a uniform prior on the entire real line (``the most natural prior'', p. 533) will yield infinite support in favor of the null hypothesis, a ``silly answer'' (p. 533). Moreover, in order to escape from the paradox, Bartlett argues that in the planning stage of an experiment, sample size may be chosen such that $\sqrt{n}$ is proportional to $1/I$ (i.e., researchers who expect small effects will collect many observations).

Based on our reading, we conclude that both Lindley and Bartlett unwittingly presented a slightly confused version of Jeffreys's earlier work. As far as Lindley is concerned, he indeed omitted the $1/I$ term that is correctly included in Jeffreys's equations (e.g., see above: \citealp[p. 417]{Jeffreys1936Further}; \citealp[pp. 250-251]{Jeffreys1937SI}; \citealp[p. 161]{Jeffreys1938}; \citealp[p. 194]{Jeffreys1939}; \citealp[pp. 221-222]{Jeffreys1948}). In addition, Lindley appears to have been unaware of Jeffreys's general approximate $\sqrt{n}$ form of the Bayes factor. Lindley does present this form at a later stage of his paper, but without the $1/I$ term, and preceding it with an attribution to Barnard: ``An alternative interpretation of the paradox was suggested to me by Prof. Barnard.'' (p. 189). Lindley then notes that this $\sqrt{n}$ form shows that ``Clearly (...) for fixed significance level the likelihood of the null hypothesis increases indefinitely with the sample size.'' (p. 189). As mentioned above, the form of this equation and its conclusion were already presented two decades earlier by \citet[p. 417]{Jeffreys1936Further}.

As far as Bartlett is concerned, his conclusion that a uniform (improper) prior leads to the ``silly answer'' of infinite support for the null hypothesis was anticipated by Jeffreys in 1935:
\begin{quotation}
``To apply this theory it is therefore necessary that we should have previous knowledge of the range of possible values of $y$. (...) Since $m$ enters only through its logarithm its effect is in any case not great in practical cases, and it does not need to be determined very accurately (...)

It may happen, however, that we have no previous information about the range of admissible values of $y$; then $m$ is effectively infinite, and it appears that no matter how many observations we have we shall never be able to infer a systematic difference.'' \citep[p. 207]{Jeffreys1935}
\end{quotation}
Jeffreys also discussed the problem of improper priors for testing in the 1948 second edition of \emph{Theory of Probability}, in the section \emph{Required properties of $f(\alpha)$}:
\begin{quotation}
``It might appear that on $q'$ the new parameter is regarded as unknown and therefore that we should use the estimation prior probability for it. But this leads to an immediate difficulty. Suppose that we are considering whether a location parameter $\alpha$ is 0. The estimation prior probability for it is uniform, and (...) we should have to take $f(\alpha) = 0$, and $K$ would always be infinite. We must instead say that the mere fact that it has been suggested that $\alpha$ is zero corresponds to some presumption that it is fairly small.'' (\citealp[p. 225]{Jeffreys1948}; \citealp[p. 251]{Jeffreys1961})
\end{quotation}
Thus, the popular belief that Bartlett was the first to point out the problem with improper priors for Bayes factor testing (e.g., \citealp[p. 78]{OHaganForster2004}) is incorrect.

Bartlett also commented on a ``more trivial slip'' in Lindley's paper, that is, ``a further dropping of a factor $1/\sigma$ in the last formula on p. 191''. This is the offending equation:
\begin{equation*}
    \sqrt{ \left(\frac{n}{2\pi}\right)} \exp{ \left\{ - \frac{n(\bar{x} - \theta_0)^2}{2\sigma^2} \right\} }. 
\end{equation*}
However, this equation is in fact similar to those presented by Jeffreys. As noted in \citet{Cousins2017}, the unit-information prior (e.g., \citealp{KassWasserman1995}) sets the range $m$ equal to the uncertainty associated with a single observation, meaning that after dividing the $m$ and the $1/\sigma$ terms, only the $\sqrt{n}$ term remains. 



Finally, Bartlett suggests to reduce the spread of the prior as $\sqrt{n}$ (see also \citealp{Andrews1994}; \citealp[pp. 106-107]{Cox2006}, as noted by \citealp{Cousins2017}). In other words, he assumes that researchers who collect a large sample do so because they expect the effect to be relatively small -- the sample size therefore provides a clue about the spread of the prior distribution for the test-relevant parameter under $\mathcal{H}_1$. There are several problems with this suggestion. First and foremost, Bartlett's scaling solution makes it impossible for the Bayes factor to produce convincing evidence in favor of the null hypothesis; as $n$ increases, the alternative hypothesis will increasingly resemble the null hypothesis, and consequently the null hypothesis can never reach a compelling level of support. This is a key objection, as a cornerstone of Jeffreys's philosophy of testing is that ``An adequate theory of scientific investigation must leave it open for any hypothesis whatever that can be clearly stated to be accepted on a moderate amount of evidence.'' \citep[p. 129]{Jeffreys1961}. This notion harks back to Jeffreys's early work with Dorothy Wrinch, in which they argued that in order for a universal generalization (e.g., propositions such as ``all ravens are black'') to attain a compelling degree of plausibility it is necessary to adjust Laplace's idea of uniform prior distributions and assign point mass to the general law (i.e., \citealp{WrinchJeffreys1921}; \citealp{LyEtAl2020Alternative}). Second, Bartlett's solution does not apply to observational studies, where the issue of sample size planning is irrelevant. Third, researchers may collect larger samples for a variety of other reasons including feasibility (e.g., the presence of sufficient funding), ease of data collection (e.g., via online surveys), scientific or societal importance of the topic under study, personality characteristics of the researcher, and so on. Finally, as indicated above, in 1937 Jeffreys already mentioned and rejected Bartlett's 1957 proposal \citep[p. 259]{Jeffreys1937SI}. 

In sum, the arguments presented in \citet{Lindley1957} and \citet{Bartlett1957} were already discussed two decades earlier by Jeffreys, in more detail and without errors. The main difference is in the evaluation of the practical ramifications of the paradox; whereas Jeffreys downplays the discrepancy between Bayes factors and $p$-values for practical data analysis (``curiously'', according to \citealp[p. 400]{Cousins2017}), Lindley stresses it. In a later article, Lindley doubles down: ``There is therefore a serious and systematic difference between the Bayesian and Fisherian calculations, in the sense that a Fisherian approach much more easily casts doubt on the null value than does Bayes. \emph{Perhaps this is why significance tests are so popular with scientists: they make effects appear so easily}. Notice that this result depends on a `sharp' prior being used, with $p(\theta = 0) > 0$.'' (\citealp[p. 502]{Lindley1986Comment}, italics added for emphasis). The reason for this difference in perspective is arguably due to the fact that Jeffreys calibrated a $p=.05$ result to a Bayes factor of 1 (reasoning that these were the watershed values in the two statistical paradigms), whereas Lindley sought to compare the \emph{p}-value and the posterior probability for the null hypothesis directly. 

\section*{The Root of the Paradox: A Summary of Jeffreys's Argument}

Jeffreys generally explained the paradox in two ways. The first way is to note that the $p$-value focuses on the predictions from $\mathcal{H}_0$, whereas the Bayes factor compares the predictions from $\mathcal{H}_0$ against those from a composite $\mathcal{H}_1$. At hand is the scenario where sample size $n$ increases but the multiple of the standard error is constant, such that $\nicefrac{\hat{\theta}}{\text{se}(\hat{\theta})} = c, \, \forall n \rightarrow \infty$. In this case the predictive adequacy of $\mathcal{H}_0$ is unaffected --and consequently the $p$-value remains constant also--, but the predictive adequacy of $\mathcal{H}_1$ gradually deteriorates. The reason for this deterioration is that, as $n$ increases, an increasingly smaller set of parameter values provides acceptable predictions. An ever increasing part of $\mathcal{H}_1$ is found wanting, and this decreases the average predictive performance across all parameter values under $\mathcal{H}_1$. This phenomenon does not occur if the predictive adequacy of $\mathcal{H}_1$ is based only on the maximum likelihood estimate $\hat{\theta}$; however, this is a cherry-picked value that is in need of a multiplicity correction, for else the null hypothesis could never be supported by any data. The correction for cherry-picking (or \emph{selection}, as Jeffreys called it) is achieved automatically through the prior distribution (see also \citealp[pp. 401-402]{Cousins2017} and \citealp[Chapter 20]{Jaynes2003}). The ``correction for selection'' explanation for the deteriorating predictive performance of $\mathcal{H}_1$ was mentioned prominently presented in the \emph{Theory of Probability}, for instance in the fragments cited above (i.e., \citealp[pp. 379-380]{Jeffreys1938Comparison2}; \citealp[pp. 359-360]{Jeffreys1939}, \citealp[pp. 399-400]{Jeffreys1948}, and \citealp[pp. 435-436]{Jeffreys1961}; see also \citealp[p. 72]{Jeffreys1953}) and also in the following: 
\begin{quotation}
``The possibility of getting actual support for the null hypothesis from the observations really comes from the fact that the value of $\alpha$ indicated by it is unique. $q'$ indicates only a range of possible values, and \emph{if we select the one that happens to fit the observations best we must allow for the fact that it is a selected value.} If $|a|$ is less than $s$, this is what we should expect on the hypothesis that $\alpha$ is 0, but if $\alpha$ was equally likely to be anywhere in a range of length $m$ it requires that an event with a probability $2s/m$ shall have come off. If $|a|$ is much larger than $s$, however, $a$ would be a very unlikely value to occur if $\alpha$ was 0, but no more unlikely than any other if $\alpha$ was not 0. In each case we adopt the less remarkable coincidence.'' (\citealp[p. 248]{Jeffreys1961}, italics added for emphasis; echoed in \citealp[pp. 194-195]{Jeffreys1939} and \citealp[p. 222]{Jeffreys1948})
\end{quotation}

Jeffreys's second, related explanation for the paradox refers to the need for consistency under $\mathcal{H}_0$. As mentioned in the above fragment, Jeffreys argues that when the estimate is of the order of the standard error, this constitutes increasingly strong evidence in favor of $\mathcal{H}_0$ as sample size grows. The idea is intuitive: for instance, 5 heads out of 10 tosses yields less evidence in favor of the fair coin hypothesis $\theta_0=\nicefrac{1}{2}$ than would 500 heads out of 1000 tosses (cf. \citealp[p. 332]{Berkson1942}). This implies, however, that the Bayes factor break-even point $\text{BF}_{01} = 1$ has to be at a multiple of the standard error that increases with $n$. This effectively creates the paradox (e.g., \citealp{WagenmakersEtAlinpressSupportInterval}).

\section*{Two Examples by Jack Good}

Across several articles, Jack Good attempted to explain why it is problematic to use a significance threshold that is a constant multiple of the standard error. A first example was presented in \citet{Good1980p}: 

\begin{quotation}
``Dr. Deborah Mayo raised the following question. How could one convince a very naive student, Simplissimus, that a given tail-area probability (P-value), say $1/100$, is weaker evidence against the null hypothesis when the sample is larger? Although this fact is familiar in Bayesian statistics the question is how to argue it without (explicit) reference to Bayesian methods.

One can achieve this aim, without even referring to
power functions, in the following manner.

Take a very concrete example, say the tossing of a coin, and count the number r of heads (``successes'') in N trials. Ask Simplissimus to specify any simple non-null hypothesis for the probability p of a head. Suppose he gives you a value $p = .5 + \epsilon$. First compute a value of N so that a $\epsilon$ value of r approximately equal to N$(.5 + \frac{\epsilon}{7})$ would imply a tail-area probability close to 1/100. Then point out that the fraction $.5 + \frac{\epsilon}{7}$ of successes is much closer to .5 than it is to $.5 + \epsilon$ and therefore must \emph{support} the null hypothesis as against the specific rival hypothesis proposed by Simplissimus. Thus, for any specified simple non-null hypothesis, N can always be made so large that a specified tail-area probability supports the null hypothesis more than the rival one. This should convince Simplissimus, if he had been listening, that the larger is N the smaller the set S of simple non-null hypotheses that can receive support (as compared with p = $\nicefrac{1}{2}$) in virtue of a specified P-value. If the tail-area probability, for example 1/100, is held constant, the set S converges upon the point p = $\nicefrac{1}{2}$ when N is made larger and larger.'' (\citealp[pp. 307--308]{Good1980}; italics in original)
\end{quotation}

As elaborated in \citet{LyWagenmakersinpressKelto}:
\begin{quotation}
``For instance, assume Simplissimus specifies their simple non-null hypothesis as \( \theta=0.57 \) with \( \epsilon=0.07 \). Then our target value for the number of successes \( s \) equals \( n (0.5 + 0.07/7) = n \times 0.51 \). So for a sample proportion of \( 0.51 \) we now seek \( n \) such that the two-sided tail area probability equals \( .01 \). We find that \( n = 16700 \) --consisting of \( 8517 \) heads, for a sample proportion of \( s = 8517/16700 = 0.51 \), as stipulated-- yields a tail-area just below \( .01 \). But the sample proportion of \( 0.51 \) is much closer to the null hypothesis (i.e., \( \theta=0.50 \)) than to the non-null hypothesis specified by Simplissimus (i.e., \( \theta=0.57 \)).''
\end{quotation}

In a later article, Good present a second example:
\begin{quotation}
``In the course of discussion of Good (1980), Dr. Golde Holtzman suggested that instead of considering a binomial model in which all values of the binomial parameter p are considered, we think of a bag known to contain exactly 1000 balls, some white and some black. The null hypothesis, by definition, is that there are 500 of each. The sampling is to be random, with replacement, with N drawings. 

For definiteness suppose that the outcome is $\nicefrac{1}{2}N + \sqrt{N}$ white and therefore $\nicefrac{1}{2}N - \sqrt{N}$ black balls. (We can suppose N is a perfect square.) Then P, taken as a double tail, is about $.05$; and the fraction of white balls drawn is $\nicefrac{1}{2} + N^{-\nicefrac{1}{2}}$. If N is large enough, the closest possible rival to $w = 500$ is $w = 501$, where $w$ is equal to the number of white balls in the bag. If therefore $N^{-\nicefrac{1}{2}}$ is much smaller than $1/1000$, that is, if $N/1,000,000$ is large, the probability of the observed outcome will be much larger assuming the null hypothesis than if any other hypothesis is assumed, even $w = 501$. Thus the tail-area probability of $.05$ will then support the null hypothesis, and the larger $N$ is (above a certain threshold) the more the support will be if the tail-area probability is the same in each case. Moreover, if we were fairly confident of our model in the first place, the tail-area probability of $.05$ would not be small enough to cause us to suspect the model.'' \citep[pp. 312--313]{Good1983b}
\end{quotation}

For simplicity, suppose the bag contains just 10 balls. Drawing \nicefrac{120}{200} white balls yields $\hat{\theta}=.60$ and gives $p\approx.006$; Drawing \nicefrac{429}{780} white balls yields $\hat{\theta}=.55$ and also gives $p\approx.006$; and drawing \nicefrac{9690}{19000} white balls yields $\hat{\theta}=.51$ and again gives $p\approx.006$. To reject $\mathcal{H}_0$ based on a sample proportion of $.55$ (exactly in between the expected proportion for 5 and 6 white balls out of 10) seems premature, and to do so for a sample proportion of $.51$ seems preposterous, as the data are much more likely under $\mathcal{H}_0: \nicefrac{5}{10}$ white balls than under even the most likely of the alternative compositions (i.e., $\nicefrac{6}{10}$ white balls; for similar examples see e.g., \citealp{Freeman1993,PericchiPereira2016}). The problem becomes even more severe when the bag contains only two balls. In this case, any sample of mixed composition, no matter how lopsided (e.g., 1 white ball and 100 black balls) decisively falsifies $\mathcal{H}_1$ and thereby proves $\mathcal{H}_0$. 

Note that for this particular example, a frequentist may argue that the details of the problem necessitate the choice of a different test statistic, such as the likelihood ratio between $\mathcal{H}_0: \theta = \nicefrac{1}{2}$ and a specific $\mathcal{H}_1$ (e.g., the one closest to $\hat\theta$).  

\section*{Frequentist Considerations}

Jeffreys demonstrated that the evidence provided by the data for a point hypothesis $\mathcal{H}_0$ vis-a-vis a composite hypothesis $\mathcal{H}_1$ scales with $\sqrt{n}$; consequently, any evidence threshold cannot be a constant multiple of the standard error. This result undercuts the popular interpretation of the classical $p$-value in terms of a fixed, sample-size independent measure of evidence against $\mathcal{H}_0$. This interpretation was promoted by Fisher himself, who argued explicitly that the interpretation of the $p$-value is independent of sample size:
\begin{quotation}
``It is not true (...) that valid conclusions cannot be drawn from small samples; if accurate methods are used in calculating the probability, we thereby make full allowance for the size of the sample, and should be influenced in our judgment only by the value of probability indicated. The great increase of certainty which accrues from increasing data is reflected in the value of P, if accurate methods are used.'' (Fisher, 1934, p. 182).
\end{quotation}
Berkson agreed with Fisher's assessment and stated that ``small $P$'s are more or less independent, in the weight of the evidence they afford, of the numbers in the sample.'' (\citealp[p. 333]{Berkson1942}; cf. \citealp[p. 70]{Royall1997}). Jeffreys's work and the associated paradox cast doubt on this evidential interpretation of the $p$-value.

However, in the Neyman-Pearson paradigm the $\sqrt{n}$ scaling of the evidence can be accommodated by reducing $\alpha$ when $n$ is high. This possibility was already suggested by Jeffreys in 1938:
\begin{quotation}
``It [the 5\% rule -- EWAL] would mean drawing the line at such a limit as to give a fixed percentage of what Neyman and E. S. Pearson call errors of the first kind, with respect to the number of cases where $q$ is true; but as the limit is at our disposal we are entitled to take it further out and reduce this percentage still further if there is no special reason to expect values of the new parameter in the range affected. To reject the null hypothesis in any cases at all where it is true is not a desirable action for its own sake. It is an evil that becomes necessary if we are to have any criterion for detecting cases where $q$ is untrue, and we are justified in taking such steps as will reduce its importance to a minimum.'' \citep[p. 379]{Jeffreys1938Comparison2}
\end{quotation}

\noindent A similar remark appears in Appendix I of the first edition of \emph{Theory of Probability}:
\begin{quotation}
``(...) if we assert a genuine departure whenever $P$ is less than $0.01$ we shall expect to be wrong in the long run in 1 per cent. of the cases where $q$ is true. According to my theory we should expect to make fewer mistakes by taking the limit further out; when $K=1$ lies above $P=0.01$ there will be a smaller risk of rejecting $q$ wrongly, partly counter-balanced by a slight increase in the risk of missing a small genuine departure.'' (\citealp[p. 360]{Jeffreys1939}, echoed in \citealp[p. 435]{Jeffreys1961})
\end{quotation}

In the main text of \emph{Theory of Probability}, Jeffreys also pointed out that --if the prior distribution for the test-relevant parameter under $\mathcal{H}_1$ is well-calibrated-- the total number of errors (i.e., $\alpha + \beta$) is minimized by using $\text{BF}_{01}=1$ as the criterion for accept/reject decisions:
\begin{quotation}
``It may, however, be interesting to see what would happen if the new parameter is needed as often as not, and if the values when it is needed are uniformly distributed over the possible range. Then the frequencies in the world would be proportional to my assessment of the prior probability. Suppose, then, that the problem is, not knowing in any particular case whether the parameter is 0 or not, to identify the cases so as to have a minimum total number of mistakes of both kinds. (...)

Hence, with world-frequencies in proportion to the prior probability used to express ignorance, the total number of mistakes will be made a minimum if the line is drawn at the critical value that makes $K=1$.

Now I do not say that this proportionality holds; all that I should say myself is that at the outset we should expect to make a minimum number of mistakes in this way, but that accumulation of information may lead to a revision of the prior probabilities for further use and the critical value may be correspondingly somewhat altered. But whatever the frequency law may be (...) $K$ would be altered by a factor independent of the number of observations. \emph{We should therefore get the best result, with any distribution (...), by some form that makes the ratio of the critical value to the standard error increase with $n$. It appears then that whatever the distribution may be, the use of a fixed $P$ limit cannot be the one that will make the smallest number of mistakes.} The absolute best is of course unknown since we do not know the distribution in question except so far as we can infer it from similar cases.'' (\citealp[pp. 326-328]{Jeffreys1939}, echoed in \citealp[pp. 396-397]{Jeffreys1961}; italics added for emphasis)
\end{quotation}

Thus, if the prior distribution is calibrated then the Bayes factor provides an optimal frequentist decision criterion. This also holds when the frequentist purpose is to minimize a weighted sum of errors, $\lambda \alpha+\beta$ \citep{Cornfield1966}. Thus, from a Neyman-Pearson perspective, the conflict with a Bayesian assessment of evidence arises specifically in the common scenario where the researcher fixes the probability $\alpha$ of a Type I error (say to 5\%) and then tries to minimize the probability $\beta$ of a Type II error. However, as pointed out above, in high-$n$ situations the researcher may prefer to sacrifice some power in order to lower the probability of a Type I error. As indicated above Egon Pearson himself judged this strategy ``quite legitimate'' \citep[p. 69]{Pearson1953}. Applying this strategy substantially reduces the discrepancy between the frequentist and the Bayesian results.\footnote{Also note that under this strategy, the frequentist results obey the likelihood principle and the stopping rule principle (e.g., \citealp{Cornfield1966,Lindley1953,PericchiPereira2016}).} For related work see for instance \citet[Chapter 9]{DeGrootSchervish2012}, \citealp{Good1992}, \citealp{KimChoi2021}, \citet[Chapter 4]{Leamer1978}, \citet{Lehmann1958}, \citet{Lindley1953}, \citet{MaierLakensinpress}, \citet{MudgeEtAl2012}, \citet{PerezPericchi2014}, \citet{PericchiPereira2016}, \citet[pp. 64-67]{SavageEtAl1962}, and \citet[Section 5]{Savage1964}.

In sum, the Jeffreys-Lindley paradox may be given a purely frequentist interpretation as a discrepancy between (a) minimizing $\beta$ for fixed $\alpha$; versus (b) minimizing the weighted sum of errors, $\lambda \alpha + \beta$.\footnote{\rev{A reviewer suggested that the above considerations are moot in case the alternative hypothesis is composite, as the test-relevant parameter cannot be averaged out in the frequentist paradigm.}} A purely Bayesian version of the paradox will be provided in the next section.


\section*{A Fully Bayesian Version of the Paradox}

It is well-known that the one-sided $p$-value is asymptotically equal to the posterior mass lower than the point of test (e.g., \citealp{CasellaBerger1987}; \citealp{Lindley1965Inference}; \citealp{Pratt1965}; \citealp{MarsmanWagenmakers2017ThreeInsights} and references therein); for some problems, the relation is exact. This means that the $p$-value can be given a Bayesian interpretation as the (approximate) probability that the observed effect has the wrong sign. Specifically, the odds form $(1-p)/p$ is an approximation for $\text{BF}_{+-}$, that is, the Bayes factor for $\mathcal{H}_+: \delta > 0$ versus $\mathcal{H}_-: \delta < 0$: a Bayesian test for the direction of an effect size $\delta$. Jeffreys considered this a problem of estimation rather than of testing:
\begin{quotation}
``It should be said that several of the $P$ integrals have a definite place in the present theory, in problems of pure estimation. For the normal law with a known standard error, or for those sampling problems that reduce to it, the total area of the tail represents the probability, given the data, that the estimated difference has the wrong sign--provided that there is no question whether the difference is zero.(...) They give the correct answer if the question is: If there is nothing to require consideration of some special values of the parameter, what is the probability distribution of that parameter given the observations?'' (\citealp[pp. 387-388]{Jeffreys1961}; see also \citealp[pp. 317-318]{Jeffreys1939})
\end{quotation}

The relation between the one-sided $p$-value and the Bayesian test for direction suggests that the Jeffreys-Lindley paradox can be given a fully Bayesian interpretation. Specifically, data may be constructed which will convince the Bayesian that the population effect is positive rather than negative (i.e., $p(\delta > 0 \given y, \mathcal{H}_1) \gg p(\delta < 0 \given y, \mathcal{H}_1$), whereas this same Bayesian will also be convinced that the population effect is absent rather than present (i.e., $p(\delta = 0 \given y) \gg p(\delta \neq 0 \given y)$). Let $\text{BF}_{+-}$ denote $p(\delta > 0 \given y, \mathcal{H}_1) \, / \, p(\delta < 0 \given y, \mathcal{H}_1)$. Suppose data are constructed such that $\text{BF}_{+-}$ is constant. As $n$ increases, the evidence that the effect is absent rather than present will increase without bound, and this ensures that, with sufficiently high $n$, the Bayesian will believe that the effect is positive rather than negative, and simultaneously believe that it is absent rather than present. This state of knowledge is not incoherent, but it may be counter-intuitive.

\begin{figure}[ph!]
\begin{tabular}{ccc}
    \begin{minipage}{.9 \textwidth}
    \centering
    \includegraphics[width= 0.6 \linewidth]{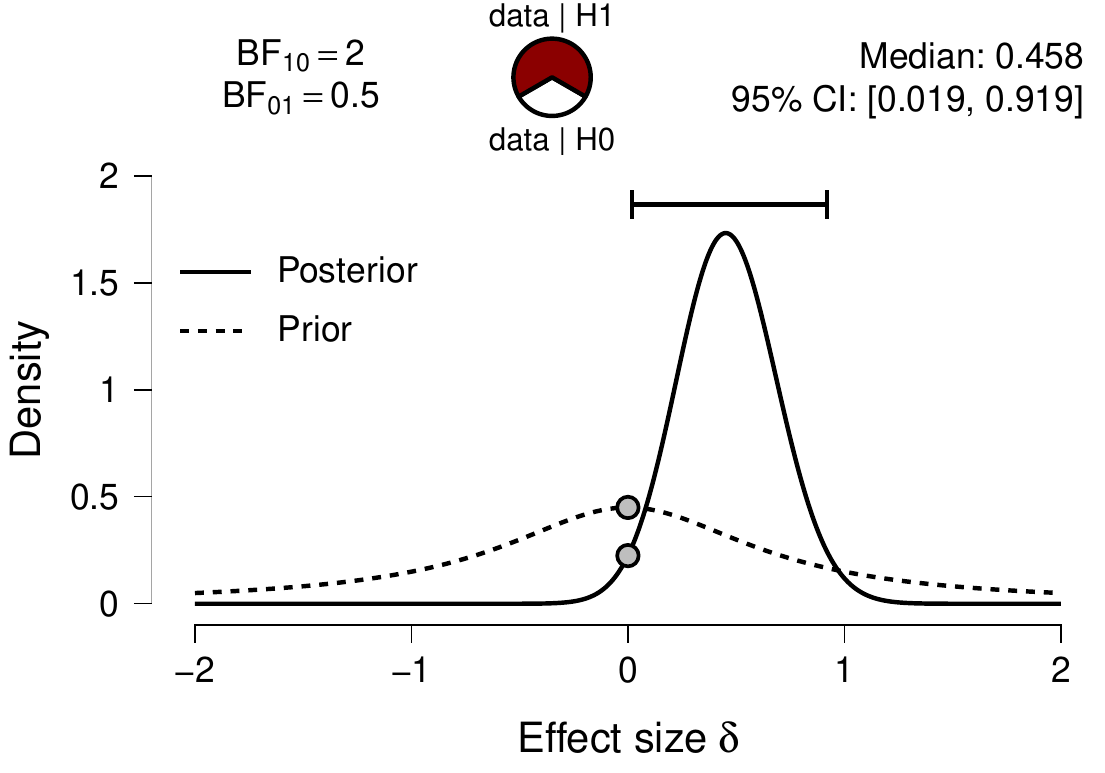} %
    \end{minipage}  %
    \\
    \begin{minipage}{.9 \textwidth}
    \centering
    \includegraphics[width= 0.6 \linewidth]{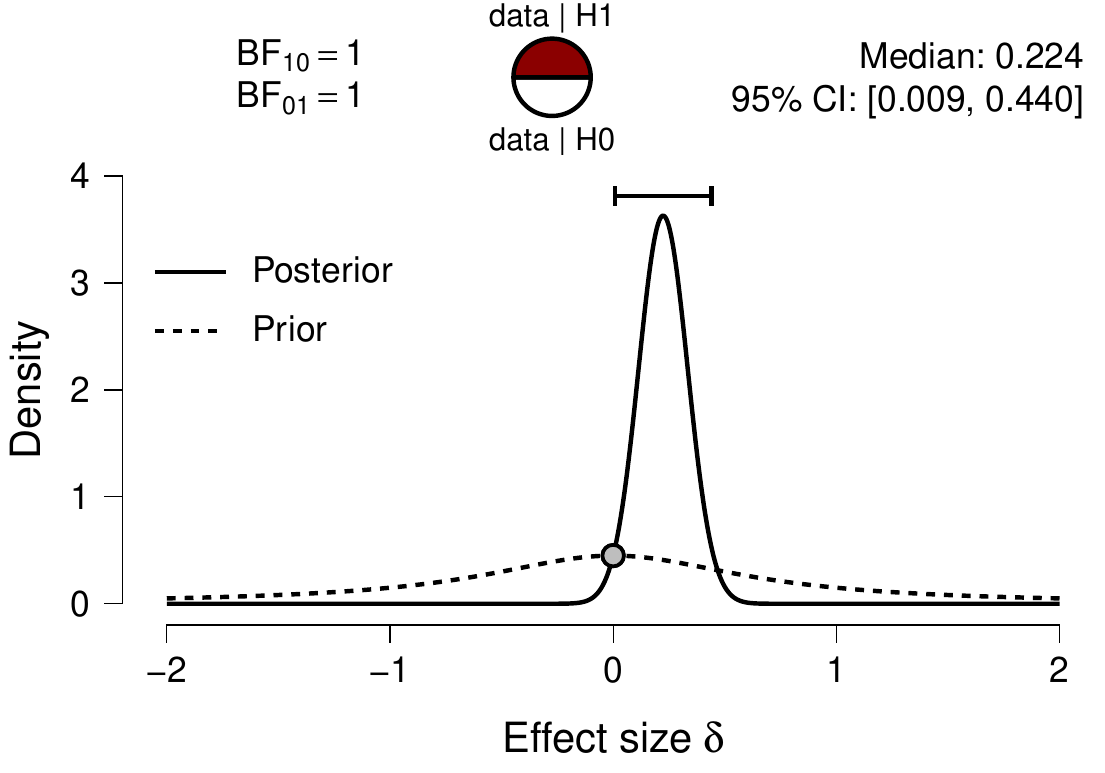} %
    \end{minipage}  %
    \\
    \begin{minipage}{.9 \textwidth}
    \centering
    \includegraphics[width= 0.6 \linewidth]{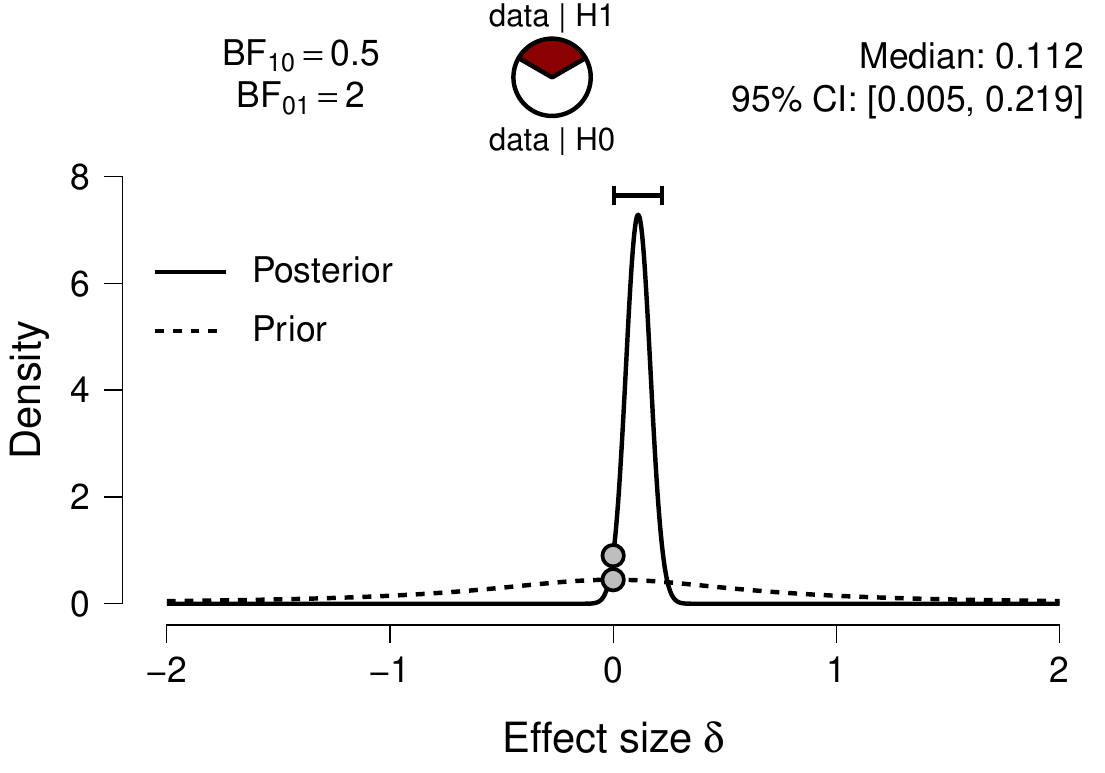} %
    \end{minipage}  %
\end{tabular}
\caption{Fully Bayesian version of the Jeffreys-Lindley paradox, illustrated with the $t$-test. All panels have the same posterior mass on negative effect size: $p(\delta < 0 \given y, \mathcal{H}_1) = 0.02041783$; thus, \( \BF_{+-}= 47.9768 \). As sample size $n$ grows, $\mathcal{H}_0$ receives increasing support from the data. Top: \( t= 2.321 \), \( n=20 \). Middle: \( t=2.113 \), \( n=82 \). Bottom: \( t=2.062 \), \( n=332 \). See text for details. Figures from JASP (\url{jasp-stats.org}).}
\label{figJLBBfOneSided}
\end{figure}

A concrete demonstration of the fully Bayesian version of the paradox is given in Figure~\ref{figJLBBfOneSided}. Each panel concerns the same Bayesian one-sample $t$-test \citep{Jeffreys1948} and shows prior and posterior distributions on effect size $\delta = \nicefrac{\mu}{\sigma}$; the prior distribution on $\delta$ is a zero-centered Cauchy with scale $\nicefrac{1}{\sqrt{2}}$ (e.g., \citealp{GronauEtAl2020Informed,MoreyRouderBayesFactorPackage}). In all three panels, the $t$-values and sample sizes were chosen such that $p(\delta < 0 \given y, \mathcal{H}_1) = 0.02041783$; thus, \( \BF_{+-}= 47.9768 \), indicating strong evidence that the population effect is positive rather than negative. 

\noindent The top, middle, and bottom panel have 20, 82, and 332 observations, respectively. As sample size increases from top to bottom, the posterior distribution narrows and shifts towards zero. As a result, the Bayes factor increasingly favors $\mathcal{H}_0$ over $\mathcal{H}_1$. In the top panel, $\text{BF}_{10} = 2$ (i.e., weak evidence in favor of the presence of an effect); in the middle panel, $\text{BF}_{10} = 1$ (i.e., complete absence of evidence); and in the bottom panel, $\text{BF}_{10} = \nicefrac{1}{2}$ (i.e., weak evidence in favor of the absence of an effect).\footnote{Top, middle, and bottom panels have a one-sided $p$-value of $.016$, $.019$, and $.020$, respectively.}    

The pattern shown in Figure~\ref{figJLBBfOneSided} can be appreciated by recourse to the Savage-Dickey density ratio (e.g., \citealp{Dickey1971,VerdinelliWasserman1995,WetzelsEtAl2010Borel}). Under mild assumptions, this density ratio states that $\text{BF}_{10} = p(\delta = 0 \given \mathcal{H}_1) \, / \, p(\delta =0 \given y, \mathcal{H}_1)$. In other words, the Bayes factor is given by the ratio of prior to posterior ordinate for $\delta$ under $\mathcal{H}_1$ at the point of test. The ordinate of the Cauchy prior distribution at $\delta=0$ equals approximately $0.45$. When $\text{BF}_{10}=2$, this implies that the posterior ordinate equals $0.45/2$. This can be confirmed by a visual inspection of the two grey dots in the top panel from Figure~\ref{figJLBBfOneSided}: the data have shifted the posterior distribution away from zero, lowering the ordinate at $\delta=0$; consequently, the data favor $\mathcal{H}_1$ over $\mathcal{H}_0$. The middle panel shows that the prior ordinate equals the posterior ordinate, for a Bayes factor of 1, whereas the bottom panel shows that the posterior ordinate at $\delta=0$ is now larger than the prior ordinate, indicating that the data favor $\mathcal{H}_0$ over $\mathcal{H}_1$.

The general rule is that, when the observations accumulate indefinitely and the posterior distribution for $\delta$ becomes more peaked, retaining the same posterior mass on negative values of effect size (i.e., keeping $\text{BF}_{+-}$ at a constant value) entails an increase of the posterior ordinate at $\delta=0$; by the Savage-Dickey density ratio, this means more evidence for $\mathcal{H}_0$ (i.e., $\text{BF}_{01}$ grows without bound). In sum, the paradox is also relevant within the framework of Bayesian statistics. 

\section*{Two Attempts to Escape from the Paradox}
The Jeffreys-Lindley paradox inconveniences many statisticians. For frequentist statisticians, the paradox suggests that an epistemic interpretation of a $p$-value requires that sample size is somehow taken into account -- with a very large sample, a $p=.01$ result may well indicate strong support \emph{in favor} of $\mathcal{H}_0$. For Bayesian statisticians, the paradox suggests that the quantification of evidence hinges on the specification of the test-relevant prior distribution under $\mathcal{H}_1$ -- this essentially prohibits the use of vague or improper priors.\footnote{As mentioned in the section on Bartlett's article, Jeffreys was well aware of this and suggested that different prior distributions be used for testing vs. estimation (cf. \citealp[p. 207]{Jeffreys1935}; \citealp[p. 225]{Jeffreys1948}).} Perhaps for this reason both frequentist and Bayesian statisticians have sought to defang the paradox by questioning Jeffreys's core assumptions. The main objections fall in two categories that will be discussed in turn; the first objection concerns the specification of $\mathcal{H}_0$, whereas the second objection finds fault with the specification of $\mathcal{H}_1$.

\subsection*{Objection 1: ``Down with Point Masses!''\footnote{\citet[p. 42]{RobertRousseau2011}.}}
In Jeffreys's original development, prior mass \nicefrac{1}{2} is assigned to the point-null hypothesis $\mathcal{H}_0$. One attempt to question the relevance of the paradox is to argue that the null hypothesis is never true exactly, and it is unwise to assign separate prior mass to a single point from a continuous distribution (e.g., \citealp[p. 109]{PereiraStern1999}). For instance, \citet{Bernardo2009} argues that
\begin{quotation}
``Jeffreys intends to obtain a posterior probability for a precise null hypothesis and, to do this, he is forced to use a mixed prior which puts a lump of probability $p = \text{Pr}(H_0)$ on the null, say $H_0 \equiv {\theta = \theta_0}$, and distributes the rest with a \emph{proper} prior $p(\theta)$ (he mostly chooses $p = 1/2$). This has a very upsetting consequence, usually known as Lindley's paradox (Lindley, 1957): for any fixed prior probability $p$ independent of the sample sixe [sic] $n$, the procedure will wrongly accept $H_0$ whenever the likelihood is concentrated around a true parameter value which lies $O(n^{-\frac{1}{2}})$ from $H_0$. I find it difficult to accept a procedure which is \emph{known} to produce the wrong answer under specific, but not controllable, circumstances (...)'' (\citealp[p. 174]{Bernardo2009}; italics in original)
\end{quotation}

Moreover, in his paradox paper, \citet[p. 188]{Lindley1957} explicitly argues that prior mass needs to be assigned to a point in order for the paradox to arise: ``...the phenomenon would persist with almost any prior probability distribution that had a concentration on the null value and no concentrations elsewhere. (...) It is, however, essential that the concentration on the null value exists, and it is this that has to be considered.''

The impression that the paradox arises because $\mathcal{H}_0$ has separate prior mass is strengthened by Jeffreys's own work. Indeed, Jeffreys argued that his major conceptual advance over Laplace was the insight that, with moderate sample sizes, a general law can only ever receive compelling evidence when that law is assigned separate mass from the outset \citep{WrinchJeffreys1921}. As summarized by Jeffreys when he was 89 years old:
\begin{quotation}
``My chief interest is in significance tests. This goes back to a remark in Pearson's \emph{Grammar of Science} and to a paper of 1918 by C. D. Broad. Broad used Laplace's theory of sampling, which supposes that if we have a population of $n$ members, $r$ of which may have a property $\varphi$, and we do not know $r$, the prior probability of any particular value of $r$ (0 to $n$) is $1/(n+1)$. Broad showed that on this assessment, if we take a sample of number $m$ and find them all with $\varphi$, the posterior probability that all $n$ are $\varphi$'s is $(m+1)/(n+1)$. A general rule would never acquire a high probability until nearly the whole of the class had been inspected. We could never be reasonably sure that apple trees would always bear apples (if anything). The result is preposterous, and started the work of Wrinch and myself in 1919-1923. Our point was that giving prior probability $1/(n+1)$ to a general law is that for $n$ large we are already expressing strong confidence that no general law is true. The way out is obvious. To make it possible to get a high probability for a general law from a finite sample the prior probability must have at least some positive value independent of $n$.'' \citep[p. 452]{Jeffreys1980}
\end{quotation}

The objection to the role of the point-null consists of two separate arguments, both of which need to hold: (1) the point-null $\mathcal{H}_0$ is never true exactly, and should therefore not be assigned separate mass; (2) only when $\mathcal{H}_0$ is assigned separate mass does the paradox manifest itself. With respect to the first argument, Jeffreys argued that assuming the falsity of the null without empirical evidence runs counter to scientific practice: ``The onus of proof is always on the advocate of the more complicated hypothesis.'' (\citealp[p. 278]{Jeffreys1939}; echoed in \citealp[p. 343]{Jeffreys1961}; but see \citealp{Gelman2009}). In addition, Jeffreys argued that assigning mass to the point-null hypothesis constitutes the best practical way of progress, yields better predictive performance, and prevents the haphazard inclusion of numerous parameters:
\begin{quotation}
``Some feeling of discomfort seems to attach itself to the assertion of the special value as \emph{right}, since it may be slightly wrong but not sufficiently to be revealed by a test on the data available; but no significance test asserts it as certainly right. We are aiming at the best way of progress, not at the unattainable ideal of immediate certainty. What happens if the null hypothesis is retained after a significance test is that the maximum likelihood solution or a solution given by some other method of estimation is rejected. The question is, When we do this, do we expect thereby to get more or less correct inferences than if we followed the rule of keeping the estimation solution regardless of any question of significance? I maintain that the only possible answer is that we expect to get more. The difference as estimated is interpreted as random error and irrelevant to future observations. In the last resort, if this interpretation is rejected, there is no escape from the admission that a new parameter may be needed for every observation, and then all combination of observations is meaningless, and the only valid presentation of data is a mere catalogue without any summaries at all.(...)

The distinction between problems of estimation and significance arises in biological applications, though I have naturally tended to speak mainly of physical ones. Suppose that a Mendelian finds in a breeding experiment 459 members of one type, 137 of the other. The expectations on the basis of a $3:1$ ratio would be 447 and 149. The difference would be declared not significant by any test. But the attitude that refuses to attach any meaning to the statement that the simple rule is right must apparently say that if any predictions are to be made from the observations the best that can be done is to make them on the basis of the ratio $459/137$, with allowance for the uncertainty of sampling. I say that the best is to use the $3/1$ rule, considering no uncertainty beyond the sampling errors of the new experiments. In fact the latter is what a geneticist would do. The observed result would be recorded and might possibly be reconsidered at a later stage if there was some question of differences of viability after many more observations had accumulated; but meanwhile it would be regarded as confirmation of the theoretical value. This is a problem of what I call significance.'' (\citealp[pp. 318-320]{Jeffreys1939}; echoed in \citealp[pp. 388-389]{Jeffreys1961}; italics in original)
\end{quotation}

With respect to the second argument --that the paradox manifests itself only when $\mathcal{H}_0$ is assigned separate mass--, it should first be noted that the paradox may be formulated not on the level of posterior probabilities but on the level of Bayes factors, as Jeffreys was wont to do. Thus, the paradox can be reformulated to state that data can always be found such that the $p$-value suggests that $\mathcal{H}_0$ should be rejected, whereas the Bayes factor indicates that the same data provide strong support in \emph{favor} of $\mathcal{H}_0$. Because the Bayes factor equals the ratio of marginal likelihoods under $\mathcal{H}_0$ and $\mathcal{H}_1$ it does not depend on the prior model probability that is assigned to $\mathcal{H}_0$ (cf. \citealp{Pericchi2011}). 

\begin{figure}
\begin{tabular}{cc}
    \begin{minipage}{.5 \textwidth}
    \centering
    \qquad \scalebox{1.25}{\( g_{0}=0.1, g_{1} = 1 \)}
    \includegraphics[width=\linewidth]{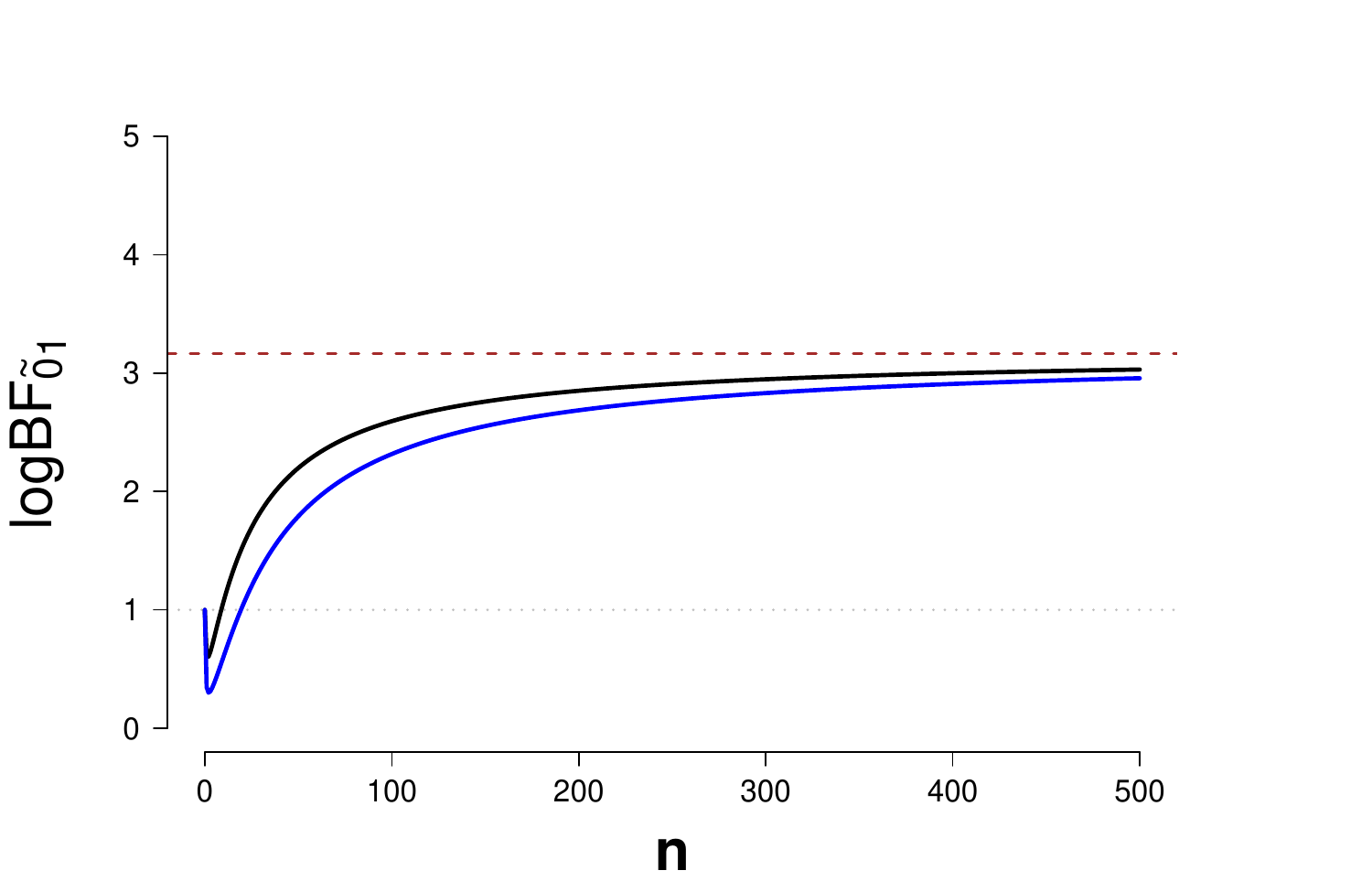}
    \end{minipage}  & %
    \begin{minipage}{.5 \textwidth}
    \centering
    \qquad \scalebox{1.25}{\( g_{0}=0.05, g_{1} = 1 \)}
    \includegraphics[width= \linewidth]{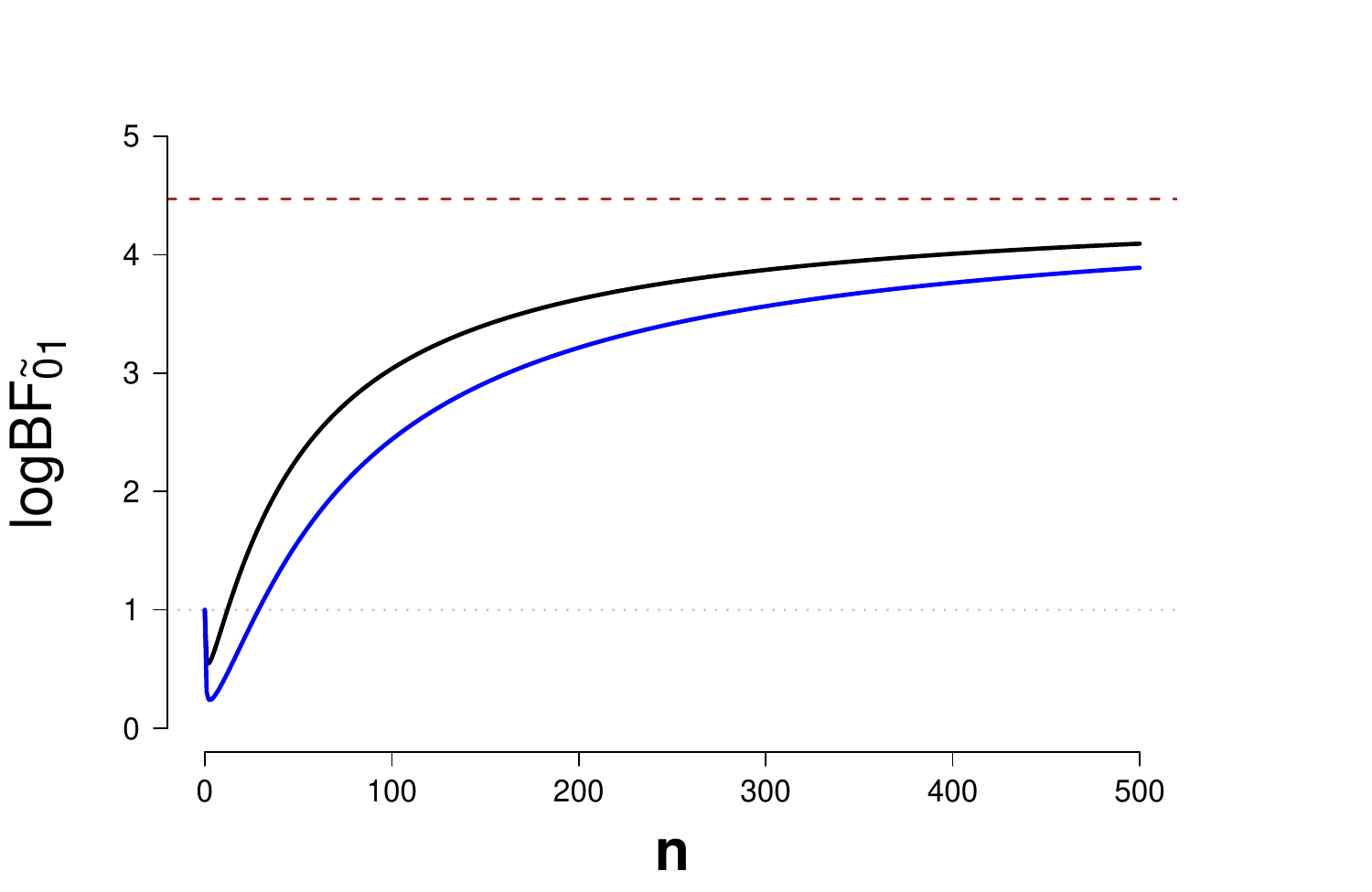}
    \end{minipage}  
\end{tabular}
\caption{Replacing the point-null hypothesis by a peri-null hypothesis does not avoid the Jeffreys-Lindley paradox. In the case of the Bayes factor $z$-test, increasing sample size $n$ for a fixed attained value of $\alpha$ inevitably results in positive evidence for the peri-null hypothesis. This evidence converges to an upper bound \( \sqrt{ g_{1} / g_{0}} \) that is indicated by the horizontal dashed brown line. The black and blue curves correspond to data that yield \( \alpha = .05 \) and \( \alpha =.01 \), respectively. Left panel: peri-null hypothesis with $g_0 = 0.1$; right panel: peri-null hypothesis with $g_0 = 0.05$.}
\label{figPerinull}
\end{figure}

The second argument can also be countered directly: as we show below, the paradox does \emph{not} require the presence of a point-null hypothesis. This fact is almost universally overlooked (for an exception see \citealp{Cousins2017}). Thus, granting that the point-null hypothesis \( \Hc_{0}: \delta = 0 \) is never true exactly, let us replace $\mathcal{H}_0$ by a peri-null hypothesis, say, \( \widetilde{\Hc}_{0}: \delta \sim \Nc ( 0, g_{0}) \) with variance $g_{0}$ small to reflect the skeptic's belief that the effect is near zero (e.g., \citealp{Lindley2011,LyWagenmakersinpressPeri,MoreyRouder2011}). The peri-null does not include point masses; yet, the Jeffreys-Lindley paradox still applies. For instance, consider the \( z \)-test with data normally distributed \( Y_{i} \overset{\text{iid}}{\sim} \Nc ( \mu, \sigma^{2}) \), where \( \sigma \) is known, say, \( \sigma = 1 \), and normal priors \( \delta = \mu/\sigma \sim \Nc ( 0, g_{k}) \) for \( k=0,1 \) with \( g_{0} < g_{1} \). The peri-null Bayes factor is then 
\begin{align}
\label{eq:PerinullBF}
\BF_{\widetilde{01}}(z,n)  = \sqrt{\frac{ 1+n g_{1} }{ 1+n g_{0} }} \exp \left ( \frac{(g_{0} - g_{1}) n z^{2} }{2 ( 1+n g_{0} ) ( 1+n g_{1} )} \right ) . 
\end{align}
Note that for the two-sided test with the \( \alpha \)-threshold fixed, we have that \( z = \Phi^{-1}(1-\alpha) \) where \( \Phi^{-1} \) is the quantile function of a standard normal distribution. By definition of the \( Z \)-statistic the fixed \( \alpha \) threshold can be expressed in terms of the sample mean and yields \( \bar{y} = \frac{\sigma}{\sqrt{n}} \Phi^{-1}(1- \alpha) \). Observe that with a fixed \( \alpha \) threshold, the value of \( \bar{y} \) at which the null is rejected goes to zero as \( n \) increases.  
Plugging \( z = \Phi^{-1}(1-\alpha) \) into Equation~\ref{eq:PerinullBF} shows that \( \lim_{n \rightarrow \infty} \BF_{\widetilde{01}}(z,n) = \sqrt{ g_{1} / g_{0}} \). Since \( g_{1} > g_{0} \) this implies that $\BF_{\widetilde{01}}$ will eventually provide evidence in favor of the peri-null hypothesis, even though \( p \leq \alpha \) suggests a rejection of the null. The limit \( \sqrt{ g_{1} / g_{0}} \) is the maximum evidence for the peri-null that can be attained, as for all \( \alpha \in (0, 1) \) the peri-null Bayes factor starts at one. Depending on \( g_{0} \) and \( g_{1} \), small values of \( n \) may result in a value of \( \BF_{\widetilde{01}}(z,n) \) that indicates some evidence for the alternative hypothesis; as $n$ increases, \( \BF_{\widetilde{01}}(z,n) \) will monotonically increase towards \( \sqrt{ g_{1} / g_{0}} \). A specific demonstration is provided in Figure~\ref{figPerinull}. %

The main effect of replacing the point-null hypothesis by a peri-null hypothesis is that for a fixed $p$-value, the evidence in favor of the null no longer grows without bound. However, with $g_0 < g_1$, the peri-null evidence bound \( \sqrt{ g_{1} / g_{0}} \) still favors the null over the alternative for any non-zero $\alpha$ attained. 

In sum, the Jeffreys-Lindley paradox does not depend on the presence of a point-null hypothesis, as is usually claimed. For fixed $p = \alpha$, the data will inevitably support a peri-null hypothesis over the alternative hypothesis as sample size grows large. The strength of this support is bounded, but in favor of the peri-null, thus leaving the conflict qualitatively intact. In other words, even when the point-null is replaced by a peri-null hypothesis, ``there would be cases, with large numbers of observations, when a new parameter is asserted on evidence that is actually against it.'' \citep[pp. 379]{Jeffreys1938Comparison2}.

\subsection*{Objection 2: The Paradox Signals that the Prior Distribution Was Too Wide}

Whenever the paradox occurs, a natural objection to the Bayes factor outcome is that the prior distribution for the test-relevant parameter under $\mathcal{H}_1$ was too wide, wasting considerable prior mass on large values of effect size that yield poor predictive performance. Thus, as implied by \citet{Bartlett1957}, the paradox reveals a fault in the specification of $\mathcal{H}_1$ rather than $\mathcal{H}_0$.

This objection is valid in the sense that --as $n$ increases and the $p$-value remains constant-- the increasingly poor predictive performance of $\mathcal{H}_1$ is indeed due to the fact that an increasing proportion of prior mass is inconsistent with the data, and this is the root of the paradox. For instance, the paradox would not arise if the predictions of $\mathcal{H}_1$ were evaluated under the maximum likelihood estimator $\hat{\theta}$ \citep{Cousins2017}. However, as mentioned above, $\hat{\theta}$ is a cherry-picked value, and using it would favor $\mathcal{H}_1$ over $\mathcal{H}_0$ regardless of the data.

In general terms, the critique that the prior was too wide is made post-hoc; after observing a near-zero effect size one may always argue that, in hindsight, the prior was too wide -- if such reasoning were allowed then the data could never undercut $\mathcal{H}_1$ and support $\mathcal{H}_0$. As long as the prior width does not shrink as a function of sample size, the paradox arises under any non-zero prior width. 

\section*{Concluding Comments}
In this paper we examined the history and nature of the Jeffreys-Lindley paradox. Our main conclusions are as follows:
\begin{enumerate}
    \item Contrary to what the current literature suggest (e.g., \citealp[p. 394]{BernardoSmith2000}; \citealp[p. 78]{OHaganForster2004}), the Jeffreys-Lindley paradox was central to Harold Jeffreys's philosophy of Bayesian testing; in Jeffreys's tests, the critical threshold is not based on a constant multiple of the standard error but instead involves a $\sqrt{n}$ term.\footnote{For what it's worth, a Google search for ``Lindley paradox'' or ``Lindley's paradox'' yields about 6,190 results, whereas ``Jeffreys-Lindley paradox'' yields about 3.530 results; the phrase ``Jeffreys's paradox'' or ``Jeffreys paradox'' or ``Jeffreys' paradox'' yields 964 results (July 5th, 2021).}
    \item From 1935 to 1936, Jeffreys had discovered, understood, published, emphasized, explained, and illustrated the paradox. It remained a recurring theme throughout his later articles and books.
    \item The articles by \citet{Lindley1957} and \citet{Bartlett1957} echo earlier work by Jeffreys. This is acknowledged by both authors, but they do not seem fully aware of the extent to which Jeffreys had already studied the issue. The two 1957 articles also introduced some mathematical errors and conceptual misunderstandings.\footnote{In his later work, Jeffreys never cited the 1957 articles, perhaps because he felt these did not offer novel insights.}
    \item The paradox is caused by the fact that, as $n$ increases and $p$ remains constant, an ever increasing set of parameter values under $\mathcal{H}_1$ is inconsistent with the observed data, decreasing $\mathcal{H}_1$'s average predictive performance (i.e., the marginal likelihood).
    \item A fully frequentist version of the paradox contrasts the inductive behavior of two frequentists, one who fixes $\alpha$ and minimizes $\beta$, the other who minimizes a weighted linear sum of $\alpha$ and $\beta$ (e.g., \citealp{Cornfield1966,Lindley1953,Lehmann1958}). As $n$ grows large, the same data that prompt the former frequentist to reject $\mathcal{H}_0$ will prompt the latter frequentist to retain $\mathcal{H}_0$. The behavior of the latter frequentist is qualitatively consistent with the tests proposed by Jeffreys.
    \item A fully Bayesian version of the paradox contrasts the beliefs of two Bayesians, one who tests $\mathcal{H}_+:\delta > 0$ versus $\mathcal{H}_-:\delta < 0$ (i.e., the direction of the effect), the other who tests $\mathcal{H}_0:\delta = 0$ versus $\mathcal{H}_1:\delta \neq 0$ (i.e., the presence of the effect). As $n$ grows large, the same data that prompt the former Bayesian to conclude that the data offer strong support for the hypothesis that the effect is positive will prompt the latter Bayesian to conclude that the data offer strong support for the hypothesis that the effect is absent.  
    \item Contrary to what the current literature suggest, the root of the paradox is not in the assignment of prior mass to a point hypothesis $\mathcal{H}_0$; the paradox is also present when the point-null hypothesis is replaced by a peri-null hypothesis (i.e., a relatively peaked continuous distribution). 
    \item The Jeffreys-Lindley paradox is relatively robust: it holds whether or not $\mathcal{H}_0$ is a point-null or a peri-null hypothesis, and it holds regardless of the width of the prior distribution for the test-relevant parameter under $\mathcal{H}_1$ -- as long as the width is larger than that of the prior distribution under the peri-null hypothesis, and as long as it does not shrink with sample size.
    \item The Jeffreys-Lindley paradox results from the discrepancy between two modes of inference: (1) evaluating a single model (e.g., fixed-$\alpha$ decision making); (2) contrasting two models, one of which is relatively simple (e.g., the skeptic's $\mathcal{H}_0$) and one which is more complex (e.g., the proponent's $\mathcal{H}_1$). \rev{In other words, the traditional frequentist test is absolute, whereas Jeffreys's Bayes factor test is relative.}
\end{enumerate}

We wish to emphasize that, when discussion the paradox, Lindley himself was always careful to credit Jeffreys (e.g., \citealp[p. 119]{Robert2013}: ``Dennis systematically refereed [sic] to Jeffreys for stating the paradox, both in his paper and his personal communications.''). However, it appears that Lindley did not fully appreciate the degree to which Jeffreys had worked on the paradox in the 1930s already. This may appear surprising, since Lindley had taken classes from Jeffreys; indeed, Lindley may be considered one of only a handful of statisticians who were keenly aware of Jeffreys's statistical methodology. A hint at the reason for this blind spot is given by Lindley himself, in a festschrift in honor of Jeffreys: 
\begin{quotation}
``There have been several occasions on which one of us statisticians has asked Jeffreys about some point, and his answer has been ``I dealt with that in the \emph{Theory}'' and he would go on to point out where. The questioner would then return to his room, take the book down from his shelf and sure enough, after some thinking, he would realize that the point was discussed there and that the discussion went some, if not the whole, way to answering the original question. In that last sentence I say ``after some thinking'' because Jeffreys's style does not give immediate comprehension. It is necessary to work at it. In my experience illumination usually appears and one wonders why it was so difficult to see at first. That is one reason why the book, although widely bought, has not been read or cited as much as it ought.'' (\citealp[p. 37]{Lindley1980}; italics in original)
\end{quotation}
We share Lindley's experience. In fact, we have studied Jeffreys's work for many years, and we have reread \emph{Theory of Probability} several times over. Only recently did it dawn on us that the paradox was a central element of Jeffreys's statistical philosophy on hypothesis testing. We cannot offer a compelling explanation for why this was so difficult for us to see at first.

It is certainly the case that Jeffreys underplayed the differences between $p$-values and Bayes factors from a pragmatic point of view. For instance, Jeffreys stated that ``The rule that a difference becomes significant at about two or three times its standard error is therefore about right for ordinary numbers of observations.'' \citep[p. 213]{Jeffreys1935} and ``Thus even though $P$ tests sometimes theoretically assert $\sim  \! q$ when the number of observations is large and my tests support $q$, the occasions will be extremely rare.'' (\citealp[p. 360]{Jeffreys1939}, echoed in \citealp[p. 435]{Jeffreys1961}). \rev{Moreover, Jeffreys felt that in such cases the data often indicate model misspecification, in the sense that both the null hypothesis and the alternative hypothesis are found wanting, and a closer consideration of the data may suggest a third alternative (e.g., \citealp[p. 310]{Jeffreys1938Continuous}; \citealp[p. 436]{Jeffreys1961}).} 

Jeffreys's assessment of the $p$-value as ``about right for ordinary numbers of observations'' conflicts with the assessment of later Bayesians (e.g., \citealp{BergerDelampady1987,Edwards1965,EdwardsEtAl1963,SellkeEtAl2001}), who have argued that $p$-values just below $.05$ do not constitute compelling evidence against $\mathcal{H}_0$. Jeffreys's relatively mild assessment is due to the fact that he calibrated $p=.05$ to $\text{BF}_{10}=1$. However, it may be argued that in order to ``reject'' the null hypothesis we need strong evidence, or at least not evidence that is ``hardly worth mentioning'' (when $1 < \text{BF}_{10} < 3$; \citealp[p. 357]{Jeffreys1939}). In addition, few researchers will consider a $p$-value of exactly $.05$ as the point where they believe the data to be entirely uninformative. 
Would Jeffreys have endorsed the recent proposal to reduce the significance level for new discoveries from $\alpha=.05$ to $\alpha=.005$ \citep{BenjaminEtAl2018}? We believe he would have had reservations. Although the proposal was motivated in part by Bayesian insights that originate from Jeffreys himself, the stricter $\alpha$ level still entails a threshold that is a constant multiple of the standard error and omits the crucial $\sqrt{n}$ term. Moreover, endorsing the $\alpha=.005$ proposal would mean an implicit admission that his repeated reassurances concerning the use of $\alpha=.05$ as ``about right'' were in fact wrong. 

A thorough understanding of the Jeffreys-Lindley paradox remains critically important for the assessment of statistical methodology, both old and new. Ultimately, the paradox may even bring about some reconciliation between the Bayesian and the frequentist frameworks -- in particular, the paradox may motivate frequentists to explore procedures that minimize the weighted sum of $\alpha$ and $\beta$, which ought to yield conclusions similar to those obtained with Jeffreys's Bayesian tests (cf. \citealp{Lindley1953,PericchiPereira2016}). We believe this manuscript provides some new historical and conceptual background to the Jeffreys-Lindley paradox, and we hope that this will be useful for statistical theory as well as statistical practice.  

\section*{Statements and Declarations}
The authors declare that they coordinate the development of the open-source
software package JASP (\url{https://jasp-stats.org}), a non-commercial,
publicly-funded effort to make Bayesian and non-Bayesian statistics
accessible to a broader group of researchers and students.

\newpage
\bibliographystyle{plainnat}
\bibliography{referenties,Alexander}

\newcommand{\noop}[1]{}
\begin{thebibliography}{}

\bibitem [\protect \citeauthoryear {%
Aitkin%
}{%
Aitkin%
}{%
{\protect \APACyear {1991}}%
}]{%
Aitkin1991}
\APACinsertmetastar {%
Aitkin1991}%
\begin{APACrefauthors}%
Aitkin, M.%
\end{APACrefauthors}%
\unskip\
\newblock
\APACrefYearMonthDay{1991}{}{}.
\newblock
{\BBOQ}\APACrefatitle {Posterior {B}ayes Factors} {Posterior {B}ayes
  factors}.{\BBCQ}
\newblock
\APACjournalVolNumPages{{Journal of the Royal Statistical Society. Series B
  (Methodological)}}{53}{}{111--142}.
\PrintBackRefs{\CurrentBib}

\bibitem [\protect \citeauthoryear {%
Andrews%
}{%
Andrews%
}{%
{\protect \APACyear {1994}}%
}]{%
Andrews1994}
\APACinsertmetastar {%
Andrews1994}%
\begin{APACrefauthors}%
Andrews, D\BPBI W\BPBI K.%
\end{APACrefauthors}%
\unskip\
\newblock
\APACrefYearMonthDay{1994}{}{}.
\newblock
{\BBOQ}\APACrefatitle {The Large Sample Correspondence between Classical
  Hypothesis Tests and {B}ayesian Posterior Odds Tests} {The large sample
  correspondence between classical hypothesis tests and {B}ayesian posterior
  odds tests}.{\BBCQ}
\newblock
\APACjournalVolNumPages{Econometrica}{62}{}{1207--1232}.
\PrintBackRefs{\CurrentBib}

\bibitem [\protect \citeauthoryear {%
Bartlett%
}{%
Bartlett%
}{%
{\protect \APACyear {1957}}%
}]{%
Bartlett1957}
\APACinsertmetastar {%
Bartlett1957}%
\begin{APACrefauthors}%
Bartlett, M\BPBI S.%
\end{APACrefauthors}%
\unskip\
\newblock
\APACrefYearMonthDay{1957}{}{}.
\newblock
{\BBOQ}\APACrefatitle {A Comment on {D. V. Lindley}'s Statistical Paradox} {A
  comment on {D. V. Lindley}'s statistical paradox}.{\BBCQ}
\newblock
\APACjournalVolNumPages{Biometrika}{44}{}{533--534}.
\PrintBackRefs{\CurrentBib}

\bibitem [\protect \citeauthoryear {%
Bayarri%
, Berger%
, Forte%
\BCBL {}\ \BBA {} {Garc\'{\i}a-Donato}%
}{%
Bayarri%
\ \protect \BOthers {.}}{%
{\protect \APACyear {2012}}%
}]{%
BayarriEtAl2012}
\APACinsertmetastar {%
BayarriEtAl2012}%
\begin{APACrefauthors}%
Bayarri, M\BPBI J.%
, Berger, J\BPBI O.%
, Forte, A.%
\BCBL {}\ \BBA {} {Garc\'{\i}a-Donato}, G.%
\end{APACrefauthors}%
\unskip\
\newblock
\APACrefYearMonthDay{2012}{}{}.
\newblock
{\BBOQ}\APACrefatitle {Criteria for {B}ayesian Model Choice With Application to
  Variable Selection} {Criteria for {B}ayesian model choice with application to
  variable selection}.{\BBCQ}
\newblock
\APACjournalVolNumPages{The Annals of Statistics}{40}{}{1550--1577}.
\PrintBackRefs{\CurrentBib}

\bibitem [\protect \citeauthoryear {%
Benjamin%
\ \protect \BOthers {.}}{%
Benjamin%
\ \protect \BOthers {.}}{%
{\protect \APACyear {2018}}%
}]{%
BenjaminEtAl2018}
\APACinsertmetastar {%
BenjaminEtAl2018}%
\begin{APACrefauthors}%
Benjamin, D\BPBI J.%
, Berger, J\BPBI O.%
, Johannesson, M.%
, Nosek, B\BPBI A.%
, Wagenmakers, E\BHBI J.%
, Berk, R.%
\BDBL {}Johnson, V\BPBI E.%
\end{APACrefauthors}%
\unskip\
\newblock
\APACrefYearMonthDay{2018}{}{}.
\newblock
{\BBOQ}\APACrefatitle {Redefine Statistical Significance} {Redefine statistical
  significance}.{\BBCQ}
\newblock
\APACjournalVolNumPages{Nature Human Behaviour}{2}{}{6--10}.
\PrintBackRefs{\CurrentBib}

\bibitem [\protect \citeauthoryear {%
Bennett%
}{%
Bennett%
}{%
{\protect \APACyear {1990}}%
}]{%
Bennett1990}
\APACinsertmetastar {%
Bennett1990}%
\begin{APACrefauthors}%
Bennett, J\BPBI H.%
\end{APACrefauthors}%
\ (\BED).
\unskip\
\newblock
\APACrefYear{1990}.
\newblock
\APACrefbtitle {Statistical Inference and Analysis: {S}elected Correspondence
  of {R. A. Fisher}} {Statistical inference and analysis: {S}elected
  correspondence of {R. A. Fisher}}.
\newblock
\APACaddressPublisher{Oxford}{Clarendon Press}.
\PrintBackRefs{\CurrentBib}

\bibitem [\protect \citeauthoryear {%
Berger%
\ \BBA {} Delampady%
}{%
Berger%
\ \BBA {} Delampady%
}{%
{\protect \APACyear {1987}}%
}]{%
BergerDelampady1987}
\APACinsertmetastar {%
BergerDelampady1987}%
\begin{APACrefauthors}%
Berger, J\BPBI O.%
\BCBT {}\ \BBA {} Delampady, M.%
\end{APACrefauthors}%
\unskip\
\newblock
\APACrefYearMonthDay{1987}{}{}.
\newblock
{\BBOQ}\APACrefatitle {Testing Precise Hypotheses} {Testing precise
  hypotheses}.{\BBCQ}
\newblock
\APACjournalVolNumPages{Statistical Science}{2}{}{317--352}.
\PrintBackRefs{\CurrentBib}

\bibitem [\protect \citeauthoryear {%
Berkson%
}{%
Berkson%
}{%
{\protect \APACyear {1942}}%
}]{%
Berkson1942}
\APACinsertmetastar {%
Berkson1942}%
\begin{APACrefauthors}%
Berkson, J.%
\end{APACrefauthors}%
\unskip\
\newblock
\APACrefYearMonthDay{1942}{}{}.
\newblock
{\BBOQ}\APACrefatitle {Tests of Significance Considered as Evidence} {Tests of
  significance considered as evidence}.{\BBCQ}
\newblock
\APACjournalVolNumPages{Journal of the American Statistical
  Association}{37}{}{325--335}.
\PrintBackRefs{\CurrentBib}

\bibitem [\protect \citeauthoryear {%
Bernardo%
}{%
Bernardo%
}{%
{\protect \APACyear {1980}}%
}]{%
Bernardo1980}
\APACinsertmetastar {%
Bernardo1980}%
\begin{APACrefauthors}%
Bernardo, J\BPBI M.%
\end{APACrefauthors}%
\unskip\
\newblock
\APACrefYearMonthDay{1980}{}{}.
\newblock
{\BBOQ}\APACrefatitle {A {B}ayesian Analysis of Classical Hypothesis Testing
  (With Discussion)} {A {B}ayesian analysis of classical hypothesis testing
  (with discussion)}.{\BBCQ}
\newblock
\APACjournalVolNumPages{{Trabajos de Estadistica y de Investigacion
  Operativa}}{31}{}{605--647}.
\PrintBackRefs{\CurrentBib}

\bibitem [\protect \citeauthoryear {%
Bernardo%
}{%
Bernardo%
}{%
{\protect \APACyear {2009}}%
}]{%
Bernardo2009}
\APACinsertmetastar {%
Bernardo2009}%
\begin{APACrefauthors}%
Bernardo, J\BPBI M.%
\end{APACrefauthors}%
\unskip\
\newblock
\APACrefYearMonthDay{2009}{}{}.
\newblock
{\BBOQ}\APACrefatitle {{[Harold Jeffreys's theory of probability revisited]:
  Comment}} {{[Harold Jeffreys's theory of probability revisited]:
  Comment}}.{\BBCQ}
\newblock
\APACjournalVolNumPages{Statistical Science}{24}{}{173--175}.
\PrintBackRefs{\CurrentBib}

\bibitem [\protect \citeauthoryear {%
Bernardo%
}{%
Bernardo%
}{%
{\protect \APACyear {2011}}%
}]{%
Bernardo2011}
\APACinsertmetastar {%
Bernardo2011}%
\begin{APACrefauthors}%
Bernardo, J\BPBI M.%
\end{APACrefauthors}%
\unskip\
\newblock
\APACrefYearMonthDay{2011}{}{}.
\newblock
{\BBOQ}\APACrefatitle {Integrated Objective {B}ayesian Estimation and
  Hypothesis Testing} {Integrated objective {B}ayesian estimation and
  hypothesis testing}.{\BBCQ}
\newblock
\BIn{} J\BPBI M.~Bernardo\ \BOthers {.}\ (\BEDS), \APACrefbtitle {{B}ayesian
  Statistics 9} {{B}ayesian statistics 9}\ (\BPGS\ 1--68).
\newblock
\APACaddressPublisher{Oxford}{Oxford University Press}.
\PrintBackRefs{\CurrentBib}

\bibitem [\protect \citeauthoryear {%
Bernardo%
\ \BBA {} Smith%
}{%
Bernardo%
\ \BBA {} Smith%
}{%
{\protect \APACyear {2000}}%
}]{%
BernardoSmith2000}
\APACinsertmetastar {%
BernardoSmith2000}%
\begin{APACrefauthors}%
Bernardo, J\BPBI M.%
\BCBT {}\ \BBA {} Smith, A\BPBI F\BPBI M.%
\end{APACrefauthors}%
\unskip\
\newblock
\APACrefYear{2000}.
\newblock
\APACrefbtitle {{B}ayesian Theory} {{B}ayesian theory}.
\newblock
\APACaddressPublisher{Chichester}{Wiley}.
\PrintBackRefs{\CurrentBib}

\bibitem [\protect \citeauthoryear {%
Berrar%
\ \BBA {} Dubitzky%
}{%
Berrar%
\ \BBA {} Dubitzky%
}{%
{\protect \APACyear {2017}}%
}]{%
berrar2017jeffreys}
\APACinsertmetastar {%
berrar2017jeffreys}%
\begin{APACrefauthors}%
Berrar, D.%
\BCBT {}\ \BBA {} Dubitzky, W.%
\end{APACrefauthors}%
\unskip\
\newblock
\APACrefYearMonthDay{2017}{}{}.
\newblock
{\BBOQ}\APACrefatitle {On the {Jeffreys-Lindley} Paradox and the looming
  reproducibility crisis in machine learning} {On the {Jeffreys-Lindley}
  paradox and the looming reproducibility crisis in machine learning}.{\BBCQ}
\newblock
\BIn{} \APACrefbtitle {2017 IEEE International Conference on Data Science and
  Advanced Analytics (DSAA)} {2017 ieee international conference on data
  science and advanced analytics (dsaa)}\ (\BPGS\ 334--340).
\PrintBackRefs{\CurrentBib}

\bibitem [\protect \citeauthoryear {%
Burnham%
\ \BBA {} Anderson%
}{%
Burnham%
\ \BBA {} Anderson%
}{%
{\protect \APACyear {2004}}%
}]{%
BurnhamAnderson2004}
\APACinsertmetastar {%
BurnhamAnderson2004}%
\begin{APACrefauthors}%
Burnham, K\BPBI P.%
\BCBT {}\ \BBA {} Anderson, D\BPBI R.%
\end{APACrefauthors}%
\unskip\
\newblock
\APACrefYearMonthDay{2004}{}{}.
\newblock
{\BBOQ}\APACrefatitle {Multimodel Inference: {U}nderstanding {AIC} and {BIC} in
  Model Selection} {Multimodel inference: {U}nderstanding {AIC} and {BIC} in
  model selection}.{\BBCQ}
\newblock
\APACjournalVolNumPages{Sociological Methods \& Research}{33}{}{261--304}.
\PrintBackRefs{\CurrentBib}

\bibitem [\protect \citeauthoryear {%
Casella%
\ \BBA {} Berger%
}{%
Casella%
\ \BBA {} Berger%
}{%
{\protect \APACyear {1987}}%
}]{%
CasellaBerger1987}
\APACinsertmetastar {%
CasellaBerger1987}%
\begin{APACrefauthors}%
Casella, G.%
\BCBT {}\ \BBA {} Berger, R\BPBI L.%
\end{APACrefauthors}%
\unskip\
\newblock
\APACrefYearMonthDay{1987}{}{}.
\newblock
{\BBOQ}\APACrefatitle {Reconciling {B}ayesian and Frequentist Evidence in the
  One--Sided Testing Problem} {Reconciling {B}ayesian and frequentist evidence
  in the one--sided testing problem}.{\BBCQ}
\newblock
\APACjournalVolNumPages{Journal of the American Statistical
  Association}{82}{}{106--111}.
\PrintBackRefs{\CurrentBib}

\bibitem [\protect \citeauthoryear {%
Colquhoun%
}{%
Colquhoun%
}{%
{\protect \APACyear {2019}}%
}]{%
Colquhoun2019}
\APACinsertmetastar {%
Colquhoun2019}%
\begin{APACrefauthors}%
Colquhoun, D.%
\end{APACrefauthors}%
\unskip\
\newblock
\APACrefYearMonthDay{2019}{}{}.
\newblock
{\BBOQ}\APACrefatitle {The False Positive Risk: {A} Proposal Concerning What to
  Do About $p$--Values} {The false positive risk: {A} proposal concerning what
  to do about $p$--values}.{\BBCQ}
\newblock
\APACjournalVolNumPages{The American Statistician}{73}{}{192--201}.
\PrintBackRefs{\CurrentBib}

\bibitem [\protect \citeauthoryear {%
Consonni%
, Fouskakis%
, Liseo%
\BCBL {}\ \BBA {} Ntzoufras%
}{%
Consonni%
\ \protect \BOthers {.}}{%
{\protect \APACyear {2018}}%
}]{%
ConsonniEtAl2018}
\APACinsertmetastar {%
ConsonniEtAl2018}%
\begin{APACrefauthors}%
Consonni, G.%
, Fouskakis, D.%
, Liseo, B.%
\BCBL {}\ \BBA {} Ntzoufras, I.%
\end{APACrefauthors}%
\unskip\
\newblock
\APACrefYearMonthDay{2018}{}{}.
\newblock
{\BBOQ}\APACrefatitle {Prior Distributions for Objective {B}ayesian Analysis}
  {Prior distributions for objective {B}ayesian analysis}.{\BBCQ}
\newblock
\APACjournalVolNumPages{Bayesian Analysis}{13}{}{627--679}.
\PrintBackRefs{\CurrentBib}

\bibitem [\protect \citeauthoryear {%
Cornfield%
}{%
Cornfield%
}{%
{\protect \APACyear {1966}}%
}]{%
Cornfield1966}
\APACinsertmetastar {%
Cornfield1966}%
\begin{APACrefauthors}%
Cornfield, J.%
\end{APACrefauthors}%
\unskip\
\newblock
\APACrefYearMonthDay{1966}{}{}.
\newblock
{\BBOQ}\APACrefatitle {Sequential Trials, Sequential Analysis, and the
  Likelihood Principle} {Sequential trials, sequential analysis, and the
  likelihood principle}.{\BBCQ}
\newblock
\APACjournalVolNumPages{The American Statistician}{20}{}{18--23}.
\PrintBackRefs{\CurrentBib}

\bibitem [\protect \citeauthoryear {%
Cousins%
}{%
Cousins%
}{%
{\protect \APACyear {2017}}%
}]{%
Cousins2017}
\APACinsertmetastar {%
Cousins2017}%
\begin{APACrefauthors}%
Cousins, R\BPBI D.%
\end{APACrefauthors}%
\unskip\
\newblock
\APACrefYearMonthDay{2017}{}{}.
\newblock
{\BBOQ}\APACrefatitle {The {J}effreys--{L}indley Paradox and Discovery Criteria
  in High Energy Physics} {The {J}effreys--{L}indley paradox and discovery
  criteria in high energy physics}.{\BBCQ}
\newblock
\APACjournalVolNumPages{Synthese}{194}{}{395--432}.
\PrintBackRefs{\CurrentBib}

\bibitem [\protect \citeauthoryear {%
Cox%
}{%
Cox%
}{%
{\protect \APACyear {2006}}%
}]{%
Cox2006}
\APACinsertmetastar {%
Cox2006}%
\begin{APACrefauthors}%
Cox, D\BPBI R.%
\end{APACrefauthors}%
\unskip\
\newblock
\APACrefYear{2006}.
\newblock
\APACrefbtitle {Principles of Statistical Inference} {Principles of statistical
  inference}.
\newblock
\APACaddressPublisher{Cambridge}{Cambridge University Press}.
\PrintBackRefs{\CurrentBib}

\bibitem [\protect \citeauthoryear {%
{de Bragan\c{c}a Pereira}%
\ \BBA {} Stern%
}{%
{de Bragan\c{c}a Pereira}%
\ \BBA {} Stern%
}{%
{\protect \APACyear {1999}}%
}]{%
PereiraStern1999}
\APACinsertmetastar {%
PereiraStern1999}%
\begin{APACrefauthors}%
{de Bragan\c{c}a Pereira}, C\BPBI A.%
\BCBT {}\ \BBA {} Stern, J\BPBI M.%
\end{APACrefauthors}%
\unskip\
\newblock
\APACrefYearMonthDay{1999}{}{}.
\newblock
{\BBOQ}\APACrefatitle {Evidence and Credibility: {F}ull {B}ayesian Significance
  Test for Precise Hypotheses} {Evidence and credibility: {F}ull {B}ayesian
  significance test for precise hypotheses}.{\BBCQ}
\newblock
\APACjournalVolNumPages{Entropy}{1}{}{99--110}.
\PrintBackRefs{\CurrentBib}

\bibitem [\protect \citeauthoryear {%
{de Bragan\c{c}a Pereira}%
, Stern%
\BCBL {}\ \BBA {} Wechsler%
}{%
{de Bragan\c{c}a Pereira}%
\ \protect \BOthers {.}}{%
{\protect \APACyear {2008}}%
}]{%
PereiraEtAl2008}
\APACinsertmetastar {%
PereiraEtAl2008}%
\begin{APACrefauthors}%
{de Bragan\c{c}a Pereira}, C\BPBI A.%
, Stern, J\BPBI M.%
\BCBL {}\ \BBA {} Wechsler, S.%
\end{APACrefauthors}%
\unskip\
\newblock
\APACrefYearMonthDay{2008}{}{}.
\newblock
{\BBOQ}\APACrefatitle {Can a Significance Test Be Genuinely Bayesian} {Can a
  significance test be genuinely bayesian}.{\BBCQ}
\newblock
\APACjournalVolNumPages{Bayesian Analysis}{3}{}{79--100}.
\PrintBackRefs{\CurrentBib}

\bibitem [\protect \citeauthoryear {%
{DeGroot}%
\ \BBA {} Schervish%
}{%
{DeGroot}%
\ \BBA {} Schervish%
}{%
{\protect \APACyear {2012}}%
}]{%
DeGrootSchervish2012}
\APACinsertmetastar {%
DeGrootSchervish2012}%
\begin{APACrefauthors}%
{DeGroot}, M\BPBI H.%
\BCBT {}\ \BBA {} Schervish, M\BPBI J.%
\end{APACrefauthors}%
\unskip\
\newblock
\APACrefYear{2012}.
\newblock
\APACrefbtitle {Probability and Statistics} {Probability and statistics}\
  (\PrintOrdinal{4}\ \BEd).
\newblock
\APACaddressPublisher{New York}{{Addison--Wesley}}.
\PrintBackRefs{\CurrentBib}

\bibitem [\protect \citeauthoryear {%
Dickey%
}{%
Dickey%
}{%
{\protect \APACyear {1971}}%
}]{%
Dickey1971}
\APACinsertmetastar {%
Dickey1971}%
\begin{APACrefauthors}%
Dickey, J\BPBI M.%
\end{APACrefauthors}%
\unskip\
\newblock
\APACrefYearMonthDay{1971}{}{}.
\newblock
{\BBOQ}\APACrefatitle {The Weighted Likelihood Ratio, Linear Hypotheses on
  Normal Location Parameters} {The weighted likelihood ratio, linear hypotheses
  on normal location parameters}.{\BBCQ}
\newblock
\APACjournalVolNumPages{The Annals of Mathematical Statistics}{42}{}{204--223}.
\PrintBackRefs{\CurrentBib}

\bibitem [\protect \citeauthoryear {%
Edwards%
}{%
Edwards%
}{%
{\protect \APACyear {1965}}%
}]{%
Edwards1965}
\APACinsertmetastar {%
Edwards1965}%
\begin{APACrefauthors}%
Edwards, W.%
\end{APACrefauthors}%
\unskip\
\newblock
\APACrefYearMonthDay{1965}{}{}.
\newblock
{\BBOQ}\APACrefatitle {Tactical Note On The Relation Between Scientific And
  Statistical Hypotheses} {Tactical note on the relation between scientific and
  statistical hypotheses}.{\BBCQ}
\newblock
\APACjournalVolNumPages{Psychological Bulletin}{63}{}{400--402}.
\PrintBackRefs{\CurrentBib}

\bibitem [\protect \citeauthoryear {%
Edwards%
, Lindman%
\BCBL {}\ \BBA {} Savage%
}{%
Edwards%
\ \protect \BOthers {.}}{%
{\protect \APACyear {1963}}%
}]{%
EdwardsEtAl1963}
\APACinsertmetastar {%
EdwardsEtAl1963}%
\begin{APACrefauthors}%
Edwards, W.%
, Lindman, H.%
\BCBL {}\ \BBA {} Savage, L\BPBI J.%
\end{APACrefauthors}%
\unskip\
\newblock
\APACrefYearMonthDay{1963}{}{}.
\newblock
{\BBOQ}\APACrefatitle {{B}ayesian Statistical Inference for Psychological
  Research} {{B}ayesian statistical inference for psychological
  research}.{\BBCQ}
\newblock
\APACjournalVolNumPages{Psychological Review}{70}{}{193--242}.
\PrintBackRefs{\CurrentBib}

\bibitem [\protect \citeauthoryear {%
Etz%
\ \BBA {} Wagenmakers%
}{%
Etz%
\ \BBA {} Wagenmakers%
}{%
{\protect \APACyear {2017}}%
}]{%
etz2017haldane}
\APACinsertmetastar {%
etz2017haldane}%
\begin{APACrefauthors}%
Etz, A.%
\BCBT {}\ \BBA {} Wagenmakers, E\BHBI J.%
\end{APACrefauthors}%
\unskip\
\newblock
\APACrefYearMonthDay{2017}{}{}.
\newblock
{\BBOQ}\APACrefatitle {{J. B. S.} {H}aldane's Contribution to the {B}ayes
  Factor Hypothesis Test} {{J. B. S.} {H}aldane's contribution to the {B}ayes
  factor hypothesis test}.{\BBCQ}
\newblock
\APACjournalVolNumPages{Statistical Science}{32}{2}{313--329}.
\PrintBackRefs{\CurrentBib}

\bibitem [\protect \citeauthoryear {%
Fienberg%
}{%
Fienberg%
}{%
{\protect \APACyear {2003}}%
}]{%
Fienberg2003}
\APACinsertmetastar {%
Fienberg2003}%
\begin{APACrefauthors}%
Fienberg, S\BPBI E.%
\end{APACrefauthors}%
\unskip\
\newblock
\APACrefYearMonthDay{2003}{}{}.
\newblock
{\BBOQ}\APACrefatitle {When did {B}ayesian Inference Become ``{B}ayesian"?}
  {When did {B}ayesian inference become ``{B}ayesian"?}{\BBCQ}
\newblock
\APACjournalVolNumPages{{B}ayesian Analysis}{1}{}{1--41}.
\PrintBackRefs{\CurrentBib}

\bibitem [\protect \citeauthoryear {%
Fisher%
}{%
Fisher%
}{%
{\protect \APACyear {1934}}%
}]{%
fisher1934statistical5}
\APACinsertmetastar {%
fisher1934statistical5}%
\begin{APACrefauthors}%
Fisher, R\BPBI A.%
\end{APACrefauthors}%
\unskip\
\newblock
\APACrefYear{1934}.
\newblock
\APACrefbtitle {Statistical Methods for Research Workers (5th ed.)}
  {Statistical methods for research workers (5th ed.)}.
\newblock
\APACaddressPublisher{London}{Oliver and Boyd}.
\PrintBackRefs{\CurrentBib}

\bibitem [\protect \citeauthoryear {%
Fisher%
}{%
Fisher%
}{%
{\protect \APACyear {1935}}%
}]{%
Fisher1935b}
\APACinsertmetastar {%
Fisher1935b}%
\begin{APACrefauthors}%
Fisher, R\BPBI A.%
\end{APACrefauthors}%
\unskip\
\newblock
\APACrefYear{1935}.
\newblock
\APACrefbtitle {The Design of Experiments} {The design of experiments}.
\newblock
\APACaddressPublisher{Edinburgh}{Oliver and Boyd}.
\PrintBackRefs{\CurrentBib}

\bibitem [\protect \citeauthoryear {%
Fisher%
}{%
Fisher%
}{%
{\protect \APACyear {1936}}%
}]{%
Fisher1936}
\APACinsertmetastar {%
Fisher1936}%
\begin{APACrefauthors}%
Fisher, R\BPBI A.%
\end{APACrefauthors}%
\unskip\
\newblock
\APACrefYear{1936}.
\newblock
\APACrefbtitle {Statistical Methods for Research Workers (6th ed.)}
  {Statistical methods for research workers (6th ed.)}.
\newblock
\APACaddressPublisher{London}{Oliver and Boyd}.
\PrintBackRefs{\CurrentBib}

\bibitem [\protect \citeauthoryear {%
Freeman%
}{%
Freeman%
}{%
{\protect \APACyear {1993}}%
}]{%
Freeman1993}
\APACinsertmetastar {%
Freeman1993}%
\begin{APACrefauthors}%
Freeman, P\BPBI R.%
\end{APACrefauthors}%
\unskip\
\newblock
\APACrefYearMonthDay{1993}{}{}.
\newblock
{\BBOQ}\APACrefatitle {The Role of $P$-Values in Analysing Trial Results} {The
  role of $p$-values in analysing trial results}.{\BBCQ}
\newblock
\APACjournalVolNumPages{Statistics in Medicine}{12}{}{1443--1452}.
\PrintBackRefs{\CurrentBib}

\bibitem [\protect \citeauthoryear {%
Gelman%
}{%
Gelman%
}{%
{\protect \APACyear {2009}}%
}]{%
Gelman2009}
\APACinsertmetastar {%
Gelman2009}%
\begin{APACrefauthors}%
Gelman, A.%
\end{APACrefauthors}%
\unskip\
\newblock
\APACrefYearMonthDay{2009}{}{}.
\newblock
{\BBOQ}\APACrefatitle {Bayes, {J}effreys, Prior Distributions and the
  Philosophy of Statistics} {Bayes, {J}effreys, prior distributions and the
  philosophy of statistics}.{\BBCQ}
\newblock
\APACjournalVolNumPages{Statistical Science}{24}{}{176--178}.
\PrintBackRefs{\CurrentBib}

\bibitem [\protect \citeauthoryear {%
Good%
}{%
Good%
}{%
{\protect \APACyear {1980}}%
{\protect \APACexlab {{\protect \BCnt {1}}}}}]{%
Good1980}
\APACinsertmetastar {%
Good1980}%
\begin{APACrefauthors}%
Good, I\BPBI J.%
\end{APACrefauthors}%
\unskip\
\newblock
\APACrefYearMonthDay{1980{\protect \BCnt {1}}}{}{}.
\newblock
{\BBOQ}\APACrefatitle {The Contributions of {J}effreys to {B}ayesian
  Statistics} {The contributions of {J}effreys to {B}ayesian
  statistics}.{\BBCQ}
\newblock
\BIn{} A.~Zellner\ (\BED), \APACrefbtitle {Bayesian Analysis in Econometrics
  and Statistics: {E}ssays in Honor of {H}arold {J}effreys} {Bayesian analysis
  in econometrics and statistics: {E}ssays in honor of {H}arold {J}effreys}\
  (\BPGS\ 21--34).
\newblock
\APACaddressPublisher{Amsterdam, The Netherlands}{North-Holland Publishing
  Company}.
\PrintBackRefs{\CurrentBib}

\bibitem [\protect \citeauthoryear {%
Good%
}{%
Good%
}{%
{\protect \APACyear {1980}}%
{\protect \APACexlab {{\protect \BCnt {2}}}}}]{%
Good1980p}
\APACinsertmetastar {%
Good1980p}%
\begin{APACrefauthors}%
Good, I\BPBI J.%
\end{APACrefauthors}%
\unskip\
\newblock
\APACrefYearMonthDay{1980{\protect \BCnt {2}}}{}{}.
\newblock
{\BBOQ}\APACrefatitle {The Diminishing Significance of a $p$--value as the
  Sample Size Increases} {The diminishing significance of a $p$--value as the
  sample size increases}.{\BBCQ}
\newblock
\APACjournalVolNumPages{Journal of Statistical Computation and
  Simulation}{11}{}{307--313}.
\PrintBackRefs{\CurrentBib}

\bibitem [\protect \citeauthoryear {%
Good%
}{%
Good%
}{%
{\protect \APACyear {1983}}%
}]{%
Good1983b}
\APACinsertmetastar {%
Good1983b}%
\begin{APACrefauthors}%
Good, I\BPBI J.%
\end{APACrefauthors}%
\unskip\
\newblock
\APACrefYearMonthDay{1983}{}{}.
\newblock
{\BBOQ}\APACrefatitle {The Diminishing Significance of a Fixed P--Value as the
  Sample Size Increases: {A} Discrete Model} {The diminishing significance of a
  fixed p--value as the sample size increases: {A} discrete model}.{\BBCQ}
\newblock
\APACjournalVolNumPages{Journal of Statistical Computation and
  Simulation}{16}{}{312--313}.
\PrintBackRefs{\CurrentBib}

\bibitem [\protect \citeauthoryear {%
Good%
}{%
Good%
}{%
{\protect \APACyear {1992}}%
}]{%
Good1992}
\APACinsertmetastar {%
Good1992}%
\begin{APACrefauthors}%
Good, I\BPBI J.%
\end{APACrefauthors}%
\unskip\
\newblock
\APACrefYearMonthDay{1992}{}{}.
\newblock
{\BBOQ}\APACrefatitle {The {B}ayes/Non-{B}ayes Compromise: {A} Brief Review}
  {The {B}ayes/non-{B}ayes compromise: {A} brief review}.{\BBCQ}
\newblock
\APACjournalVolNumPages{Journal of the American Statistical
  Association}{87}{}{597--606}.
\PrintBackRefs{\CurrentBib}

\bibitem [\protect \citeauthoryear {%
Gronau%
, Ly%
\BCBL {}\ \BBA {} Wagenmakers%
}{%
Gronau%
\ \protect \BOthers {.}}{%
{\protect \APACyear {2020}}%
}]{%
GronauEtAl2020Informed}
\APACinsertmetastar {%
GronauEtAl2020Informed}%
\begin{APACrefauthors}%
Gronau, Q\BPBI F.%
, Ly, A.%
\BCBL {}\ \BBA {} Wagenmakers, E\BHBI J.%
\end{APACrefauthors}%
\unskip\
\newblock
\APACrefYearMonthDay{2020}{}{}.
\newblock
{\BBOQ}\APACrefatitle {Informed {B}ayesian $t$-tests} {Informed {B}ayesian
  $t$-tests}.{\BBCQ}
\newblock
\APACjournalVolNumPages{The American Statistician}{74}{}{137--143}.
\PrintBackRefs{\CurrentBib}

\bibitem [\protect \citeauthoryear {%
Howie%
}{%
Howie%
}{%
{\protect \APACyear {2002}}%
}]{%
Howie2002}
\APACinsertmetastar {%
Howie2002}%
\begin{APACrefauthors}%
Howie, D.%
\end{APACrefauthors}%
\unskip\
\newblock
\APACrefYear{2002}.
\newblock
\APACrefbtitle {Interpreting Probability: {C}ontroversies and Developments in
  the Early Twentieth Century} {Interpreting probability: {C}ontroversies and
  developments in the early twentieth century}.
\newblock
\APACaddressPublisher{Cambridge}{Cambridge University Press}.
\PrintBackRefs{\CurrentBib}

\bibitem [\protect \citeauthoryear {%
Jaynes%
}{%
Jaynes%
}{%
{\protect \APACyear {2003}}%
}]{%
Jaynes2003}
\APACinsertmetastar {%
Jaynes2003}%
\begin{APACrefauthors}%
Jaynes, E\BPBI T.%
\end{APACrefauthors}%
\unskip\
\newblock
\APACrefYear{2003}.
\newblock
\APACrefbtitle {Probability Theory: {T}he Logic of Science} {Probability
  theory: {T}he logic of science}.
\newblock
\APACaddressPublisher{Cambridge}{Cambridge University Press}.
\PrintBackRefs{\CurrentBib}

\bibitem [\protect \citeauthoryear {%
Jefferys%
}{%
Jefferys%
}{%
{\protect \APACyear {1990}}%
}]{%
Jefferys1990}
\APACinsertmetastar {%
Jefferys1990}%
\begin{APACrefauthors}%
Jefferys, W\BPBI H.%
\end{APACrefauthors}%
\unskip\
\newblock
\APACrefYearMonthDay{1990}{}{}.
\newblock
{\BBOQ}\APACrefatitle {{B}ayesian Analysis of Random Event Generator Data}
  {{B}ayesian analysis of random event generator data}.{\BBCQ}
\newblock
\APACjournalVolNumPages{Journal of Scientific Exploration}{4}{}{153--169}.
\PrintBackRefs{\CurrentBib}

\bibitem [\protect \citeauthoryear {%
Jeffreys%
}{%
Jeffreys%
}{%
{\protect \APACyear {1935}}%
}]{%
Jeffreys1935}
\APACinsertmetastar {%
Jeffreys1935}%
\begin{APACrefauthors}%
Jeffreys, H.%
\end{APACrefauthors}%
\unskip\
\newblock
\APACrefYearMonthDay{1935}{}{}.
\newblock
{\BBOQ}\APACrefatitle {Some Tests of Significance, Treated by the Theory of
  Probability} {Some tests of significance, treated by the theory of
  probability}.{\BBCQ}
\newblock
\APACjournalVolNumPages{Proceedings of the Cambridge Philosophy
  Society}{31}{}{203--222}.
\PrintBackRefs{\CurrentBib}

\bibitem [\protect \citeauthoryear {%
Jeffreys%
}{%
Jeffreys%
}{%
{\protect \APACyear {1936}}%
{\protect \APACexlab {{\protect \BCnt {1}}}}}]{%
Jeffreys1936}
\APACinsertmetastar {%
Jeffreys1936}%
\begin{APACrefauthors}%
Jeffreys, H.%
\end{APACrefauthors}%
\unskip\
\newblock
\APACrefYearMonthDay{1936{\protect \BCnt {1}}}{}{}.
\newblock
{\BBOQ}\APACrefatitle {{\SR{19361}}{O}n Some Criticisms of the Theory of
  Probability} {{\SR{19361}}{O}n some criticisms of the theory of
  probability}.{\BBCQ}
\newblock
\APACjournalVolNumPages{The London, Edinburgh, and Dublin Philosophical
  Magazine and Journal of Science}{22}{}{337--359}.
\PrintBackRefs{\CurrentBib}

\bibitem [\protect \citeauthoryear {%
Jeffreys%
}{%
Jeffreys%
}{%
{\protect \APACyear {1936}}%
{\protect \APACexlab {{\protect \BCnt {2}}}}}]{%
Jeffreys1936Further}
\APACinsertmetastar {%
Jeffreys1936Further}%
\begin{APACrefauthors}%
Jeffreys, H.%
\end{APACrefauthors}%
\unskip\
\newblock
\APACrefYearMonthDay{1936{\protect \BCnt {2}}}{}{}.
\newblock
{\BBOQ}\APACrefatitle {{\SR{19364}}{F}urther Significance Tests}
  {{\SR{19364}}{F}urther significance tests}.{\BBCQ}
\newblock
\APACjournalVolNumPages{Mathematical Proceedings of the Cambridge Philosophical
  Society}{32}{}{416--445}.
\PrintBackRefs{\CurrentBib}

\bibitem [\protect \citeauthoryear {%
Jeffreys%
}{%
Jeffreys%
}{%
{\protect \APACyear {1937}}%
{\protect \APACexlab {{\protect \BCnt {1}}}}}]{%
Jeffreys1937Tests}
\APACinsertmetastar {%
Jeffreys1937Tests}%
\begin{APACrefauthors}%
Jeffreys, H.%
\end{APACrefauthors}%
\unskip\
\newblock
\APACrefYearMonthDay{1937{\protect \BCnt {1}}}{}{}.
\newblock
{\BBOQ}\APACrefatitle {{\SR{19371}}{T}he Tests for Sampling Differences and
  Contingency} {{\SR{19371}}{T}he tests for sampling differences and
  contingency}.{\BBCQ}
\newblock
\APACjournalVolNumPages{Proceedings of the Royal Society of London. Series {A},
  Mathematical and Physical Sciences}{162}{}{479--495}.
\PrintBackRefs{\CurrentBib}

\bibitem [\protect \citeauthoryear {%
Jeffreys%
}{%
Jeffreys%
}{%
{\protect \APACyear {1937}}%
{\protect \APACexlab {{\protect \BCnt {2}}}}}]{%
Jeffreys1937SI}
\APACinsertmetastar {%
Jeffreys1937SI}%
\begin{APACrefauthors}%
Jeffreys, H.%
\end{APACrefauthors}%
\unskip\
\newblock
\APACrefYear{1937{\protect \BCnt {2}}}.
\newblock
\APACrefbtitle {{\SR{19372}}{S}cientific Inference} {{\SR{19372}}{S}cientific
  inference}\ (\PrintOrdinal{1}\ \BEd).
\newblock
\APACaddressPublisher{Cambridge, UK}{Cambridge University Press}.
\PrintBackRefs{\CurrentBib}

\bibitem [\protect \citeauthoryear {%
Jeffreys%
}{%
Jeffreys%
}{%
{\protect \APACyear {1937}}%
{\protect \APACexlab {{\protect \BCnt {3}}}}}]{%
Jeffreys1937Nature}
\APACinsertmetastar {%
Jeffreys1937Nature}%
\begin{APACrefauthors}%
Jeffreys, H.%
\end{APACrefauthors}%
\unskip\
\newblock
\APACrefYearMonthDay{1937{\protect \BCnt {3}}}{}{}.
\newblock
{\BBOQ}\APACrefatitle {Modern {A}ristotelianism: {C}ontribution to Discussion}
  {Modern {A}ristotelianism: {C}ontribution to discussion}.{\BBCQ}
\newblock
\APACjournalVolNumPages{Nature}{139}{}{1004}.
\PrintBackRefs{\CurrentBib}

\bibitem [\protect \citeauthoryear {%
Jeffreys%
}{%
Jeffreys%
}{%
{\protect \APACyear {1938}}%
{\protect \APACexlab {{\protect \BCnt {1}}}}}]{%
Jeffreys1938Comparison2}
\APACinsertmetastar {%
Jeffreys1938Comparison2}%
\begin{APACrefauthors}%
Jeffreys, H.%
\end{APACrefauthors}%
\unskip\
\newblock
\APACrefYearMonthDay{1938{\protect \BCnt {1}}}{}{}.
\newblock
{\BBOQ}\APACrefatitle {{\SR{19381}}{T}he Comparison of Series of Measures on
  Different Hypotheses concerning the Standard Errors} {{\SR{19381}}{T}he
  comparison of series of measures on different hypotheses concerning the
  standard errors}.{\BBCQ}
\newblock
\APACjournalVolNumPages{Proceedings of the Royal Society of London. Series {A},
  Mathematical and Physical Sciences}{167}{}{367--384}.
\PrintBackRefs{\CurrentBib}

\bibitem [\protect \citeauthoryear {%
Jeffreys%
}{%
Jeffreys%
}{%
{\protect \APACyear {1938}}%
{\protect \APACexlab {{\protect \BCnt {2}}}}}]{%
Jeffreys1938}
\APACinsertmetastar {%
Jeffreys1938}%
\begin{APACrefauthors}%
Jeffreys, H.%
\end{APACrefauthors}%
\unskip\
\newblock
\APACrefYearMonthDay{1938{\protect \BCnt {2}}}{}{}.
\newblock
{\BBOQ}\APACrefatitle {{\SR{19382}}{S}ignificance Tests When Several Degrees of
  Freedom Arise Simultaneously} {{\SR{19382}}{S}ignificance tests when several
  degrees of freedom arise simultaneously}.{\BBCQ}
\newblock
\APACjournalVolNumPages{Proceedings of the Royal Society of London. Series {A},
  Mathematical and Physical Sciences}{165}{}{161--198}.
\PrintBackRefs{\CurrentBib}

\bibitem [\protect \citeauthoryear {%
Jeffreys%
}{%
Jeffreys%
}{%
{\protect \APACyear {1938}}%
{\protect \APACexlab {{\protect \BCnt {3}}}}}]{%
Jeffreys1938MLE}
\APACinsertmetastar {%
Jeffreys1938MLE}%
\begin{APACrefauthors}%
Jeffreys, H.%
\end{APACrefauthors}%
\unskip\
\newblock
\APACrefYearMonthDay{1938{\protect \BCnt {3}}}{}{}.
\newblock
{\BBOQ}\APACrefatitle {{\SR{19383}}{M}aximum Likelihood, Inverse Probability
  and the Method of Moments} {{\SR{19383}}{M}aximum likelihood, inverse
  probability and the method of moments}.{\BBCQ}
\newblock
\APACjournalVolNumPages{Annals of Eugenics}{8}{}{146--151}.
\PrintBackRefs{\CurrentBib}

\bibitem [\protect \citeauthoryear {%
Jeffreys%
}{%
Jeffreys%
}{%
{\protect \APACyear {1938}}%
{\protect \APACexlab {{\protect \BCnt {4}}}}}]{%
Jeffreys1938Continuous}
\APACinsertmetastar {%
Jeffreys1938Continuous}%
\begin{APACrefauthors}%
Jeffreys, H.%
\end{APACrefauthors}%
\unskip\
\newblock
\APACrefYearMonthDay{1938{\protect \BCnt {4}}}{}{}.
\newblock
{\BBOQ}\APACrefatitle {{\SR{19384}}{S}ignificance Tests for Continuous
  Departures from Suggested Distributions of Chance}
  {{\SR{19384}}{S}ignificance tests for continuous departures from suggested
  distributions of chance}.{\BBCQ}
\newblock
\APACjournalVolNumPages{Proceedings of the Royal Society of London. Series {A},
  Mathematical and Physical Sciences}{164}{}{307--315}.
\PrintBackRefs{\CurrentBib}

\bibitem [\protect \citeauthoryear {%
Jeffreys%
}{%
Jeffreys%
}{%
{\protect \APACyear {1938}}%
{\protect \APACexlab {{\protect \BCnt {5}}}}}]{%
Jeffreys1938aftershock}
\APACinsertmetastar {%
Jeffreys1938aftershock}%
\begin{APACrefauthors}%
Jeffreys, H.%
\end{APACrefauthors}%
\unskip\
\newblock
\APACrefYearMonthDay{1938{\protect \BCnt {5}}}{}{}.
\newblock
{\BBOQ}\APACrefatitle {{\SR{19385}}{A}ftershocks and Periodicity in
  Earthquakes} {{\SR{19385}}{A}ftershocks and periodicity in
  earthquakes}.{\BBCQ}
\newblock
\APACjournalVolNumPages{Gerlands Beitr\"{a}ge zur Geophysik}{53}{}{111--139}.
\PrintBackRefs{\CurrentBib}

\bibitem [\protect \citeauthoryear {%
Jeffreys%
}{%
Jeffreys%
}{%
{\protect \APACyear {1939}}%
}]{%
Jeffreys1939}
\APACinsertmetastar {%
Jeffreys1939}%
\begin{APACrefauthors}%
Jeffreys, H.%
\end{APACrefauthors}%
\unskip\
\newblock
\APACrefYear{1939}.
\newblock
\APACrefbtitle {Theory of Probability} {Theory of probability}\
  (\PrintOrdinal{1}\ \BEd).
\newblock
\APACaddressPublisher{Oxford, UK}{Oxford University Press}.
\PrintBackRefs{\CurrentBib}

\bibitem [\protect \citeauthoryear {%
Jeffreys%
}{%
Jeffreys%
}{%
{\protect \APACyear {1940}}%
}]{%
Jeffreys1940}
\APACinsertmetastar {%
Jeffreys1940}%
\begin{APACrefauthors}%
Jeffreys, H.%
\end{APACrefauthors}%
\unskip\
\newblock
\APACrefYearMonthDay{1940}{}{}.
\newblock
{\BBOQ}\APACrefatitle {Note on the {Behrens--Fisher} Formula} {Note on the
  {Behrens--Fisher} formula}.{\BBCQ}
\newblock
\APACjournalVolNumPages{Annals of Eugenics}{10}{}{48--51}.
\PrintBackRefs{\CurrentBib}

\bibitem [\protect \citeauthoryear {%
Jeffreys%
}{%
Jeffreys%
}{%
{\protect \APACyear {1942}}%
}]{%
Jeffreys1942Functions}
\APACinsertmetastar {%
Jeffreys1942Functions}%
\begin{APACrefauthors}%
Jeffreys, H.%
\end{APACrefauthors}%
\unskip\
\newblock
\APACrefYearMonthDay{1942}{}{}.
\newblock
{\BBOQ}\APACrefatitle {On the Significance Tests for the Introduction of New
  Functions to Represent Measures} {On the significance tests for the
  introduction of new functions to represent measures}.{\BBCQ}
\newblock
\APACjournalVolNumPages{Proceedings of the Royal Society of London. Series {A},
  Mathematical and Physical Sciences}{180}{}{256--268}.
\PrintBackRefs{\CurrentBib}

\bibitem [\protect \citeauthoryear {%
Jeffreys%
}{%
Jeffreys%
}{%
{\protect \APACyear {1948}}%
}]{%
Jeffreys1948}
\APACinsertmetastar {%
Jeffreys1948}%
\begin{APACrefauthors}%
Jeffreys, H.%
\end{APACrefauthors}%
\unskip\
\newblock
\APACrefYear{1948}.
\newblock
\APACrefbtitle {Theory of Probability} {Theory of probability}\
  (\PrintOrdinal{2}\ \BEd).
\newblock
\APACaddressPublisher{Oxford, UK}{Oxford University Press}.
\PrintBackRefs{\CurrentBib}

\bibitem [\protect \citeauthoryear {%
Jeffreys%
}{%
Jeffreys%
}{%
{\protect \APACyear {1950}}%
}]{%
Jeffreys1950Russell}
\APACinsertmetastar {%
Jeffreys1950Russell}%
\begin{APACrefauthors}%
Jeffreys, H.%
\end{APACrefauthors}%
\unskip\
\newblock
\APACrefYearMonthDay{1950}{}{}.
\newblock
{\BBOQ}\APACrefatitle {Bertrand Russell on Probability} {Bertrand russell on
  probability}.{\BBCQ}
\newblock
\APACjournalVolNumPages{Mind: A Quarterly Review of Psychology and
  Philosophy}{59}{}{313--319}.
\PrintBackRefs{\CurrentBib}

\bibitem [\protect \citeauthoryear {%
Jeffreys%
}{%
Jeffreys%
}{%
{\protect \APACyear {1953}}%
}]{%
Jeffreys1953}
\APACinsertmetastar {%
Jeffreys1953}%
\begin{APACrefauthors}%
Jeffreys, H.%
\end{APACrefauthors}%
\unskip\
\newblock
\APACrefYearMonthDay{1953}{}{}.
\newblock
{\BBOQ}\APACrefatitle {Comment on ``Statistical Inference'' by {Dennis
  Lindley}} {Comment on ``statistical inference'' by {Dennis Lindley}}.{\BBCQ}
\newblock
\APACjournalVolNumPages{Journal of the Royal Statistical Society Series B
  (Methodological)}{15}{}{72}.
\PrintBackRefs{\CurrentBib}

\bibitem [\protect \citeauthoryear {%
Jeffreys%
}{%
Jeffreys%
}{%
{\protect \APACyear {1955}}%
}]{%
Jeffreys1955}
\APACinsertmetastar {%
Jeffreys1955}%
\begin{APACrefauthors}%
Jeffreys, H.%
\end{APACrefauthors}%
\unskip\
\newblock
\APACrefYearMonthDay{1955}{}{}.
\newblock
{\BBOQ}\APACrefatitle {The Present Position in Probability Theory} {The present
  position in probability theory}.{\BBCQ}
\newblock
\APACjournalVolNumPages{The British Journal for the Philosophy of
  Science}{5}{}{275--289}.
\PrintBackRefs{\CurrentBib}

\bibitem [\protect \citeauthoryear {%
Jeffreys%
}{%
Jeffreys%
}{%
{\protect \APACyear {1957}}%
{\protect \APACexlab {{\protect \BCnt {1}}}}}]{%
Jeffreys1957SI}
\APACinsertmetastar {%
Jeffreys1957SI}%
\begin{APACrefauthors}%
Jeffreys, H.%
\end{APACrefauthors}%
\unskip\
\newblock
\APACrefYear{1957{\protect \BCnt {1}}}.
\newblock
\APACrefbtitle {{\SR{19571}}Scientific Inference} {{\SR{19571}}scientific
  inference}\ (\PrintOrdinal{2}\ \BEd).
\newblock
\APACaddressPublisher{Cambridge, UK}{Cambridge University Press}.
\PrintBackRefs{\CurrentBib}

\bibitem [\protect \citeauthoryear {%
Jeffreys%
}{%
Jeffreys%
}{%
{\protect \APACyear {1957}}%
{\protect \APACexlab {{\protect \BCnt {2}}}}}]{%
Jeffreys1957}
\APACinsertmetastar {%
Jeffreys1957}%
\begin{APACrefauthors}%
Jeffreys, H.%
\end{APACrefauthors}%
\unskip\
\newblock
\APACrefYearMonthDay{1957{\protect \BCnt {2}}}{}{}.
\newblock
{\BBOQ}\APACrefatitle {{\SR{19572}}Probability Theory in Astronomy}
  {{\SR{19572}}probability theory in astronomy}.{\BBCQ}
\newblock
\APACjournalVolNumPages{Monthly Notices of the Royal Astronomical
  Society}{117}{}{347--355}.
\PrintBackRefs{\CurrentBib}

\bibitem [\protect \citeauthoryear {%
Jeffreys%
}{%
Jeffreys%
}{%
{\protect \APACyear {1961}}%
}]{%
Jeffreys1961}
\APACinsertmetastar {%
Jeffreys1961}%
\begin{APACrefauthors}%
Jeffreys, H.%
\end{APACrefauthors}%
\unskip\
\newblock
\APACrefYear{1961}.
\newblock
\APACrefbtitle {Theory of Probability} {Theory of probability}\
  (\PrintOrdinal{3}\ \BEd).
\newblock
\APACaddressPublisher{Oxford, UK}{Oxford University Press}.
\PrintBackRefs{\CurrentBib}

\bibitem [\protect \citeauthoryear {%
Jeffreys%
}{%
Jeffreys%
}{%
{\protect \APACyear {1973}}%
}]{%
Jeffreys1973}
\APACinsertmetastar {%
Jeffreys1973}%
\begin{APACrefauthors}%
Jeffreys, H.%
\end{APACrefauthors}%
\unskip\
\newblock
\APACrefYear{1973}.
\newblock
\APACrefbtitle {Scientific Inference} {Scientific inference}\
  (\PrintOrdinal{3}\ \BEd).
\newblock
\APACaddressPublisher{Cambridge, UK}{Cambridge University Press}.
\PrintBackRefs{\CurrentBib}

\bibitem [\protect \citeauthoryear {%
Jeffreys%
}{%
Jeffreys%
}{%
{\protect \APACyear {1974}}%
}]{%
Jeffreys1974}
\APACinsertmetastar {%
Jeffreys1974}%
\begin{APACrefauthors}%
Jeffreys, H.%
\end{APACrefauthors}%
\unskip\
\newblock
\APACrefYearMonthDay{1974}{}{}.
\newblock
{\BBOQ}\APACrefatitle {Fisher and Inverse Probability} {Fisher and inverse
  probability}.{\BBCQ}
\newblock
\APACjournalVolNumPages{International Statistical Review}{42}{}{1--3}.
\PrintBackRefs{\CurrentBib}

\bibitem [\protect \citeauthoryear {%
Jeffreys%
}{%
Jeffreys%
}{%
{\protect \APACyear {1977}}%
}]{%
Jeffreys1977}
\APACinsertmetastar {%
Jeffreys1977}%
\begin{APACrefauthors}%
Jeffreys, H.%
\end{APACrefauthors}%
\unskip\
\newblock
\APACrefYearMonthDay{1977}{}{}.
\newblock
{\BBOQ}\APACrefatitle {Probability Theory in Geophysics} {Probability theory in
  geophysics}.{\BBCQ}
\newblock
\APACjournalVolNumPages{Journal of the Institute of Mathematics and its
  Applications}{19}{}{87--96}.
\PrintBackRefs{\CurrentBib}

\bibitem [\protect \citeauthoryear {%
Jeffreys%
}{%
Jeffreys%
}{%
{\protect \APACyear {1980}}%
}]{%
Jeffreys1980}
\APACinsertmetastar {%
Jeffreys1980}%
\begin{APACrefauthors}%
Jeffreys, H.%
\end{APACrefauthors}%
\unskip\
\newblock
\APACrefYearMonthDay{1980}{}{}.
\newblock
{\BBOQ}\APACrefatitle {Some General Points in Probability Theory} {Some general
  points in probability theory}.{\BBCQ}
\newblock
\BIn{} A.~Zellner\ (\BED), \APACrefbtitle {Bayesian Analysis in Econometrics
  and Statistics: {E}ssays in Honor of {H}arold {J}effreys} {Bayesian analysis
  in econometrics and statistics: {E}ssays in honor of {H}arold {J}effreys}\
  (\BPGS\ 451--453).
\newblock
\APACaddressPublisher{Amsterdam, The Netherlands}{North-Holland Publishing
  Company}.
\PrintBackRefs{\CurrentBib}

\bibitem [\protect \citeauthoryear {%
Kamary%
, Mengersen%
, Robert%
\BCBL {}\ \BBA {} Rousseau%
}{%
Kamary%
\ \protect \BOthers {.}}{%
{\protect \APACyear {2014}}%
}]{%
KamaryEtAl2014}
\APACinsertmetastar {%
KamaryEtAl2014}%
\begin{APACrefauthors}%
Kamary, K.%
, Mengersen, K.%
, Robert, C\BPBI P.%
\BCBL {}\ \BBA {} Rousseau, J.%
\end{APACrefauthors}%
\unskip\
\newblock
\APACrefYearMonthDay{2014}{}{}.
\newblock
{\BBOQ}\APACrefatitle {Testing Hypotheses via a Mixture Estimation Model}
  {Testing hypotheses via a mixture estimation model}.{\BBCQ}
\newblock
\APACjournalVolNumPages{ArXiv}{}{}{}.
\newblock
\begin{APACrefURL} \url{https://arxiv.org/abs/1412.2044} \end{APACrefURL}
\PrintBackRefs{\CurrentBib}

\bibitem [\protect \citeauthoryear {%
Kass%
\ \BBA {} Raftery%
}{%
Kass%
\ \BBA {} Raftery%
}{%
{\protect \APACyear {1995}}%
}]{%
KassRaftery1995}
\APACinsertmetastar {%
KassRaftery1995}%
\begin{APACrefauthors}%
Kass, R\BPBI E.%
\BCBT {}\ \BBA {} Raftery, A\BPBI E.%
\end{APACrefauthors}%
\unskip\
\newblock
\APACrefYearMonthDay{1995}{}{}.
\newblock
{\BBOQ}\APACrefatitle {{B}ayes Factors} {{B}ayes factors}.{\BBCQ}
\newblock
\APACjournalVolNumPages{Journal of the American Statistical
  Association}{90}{}{773--795}.
\PrintBackRefs{\CurrentBib}

\bibitem [\protect \citeauthoryear {%
Kass%
\ \BBA {} Wasserman%
}{%
Kass%
\ \BBA {} Wasserman%
}{%
{\protect \APACyear {1995}}%
}]{%
KassWasserman1995}
\APACinsertmetastar {%
KassWasserman1995}%
\begin{APACrefauthors}%
Kass, R\BPBI E.%
\BCBT {}\ \BBA {} Wasserman, L.%
\end{APACrefauthors}%
\unskip\
\newblock
\APACrefYearMonthDay{1995}{}{}.
\newblock
{\BBOQ}\APACrefatitle {A Reference {B}ayesian test for Nested Hypotheses and
  Its Relationship to the {S}chwarz Criterion} {A reference {B}ayesian test for
  nested hypotheses and its relationship to the {S}chwarz criterion}.{\BBCQ}
\newblock
\APACjournalVolNumPages{Journal of the American Statistical
  Association}{90}{}{928--934}.
\PrintBackRefs{\CurrentBib}

\bibitem [\protect \citeauthoryear {%
Keysers%
, Gazzola%
\BCBL {}\ \BBA {} Wagenmakers%
}{%
Keysers%
\ \protect \BOthers {.}}{%
{\protect \APACyear {2020}}%
}]{%
KeysersEtAl2020}
\APACinsertmetastar {%
KeysersEtAl2020}%
\begin{APACrefauthors}%
Keysers, C.%
, Gazzola, V.%
\BCBL {}\ \BBA {} Wagenmakers, E\BHBI J.%
\end{APACrefauthors}%
\unskip\
\newblock
\APACrefYearMonthDay{2020}{}{}.
\newblock
{\BBOQ}\APACrefatitle {Using {B}ayes Factor Hypothesis Testing in Neuroscience
  to Establish Evidence of Absence} {Using {B}ayes factor hypothesis testing in
  neuroscience to establish evidence of absence}.{\BBCQ}
\newblock
\APACjournalVolNumPages{Nature Neuroscience}{23}{}{788--799}.
\PrintBackRefs{\CurrentBib}

\bibitem [\protect \citeauthoryear {%
Kim%
\ \BBA {} Choi%
}{%
Kim%
\ \BBA {} Choi%
}{%
{\protect \APACyear {2021}}%
}]{%
KimChoi2021}
\APACinsertmetastar {%
KimChoi2021}%
\begin{APACrefauthors}%
Kim, J\BPBI H.%
\BCBT {}\ \BBA {} Choi, I.%
\end{APACrefauthors}%
\unskip\
\newblock
\APACrefYearMonthDay{2021}{}{}.
\newblock
{\BBOQ}\APACrefatitle {Choosing the Level of Significance: {A}
  Decision--Theoretic Approach} {Choosing the level of significance: {A}
  decision--theoretic approach}.{\BBCQ}
\newblock
\APACjournalVolNumPages{Abacus}{57}{}{27--71}.
\PrintBackRefs{\CurrentBib}

\bibitem [\protect \citeauthoryear {%
Leamer%
}{%
Leamer%
}{%
{\protect \APACyear {1978}}%
}]{%
Leamer1978}
\APACinsertmetastar {%
Leamer1978}%
\begin{APACrefauthors}%
Leamer, E.%
\end{APACrefauthors}%
\unskip\
\newblock
\APACrefYear{1978}.
\newblock
\APACrefbtitle {Specification Searches: {A}d Hoc Inference with Nonexperimental
  Data} {Specification searches: {A}d hoc inference with nonexperimental data}.
\newblock
\APACaddressPublisher{New York}{Wiley}.
\PrintBackRefs{\CurrentBib}

\bibitem [\protect \citeauthoryear {%
Lehmann%
}{%
Lehmann%
}{%
{\protect \APACyear {1958}}%
}]{%
Lehmann1958}
\APACinsertmetastar {%
Lehmann1958}%
\begin{APACrefauthors}%
Lehmann, E\BPBI L.%
\end{APACrefauthors}%
\unskip\
\newblock
\APACrefYearMonthDay{1958}{}{}.
\newblock
{\BBOQ}\APACrefatitle {Significance Level and Power} {Significance level and
  power}.{\BBCQ}
\newblock
\APACjournalVolNumPages{The Annals of Mathematical
  Statistics}{29}{}{1167--1176}.
\PrintBackRefs{\CurrentBib}

\bibitem [\protect \citeauthoryear {%
Lindley%
}{%
Lindley%
}{%
{\protect \APACyear {1953}}%
}]{%
Lindley1953}
\APACinsertmetastar {%
Lindley1953}%
\begin{APACrefauthors}%
Lindley, D\BPBI V.%
\end{APACrefauthors}%
\unskip\
\newblock
\APACrefYearMonthDay{1953}{}{}.
\newblock
{\BBOQ}\APACrefatitle {Statistical Inference} {Statistical inference}.{\BBCQ}
\newblock
\APACjournalVolNumPages{Journal of the Royal Statistical Society Series B
  (Methodological)}{15}{}{30--76}.
\PrintBackRefs{\CurrentBib}

\bibitem [\protect \citeauthoryear {%
Lindley%
}{%
Lindley%
}{%
{\protect \APACyear {1957}}%
}]{%
Lindley1957}
\APACinsertmetastar {%
Lindley1957}%
\begin{APACrefauthors}%
Lindley, D\BPBI V.%
\end{APACrefauthors}%
\unskip\
\newblock
\APACrefYearMonthDay{1957}{}{}.
\newblock
{\BBOQ}\APACrefatitle {A Statistical Paradox} {A statistical paradox}.{\BBCQ}
\newblock
\APACjournalVolNumPages{Biometrika}{44}{}{187--192}.
\PrintBackRefs{\CurrentBib}

\bibitem [\protect \citeauthoryear {%
Lindley%
}{%
Lindley%
}{%
{\protect \APACyear {1965}}%
}]{%
Lindley1965Inference}
\APACinsertmetastar {%
Lindley1965Inference}%
\begin{APACrefauthors}%
Lindley, D\BPBI V.%
\end{APACrefauthors}%
\unskip\
\newblock
\APACrefYear{1965}.
\newblock
\APACrefbtitle {Introduction to Probability \& Statistics from a {B}ayesian
  Viewpoint. {P}art 2. {I}nference} {Introduction to probability \& statistics
  from a {B}ayesian viewpoint. {P}art 2. {I}nference}.
\newblock
\APACaddressPublisher{Cambridge}{Cambridge University Press}.
\PrintBackRefs{\CurrentBib}

\bibitem [\protect \citeauthoryear {%
Lindley%
}{%
Lindley%
}{%
{\protect \APACyear {1980}}%
}]{%
Lindley1980}
\APACinsertmetastar {%
Lindley1980}%
\begin{APACrefauthors}%
Lindley, D\BPBI V.%
\end{APACrefauthors}%
\unskip\
\newblock
\APACrefYearMonthDay{1980}{}{}.
\newblock
{\BBOQ}\APACrefatitle {{J}effreys's Contribution to Modern Statistical Thought}
  {{J}effreys's contribution to modern statistical thought}.{\BBCQ}
\newblock
\BIn{} A.~Zellner\ (\BED), \APACrefbtitle {Bayesian Analysis in Econometrics
  and Statistics: {E}ssays in Honor of {H}arold {J}effreys} {Bayesian analysis
  in econometrics and statistics: {E}ssays in honor of {H}arold {J}effreys}\
  (\BPGS\ 35--39).
\newblock
\APACaddressPublisher{Amsterdam, The Netherlands}{North-Holland Publishing
  Company}.
\PrintBackRefs{\CurrentBib}

\bibitem [\protect \citeauthoryear {%
Lindley%
}{%
Lindley%
}{%
{\protect \APACyear {1986}}%
}]{%
Lindley1986Comment}
\APACinsertmetastar {%
Lindley1986Comment}%
\begin{APACrefauthors}%
Lindley, D\BPBI V.%
\end{APACrefauthors}%
\unskip\
\newblock
\APACrefYearMonthDay{1986}{}{}.
\newblock
{\BBOQ}\APACrefatitle {Comment on ``Tests of Significance in Theory and
  Practice'' by {D}. {J}. {J}ohnstone} {Comment on ``tests of significance in
  theory and practice'' by {D}. {J}. {J}ohnstone}.{\BBCQ}
\newblock
\APACjournalVolNumPages{Journal of the Royal Statistical Society, Series D (The
  Statistician)}{35}{}{502--504}.
\PrintBackRefs{\CurrentBib}

\bibitem [\protect \citeauthoryear {%
Lindley%
}{%
Lindley%
}{%
{\protect \APACyear {1989}}%
}]{%
Lindley1989}
\APACinsertmetastar {%
Lindley1989}%
\begin{APACrefauthors}%
Lindley, D\BPBI V.%
\end{APACrefauthors}%
\unskip\
\newblock
\APACrefYearMonthDay{1989}{}{}.
\newblock
{\BBOQ}\APACrefatitle {{Obituary: Harold Jeffreys, 1891--1989}} {{Obituary:
  Harold Jeffreys, 1891--1989}}.{\BBCQ}
\newblock
\APACjournalVolNumPages{Journal of the Royal Statistical Society Series
  A}{152}{}{417--419}.
\PrintBackRefs{\CurrentBib}

\bibitem [\protect \citeauthoryear {%
Lindley%
}{%
Lindley%
}{%
{\protect \APACyear {2000}}%
}]{%
Lindley2000ISBA}
\APACinsertmetastar {%
Lindley2000ISBA}%
\begin{APACrefauthors}%
Lindley, D\BPBI V.%
\end{APACrefauthors}%
\unskip\
\newblock
\APACrefYearMonthDay{2000}{}{}.
\newblock
{\BBOQ}\APACrefatitle {What is a {B}ayesian?} {What is a {B}ayesian?}{\BBCQ}
\newblock
\APACjournalVolNumPages{The {ISBA} Bulletin}{7}{}{7--9}.
\PrintBackRefs{\CurrentBib}

\bibitem [\protect \citeauthoryear {%
Lindley%
}{%
Lindley%
}{%
{\protect \APACyear {2011}}%
}]{%
Lindley2011}
\APACinsertmetastar {%
Lindley2011}%
\begin{APACrefauthors}%
Lindley, D\BPBI V.%
\end{APACrefauthors}%
\unskip\
\newblock
\APACrefYearMonthDay{2011}{}{}.
\newblock
{\BBOQ}\APACrefatitle {Comment on ``Integrated Objective {B}ayesian Estimation
  and Hypothesis Testing'' by {J. M. Bernardo}} {Comment on ``integrated
  objective {B}ayesian estimation and hypothesis testing'' by {J. M.
  Bernardo}}.{\BBCQ}
\newblock
\BIn{} J\BPBI M.~Bernardo\ \BOthers {.}\ (\BEDS), \APACrefbtitle {{B}ayesian
  Statistics 9} {{B}ayesian statistics 9}\ (\BPGS\ 37--38).
\newblock
\APACaddressPublisher{Oxford}{Oxford University Press}.
\PrintBackRefs{\CurrentBib}

\bibitem [\protect \citeauthoryear {%
Ly%
\ \protect \BOthers {.}}{%
Ly%
\ \protect \BOthers {.}}{%
{\protect \APACyear {2020}}%
}]{%
LyEtAl2020Alternative}
\APACinsertmetastar {%
LyEtAl2020Alternative}%
\begin{APACrefauthors}%
Ly, A.%
, Stefan, A.%
, {van Doorn}, J.%
, Dablander, F.%
, {van den Bergh}, D.%
, Sarafoglou, A.%
\BDBL {}Wagenmakers, E\BHBI J.%
\end{APACrefauthors}%
\unskip\
\newblock
\APACrefYearMonthDay{2020}{}{}.
\newblock
{\BBOQ}\APACrefatitle {The {B}ayesian Methodology of {S}ir {H}arold {J}effreys
  as a Practical Alternative to the P--value Hypothesis Test} {The {B}ayesian
  methodology of {S}ir {H}arold {J}effreys as a practical alternative to the
  p--value hypothesis test}.{\BBCQ}
\newblock
\APACjournalVolNumPages{Computational Brain \& Behavior}{3}{}{153--161}.
\PrintBackRefs{\CurrentBib}

\bibitem [\protect \citeauthoryear {%
Ly%
, Verhagen%
\BCBL {}\ \BBA {} Wagenmakers%
}{%
Ly%
\ \protect \BOthers {.}}{%
{\protect \APACyear {2016}}%
{\protect \APACexlab {{\protect \BCnt {1}}}}}]{%
LyEtAl2016}
\APACinsertmetastar {%
LyEtAl2016}%
\begin{APACrefauthors}%
Ly, A.%
, Verhagen, A\BPBI J.%
\BCBL {}\ \BBA {} Wagenmakers, E\BHBI J.%
\end{APACrefauthors}%
\unskip\
\newblock
\APACrefYearMonthDay{2016{\protect \BCnt {1}}}{}{}.
\newblock
{\BBOQ}\APACrefatitle {{\SR{20161}} {H}arold {J}effreys's Default {B}ayes
  Factor Hypothesis Tests: {E}xplanation, Extension, and Application in
  Psychology} {{\SR{20161}} {H}arold {J}effreys's default {B}ayes factor
  hypothesis tests: {E}xplanation, extension, and application in
  psychology}.{\BBCQ}
\newblock
\APACjournalVolNumPages{Journal of Mathematical Psychology}{72}{}{19--32}.
\PrintBackRefs{\CurrentBib}

\bibitem [\protect \citeauthoryear {%
Ly%
, Verhagen%
\BCBL {}\ \BBA {} Wagenmakers%
}{%
Ly%
\ \protect \BOthers {.}}{%
{\protect \APACyear {2016}}%
{\protect \APACexlab {{\protect \BCnt {2}}}}}]{%
LyEtAl2016Rejoinder}
\APACinsertmetastar {%
LyEtAl2016Rejoinder}%
\begin{APACrefauthors}%
Ly, A.%
, Verhagen, A\BPBI J.%
\BCBL {}\ \BBA {} Wagenmakers, E\BHBI J.%
\end{APACrefauthors}%
\unskip\
\newblock
\APACrefYearMonthDay{2016{\protect \BCnt {2}}}{}{}.
\newblock
{\BBOQ}\APACrefatitle {{\SR{20162}} {A}n Evaluation of Alternative Methods for
  Testing Hypotheses, from the Perspective of {H}arold {J}effreys}
  {{\SR{20162}} {A}n evaluation of alternative methods for testing hypotheses,
  from the perspective of {H}arold {J}effreys}.{\BBCQ}
\newblock
\APACjournalVolNumPages{Journal of Mathematical Psychology}{72}{}{43--55}.
\PrintBackRefs{\CurrentBib}

\bibitem [\protect \citeauthoryear {%
Ly%
\ \BBA {} Wagenmakers%
}{%
Ly%
\ \BBA {} Wagenmakers%
}{%
{\protect \APACyear {{\protect \BIP {}}}}%
{\protect \APACexlab {{\protect \BCntIP {1}}}}}]{%
LyWagenmakersinpressPeri}
\APACinsertmetastar {%
LyWagenmakersinpressPeri}%
\begin{APACrefauthors}%
Ly, A.%
\BCBT {}\ \BBA {} Wagenmakers, E\BHBI J.%
\end{APACrefauthors}%
\unskip\
\newblock
\APACrefYearMonthDay{{\protect \BIP {}}{\protect \BCntIP {1}}}{}{}.
\newblock
{\BBOQ}\APACrefatitle {Bayes Factors for Peri--null Hypotheses} {Bayes factors
  for peri--null hypotheses}.{\BBCQ}
\newblock
\APACjournalVolNumPages{{TEST}}{}{}{}.
\newblock
\begin{APACrefURL} \url{https://arxiv.org/abs/2102.07162} \end{APACrefURL}
\PrintBackRefs{\CurrentBib}

\bibitem [\protect \citeauthoryear {%
Ly%
\ \BBA {} Wagenmakers%
}{%
Ly%
\ \BBA {} Wagenmakers%
}{%
{\protect \APACyear {{\protect \BIP {}}}}%
{\protect \APACexlab {{\protect \BCntIP {2}}}}}]{%
LyWagenmakersinpressKelto}
\APACinsertmetastar {%
LyWagenmakersinpressKelto}%
\begin{APACrefauthors}%
Ly, A.%
\BCBT {}\ \BBA {} Wagenmakers, E\BHBI J.%
\end{APACrefauthors}%
\unskip\
\newblock
\APACrefYearMonthDay{{\protect \BIP {}}{\protect \BCntIP {2}}}{}{}.
\newblock
{\BBOQ}\APACrefatitle {A Critical Evaluation of the {FBST} $ev$ for {B}ayesian
  Hypothesis Testing} {A critical evaluation of the {FBST} $ev$ for {B}ayesian
  hypothesis testing}.{\BBCQ}
\newblock
\APACjournalVolNumPages{Computational Brain \& Behavior}{}{}{}.
\newblock
\begin{APACrefURL} \url{https://psyarxiv.com/x9t6n/} \end{APACrefURL}
\PrintBackRefs{\CurrentBib}

\bibitem [\protect \citeauthoryear {%
Maier%
\ \BBA {} Lakens%
}{%
Maier%
\ \BBA {} Lakens%
}{%
{\protect \APACyear {{in press}}}%
}]{%
MaierLakensinpress}
\APACinsertmetastar {%
MaierLakensinpress}%
\begin{APACrefauthors}%
Maier, M.%
\BCBT {}\ \BBA {} Lakens, D.%
\end{APACrefauthors}%
\unskip\
\newblock
\APACrefYearMonthDay{{in press}}{}{}.
\newblock
{\BBOQ}\APACrefatitle {Justify Your Alpha: {A} Primer on Two Practical
  Approaches} {Justify your alpha: {A} primer on two practical
  approaches}.{\BBCQ}
\newblock
\APACjournalVolNumPages{Advances in Methods and Practices in Psychological
  Science}{}{}{}.
\newblock
\begin{APACrefURL} \url{https://psyarxiv.com/ts4r6} \end{APACrefURL}
\PrintBackRefs{\CurrentBib}

\bibitem [\protect \citeauthoryear {%
Marsman%
\ \BBA {} Wagenmakers%
}{%
Marsman%
\ \BBA {} Wagenmakers%
}{%
{\protect \APACyear {2017}}%
}]{%
MarsmanWagenmakers2017ThreeInsights}
\APACinsertmetastar {%
MarsmanWagenmakers2017ThreeInsights}%
\begin{APACrefauthors}%
Marsman, M.%
\BCBT {}\ \BBA {} Wagenmakers, E\BHBI J.%
\end{APACrefauthors}%
\unskip\
\newblock
\APACrefYearMonthDay{2017}{}{}.
\newblock
{\BBOQ}\APACrefatitle {Three insights From a {B}ayesian Interpretation of the
  One--sided $P$ value} {Three insights from a {B}ayesian interpretation of the
  one--sided $p$ value}.{\BBCQ}
\newblock
\APACjournalVolNumPages{Educational and Psychological
  Measurement}{77}{}{529--539}.
\PrintBackRefs{\CurrentBib}

\bibitem [\protect \citeauthoryear {%
Morey%
\ \BBA {} Rouder%
}{%
Morey%
\ \BBA {} Rouder%
}{%
{\protect \APACyear {2011}}%
}]{%
MoreyRouder2011}
\APACinsertmetastar {%
MoreyRouder2011}%
\begin{APACrefauthors}%
Morey, R\BPBI D.%
\BCBT {}\ \BBA {} Rouder, J\BPBI N.%
\end{APACrefauthors}%
\unskip\
\newblock
\APACrefYearMonthDay{2011}{}{}.
\newblock
{\BBOQ}\APACrefatitle {Bayes Factor Approaches for Testing Interval Null
  Hypotheses} {Bayes factor approaches for testing interval null
  hypotheses}.{\BBCQ}
\newblock
\APACjournalVolNumPages{Psychological Methods}{16}{}{406--419}.
\PrintBackRefs{\CurrentBib}

\bibitem [\protect \citeauthoryear {%
Morey%
\ \BBA {} Rouder%
}{%
Morey%
\ \BBA {} Rouder%
}{%
{\protect \APACyear {2018}}%
}]{%
MoreyRouderBayesFactorPackage}
\APACinsertmetastar {%
MoreyRouderBayesFactorPackage}%
\begin{APACrefauthors}%
Morey, R\BPBI D.%
\BCBT {}\ \BBA {} Rouder, J\BPBI N.%
\end{APACrefauthors}%
\unskip\
\newblock
\APACrefYearMonthDay{2018}{}{}.
\newblock
\APACrefbtitle {{BayesFactor} 0.9.12-4.2.} {{BayesFactor} 0.9.12-4.2.}
\newblock
\APAChowpublished {Comprehensive R Archive Network}.
\newblock
\begin{APACrefURL}
  \url{http://cran.r-project.org/web/packages/BayesFactor/index.html}
  \end{APACrefURL}
\PrintBackRefs{\CurrentBib}

\bibitem [\protect \citeauthoryear {%
Mudge%
, Baker%
, Edge%
\BCBL {}\ \BBA {} Houlahan%
}{%
Mudge%
\ \protect \BOthers {.}}{%
{\protect \APACyear {2012}}%
}]{%
MudgeEtAl2012}
\APACinsertmetastar {%
MudgeEtAl2012}%
\begin{APACrefauthors}%
Mudge, J\BPBI F.%
, Baker, L\BPBI F.%
, Edge, C\BPBI B.%
\BCBL {}\ \BBA {} Houlahan, J\BPBI E.%
\end{APACrefauthors}%
\unskip\
\newblock
\APACrefYearMonthDay{2012}{}{}.
\newblock
{\BBOQ}\APACrefatitle {Setting an Optimal $\alpha$ That Minimizes Errors in
  Null Hypothesis Significance Tests} {Setting an optimal $\alpha$ that
  minimizes errors in null hypothesis significance tests}.{\BBCQ}
\newblock
\APACjournalVolNumPages{{PLoS ONE}}{7}{}{{e32734}}.
\PrintBackRefs{\CurrentBib}

\bibitem [\protect \citeauthoryear {%
Nasir%
, Soliman%
, Shahbaz%
\BCBL {}\ \protect \BOthers {.}}{%
Nasir%
\ \protect \BOthers {.}}{%
{\protect \APACyear {2020}}%
}]{%
nasir2020operational}
\APACinsertmetastar {%
nasir2020operational}%
\begin{APACrefauthors}%
Nasir, M\BPBI A.%
, Soliman, A\BPBI M.%
, Shahbaz, M.%
\BCBL {}\ \BOthersPeriod {.}\end{APACrefauthors}%
\unskip\
\newblock
\APACrefYearMonthDay{2020}{}{}.
\newblock
{\BBOQ}\APACrefatitle {Operational aspect of the policy coordination for
  financial stability: role of {Jeffreys--Lindley}’s paradox in operations
  research} {Operational aspect of the policy coordination for financial
  stability: role of {Jeffreys--Lindley}’s paradox in operations
  research}.{\BBCQ}
\newblock
\APACjournalVolNumPages{Annals of Operations Research}{}{}{1--25}.
\PrintBackRefs{\CurrentBib}

\bibitem [\protect \citeauthoryear {%
O'Hagan%
\ \BBA {} Forster%
}{%
O'Hagan%
\ \BBA {} Forster%
}{%
{\protect \APACyear {2004}}%
}]{%
OHaganForster2004}
\APACinsertmetastar {%
OHaganForster2004}%
\begin{APACrefauthors}%
O'Hagan, A.%
\BCBT {}\ \BBA {} Forster, J.%
\end{APACrefauthors}%
\unskip\
\newblock
\APACrefYear{2004}.
\newblock
\APACrefbtitle {{K}endall's Advanced Theory of Statistics Vol. 2{B}: {B}ayesian
  Inference (2nd ed.)} {{K}endall's advanced theory of statistics vol. 2{B}:
  {B}ayesian inference (2nd ed.)}.
\newblock
\APACaddressPublisher{London}{Arnold}.
\PrintBackRefs{\CurrentBib}

\bibitem [\protect \citeauthoryear {%
Ormerod%
, Stewart%
, Yu%
\BCBL {}\ \BBA {} Romanes%
}{%
Ormerod%
\ \protect \BOthers {.}}{%
{\protect \APACyear {2017}}%
}]{%
OrmerodEtAl2017}
\APACinsertmetastar {%
OrmerodEtAl2017}%
\begin{APACrefauthors}%
Ormerod, J\BPBI T.%
, Stewart, M.%
, Yu, W.%
\BCBL {}\ \BBA {} Romanes, S\BPBI E.%
\end{APACrefauthors}%
\unskip\
\newblock
\APACrefYearMonthDay{2017}{}{}.
\newblock
{\BBOQ}\APACrefatitle {Bayesian Hypothesis Tests with Diffuse Priors: {C}an We
  Have Our Cake and Eat it Too?} {Bayesian hypothesis tests with diffuse
  priors: {C}an we have our cake and eat it too?}{\BBCQ}
\newblock
\APACjournalVolNumPages{{Manuscript submitted for publication}}{}{}{}.
\newblock
\begin{APACrefURL} \url{https://arxiv.org/pdf/1710.09146.pdf} \end{APACrefURL}
\PrintBackRefs{\CurrentBib}

\bibitem [\protect \citeauthoryear {%
Pearson%
}{%
Pearson%
}{%
{\protect \APACyear {1953}}%
}]{%
Pearson1953}
\APACinsertmetastar {%
Pearson1953}%
\begin{APACrefauthors}%
Pearson, E\BPBI S.%
\end{APACrefauthors}%
\unskip\
\newblock
\APACrefYearMonthDay{1953}{}{}.
\newblock
{\BBOQ}\APACrefatitle {Comment on ``Statistical Inference'' by {Dennis
  Lindley}} {Comment on ``statistical inference'' by {Dennis Lindley}}.{\BBCQ}
\newblock
\APACjournalVolNumPages{Journal of the Royal Statistical Society Series B
  (Methodological)}{15}{}{68--69}.
\PrintBackRefs{\CurrentBib}

\bibitem [\protect \citeauthoryear {%
P\'{e}rez%
\ \BBA {} Pericchi%
}{%
P\'{e}rez%
\ \BBA {} Pericchi%
}{%
{\protect \APACyear {2014}}%
}]{%
PerezPericchi2014}
\APACinsertmetastar {%
PerezPericchi2014}%
\begin{APACrefauthors}%
P\'{e}rez, M\BHBI E.%
\BCBT {}\ \BBA {} Pericchi, L\BPBI R.%
\end{APACrefauthors}%
\unskip\
\newblock
\APACrefYearMonthDay{2014}{}{}.
\newblock
{\BBOQ}\APACrefatitle {Changing Statistical Significance with the Amount of
  Information: {T}he Adaptive $\alpha$ Significance Level} {Changing
  statistical significance with the amount of information: {T}he adaptive
  $\alpha$ significance level}.{\BBCQ}
\newblock
\APACjournalVolNumPages{Statistics and Probability Letters}{85}{}{20--24}.
\PrintBackRefs{\CurrentBib}

\bibitem [\protect \citeauthoryear {%
Pericchi%
}{%
Pericchi%
}{%
{\protect \APACyear {2011}}%
}]{%
Pericchi2011}
\APACinsertmetastar {%
Pericchi2011}%
\begin{APACrefauthors}%
Pericchi, L\BPBI R.%
\end{APACrefauthors}%
\unskip\
\newblock
\APACrefYearMonthDay{2011}{}{}.
\newblock
{\BBOQ}\APACrefatitle {Comment on ``Integrated Objective {B}ayesian Estimation
  and Hypothesis Testing'' by {J. M. Bernardo}} {Comment on ``integrated
  objective {B}ayesian estimation and hypothesis testing'' by {J. M.
  Bernardo}}.{\BBCQ}
\newblock
\BIn{} J\BPBI M.~Bernardo\ \BOthers {.}\ (\BEDS), \APACrefbtitle {{B}ayesian
  Statistics 9} {{B}ayesian statistics 9}\ (\BPGS\ 25--29).
\newblock
\APACaddressPublisher{Oxford}{Oxford University Press}.
\PrintBackRefs{\CurrentBib}

\bibitem [\protect \citeauthoryear {%
Pericchi%
\ \BBA {} Pereira%
}{%
Pericchi%
\ \BBA {} Pereira%
}{%
{\protect \APACyear {2016}}%
}]{%
PericchiPereira2016}
\APACinsertmetastar {%
PericchiPereira2016}%
\begin{APACrefauthors}%
Pericchi, L\BPBI R.%
\BCBT {}\ \BBA {} Pereira, C.%
\end{APACrefauthors}%
\unskip\
\newblock
\APACrefYearMonthDay{2016}{}{}.
\newblock
{\BBOQ}\APACrefatitle {Adaptative Significance Levels Using Optimal Decision
  Rules: {B}alancing by Weighting the Error Probabilities} {Adaptative
  significance levels using optimal decision rules: {B}alancing by weighting
  the error probabilities}.{\BBCQ}
\newblock
\APACjournalVolNumPages{Brazilian Journal of Probability and
  Statistics}{30}{}{70--90}.
\PrintBackRefs{\CurrentBib}

\bibitem [\protect \citeauthoryear {%
Pratt%
}{%
Pratt%
}{%
{\protect \APACyear {1965}}%
}]{%
Pratt1965}
\APACinsertmetastar {%
Pratt1965}%
\begin{APACrefauthors}%
Pratt, J\BPBI W.%
\end{APACrefauthors}%
\unskip\
\newblock
\APACrefYearMonthDay{1965}{}{}.
\newblock
{\BBOQ}\APACrefatitle {{B}ayesian Interpretation of Standard Inference
  Statements} {{B}ayesian interpretation of standard inference
  statements}.{\BBCQ}
\newblock
\APACjournalVolNumPages{Journal of the Royal Statistical Society
  B}{27}{}{169--203}.
\PrintBackRefs{\CurrentBib}

\bibitem [\protect \citeauthoryear {%
Robert%
}{%
Robert%
}{%
{\protect \APACyear {1993}}%
}]{%
Robert1993}
\APACinsertmetastar {%
Robert1993}%
\begin{APACrefauthors}%
Robert, C\BPBI P.%
\end{APACrefauthors}%
\unskip\
\newblock
\APACrefYearMonthDay{1993}{}{}.
\newblock
{\BBOQ}\APACrefatitle {A Note on {Jeffreys--Lindley} Paradox} {A note on
  {Jeffreys--Lindley} paradox}.{\BBCQ}
\newblock
\APACjournalVolNumPages{Statistica Sinica}{3}{}{601--608}.
\PrintBackRefs{\CurrentBib}

\bibitem [\protect \citeauthoryear {%
Robert%
}{%
Robert%
}{%
{\protect \APACyear {2013}}%
}]{%
Robert2013}
\APACinsertmetastar {%
Robert2013}%
\begin{APACrefauthors}%
Robert, C\BPBI P.%
\end{APACrefauthors}%
\unskip\
\newblock
\APACrefYearMonthDay{2013}{}{}.
\newblock
{\BBOQ}\APACrefatitle {On the {Lindley--Jeffreys} Paradox} {On the
  {Lindley--Jeffreys} paradox}.{\BBCQ}
\newblock
\BIn{} A.~{O'Hagan}\ (\BED), \APACrefbtitle {A Book for {Dennis}} {A book for
  {Dennis}}\ (\BPGS\ 118--122).
\newblock
\APACaddressPublisher{}{Blurb}.
\PrintBackRefs{\CurrentBib}

\bibitem [\protect \citeauthoryear {%
Robert%
}{%
Robert%
}{%
{\protect \APACyear {2014}}%
}]{%
Robert2014}
\APACinsertmetastar {%
Robert2014}%
\begin{APACrefauthors}%
Robert, C\BPBI P.%
\end{APACrefauthors}%
\unskip\
\newblock
\APACrefYearMonthDay{2014}{}{}.
\newblock
{\BBOQ}\APACrefatitle {On the {Lindley--Jeffreys} Paradox} {On the
  {Lindley--Jeffreys} paradox}.{\BBCQ}
\newblock
\APACjournalVolNumPages{Philosophy of Science}{81}{}{216--232}.
\PrintBackRefs{\CurrentBib}

\bibitem [\protect \citeauthoryear {%
Robert%
, Chopin%
\BCBL {}\ \BBA {} Rousseau%
}{%
Robert%
\ \protect \BOthers {.}}{%
{\protect \APACyear {2009}}%
}]{%
RobertEtAl2009}
\APACinsertmetastar {%
RobertEtAl2009}%
\begin{APACrefauthors}%
Robert, C\BPBI P.%
, Chopin, N.%
\BCBL {}\ \BBA {} Rousseau, J.%
\end{APACrefauthors}%
\unskip\
\newblock
\APACrefYearMonthDay{2009}{}{}.
\newblock
{\BBOQ}\APACrefatitle {Harold {J}effreys's {Theory of Probability} Revisited}
  {Harold {J}effreys's {Theory of Probability} revisited}.{\BBCQ}
\newblock
\APACjournalVolNumPages{Statistical Science}{24}{}{141--172}.
\PrintBackRefs{\CurrentBib}

\bibitem [\protect \citeauthoryear {%
Robert%
\ \BBA {} Rousseau%
}{%
Robert%
\ \BBA {} Rousseau%
}{%
{\protect \APACyear {2011}}%
}]{%
RobertRousseau2011}
\APACinsertmetastar {%
RobertRousseau2011}%
\begin{APACrefauthors}%
Robert, C\BPBI P.%
\BCBT {}\ \BBA {} Rousseau, J.%
\end{APACrefauthors}%
\unskip\
\newblock
\APACrefYearMonthDay{2011}{}{}.
\newblock
{\BBOQ}\APACrefatitle {Comment on ``Integrated Objective {B}ayesian Estimation
  and Hypothesis Testing'' by {J. M. Bernardo}} {Comment on ``integrated
  objective {B}ayesian estimation and hypothesis testing'' by {J. M.
  Bernardo}}.{\BBCQ}
\newblock
\BIn{} J\BPBI M.~Bernardo\ \BOthers {.}\ (\BEDS), \APACrefbtitle {{B}ayesian
  Statistics 9} {{B}ayesian statistics 9}\ (\BPGS\ 41--44).
\newblock
\APACaddressPublisher{Oxford}{Oxford University Press}.
\PrintBackRefs{\CurrentBib}

\bibitem [\protect \citeauthoryear {%
R.~Royall%
}{%
R.~Royall%
}{%
{\protect \APACyear {1986}}%
}]{%
Royall1986}
\APACinsertmetastar {%
Royall1986}%
\begin{APACrefauthors}%
Royall, R.%
\end{APACrefauthors}%
\unskip\
\newblock
\APACrefYearMonthDay{1986}{}{}.
\newblock
{\BBOQ}\APACrefatitle {The Effect of Sample Size on the Meaning of Significance
  Tests} {The effect of sample size on the meaning of significance
  tests}.{\BBCQ}
\newblock
\APACjournalVolNumPages{The American Statistician}{40}{}{313--315}.
\PrintBackRefs{\CurrentBib}

\bibitem [\protect \citeauthoryear {%
R\BPBI M.~Royall%
}{%
R\BPBI M.~Royall%
}{%
{\protect \APACyear {1997}}%
}]{%
Royall1997}
\APACinsertmetastar {%
Royall1997}%
\begin{APACrefauthors}%
Royall, R\BPBI M.%
\end{APACrefauthors}%
\unskip\
\newblock
\APACrefYear{1997}.
\newblock
\APACrefbtitle {Statistical Evidence: {A} Likelihood Paradigm} {Statistical
  evidence: {A} likelihood paradigm}.
\newblock
\APACaddressPublisher{London}{{C}hapman \& {H}all}.
\PrintBackRefs{\CurrentBib}

\bibitem [\protect \citeauthoryear {%
Savage%
}{%
Savage%
}{%
{\protect \APACyear {1964}}%
}]{%
Savage1964}
\APACinsertmetastar {%
Savage1964}%
\begin{APACrefauthors}%
Savage, L\BPBI J.%
\end{APACrefauthors}%
\unskip\
\newblock
\APACrefYearMonthDay{1964}{}{}.
\newblock
{\BBOQ}\APACrefatitle {The Foundations of Statistics Reconsidered} {The
  foundations of statistics reconsidered}.{\BBCQ}
\newblock
\BIn{} H\BPBI E.~Kyburg\ \BBA {} H\BPBI E.~Smokler\ (\BEDS), \APACrefbtitle
  {Studies in Subjective Probability} {Studies in subjective probability}\
  (\BPGS\ 173--188).
\newblock
\APACaddressPublisher{New York}{John Wiley}.
\PrintBackRefs{\CurrentBib}

\bibitem [\protect \citeauthoryear {%
Savage%
\ \protect \BOthers {.}}{%
Savage%
\ \protect \BOthers {.}}{%
{\protect \APACyear {1962}}%
}]{%
SavageEtAl1962}
\APACinsertmetastar {%
SavageEtAl1962}%
\begin{APACrefauthors}%
Savage, L\BPBI J.%
, Bartlett, M\BPBI S.%
, Barnard, G\BPBI A.%
, Cox, D\BPBI R.%
, Pearson, E\BPBI S.%
, Smith, C\BPBI A\BPBI B.%
\BDBL {}Winsten, C\BPBI B.%
\end{APACrefauthors}%
\unskip\
\newblock
\APACrefYear{1962}.
\newblock
\APACrefbtitle {The Foundations of Statistical Inference} {The foundations of
  statistical inference}.
\newblock
\APACaddressPublisher{London}{Methuen}.
\PrintBackRefs{\CurrentBib}

\bibitem [\protect \citeauthoryear {%
Sellke%
, Bayarri%
\BCBL {}\ \BBA {} Berger%
}{%
Sellke%
\ \protect \BOthers {.}}{%
{\protect \APACyear {2001}}%
}]{%
SellkeEtAl2001}
\APACinsertmetastar {%
SellkeEtAl2001}%
\begin{APACrefauthors}%
Sellke, T.%
, Bayarri, M\BPBI J.%
\BCBL {}\ \BBA {} Berger, J\BPBI O.%
\end{APACrefauthors}%
\unskip\
\newblock
\APACrefYearMonthDay{2001}{}{}.
\newblock
{\BBOQ}\APACrefatitle {Calibration of $p$ Values for Testing Precise Null
  Hypotheses} {Calibration of $p$ values for testing precise null
  hypotheses}.{\BBCQ}
\newblock
\APACjournalVolNumPages{The American Statistician}{55}{}{62--71}.
\PrintBackRefs{\CurrentBib}

\bibitem [\protect \citeauthoryear {%
Senn%
}{%
Senn%
}{%
{\protect \APACyear {2001}}%
}]{%
Senn2001}
\APACinsertmetastar {%
Senn2001}%
\begin{APACrefauthors}%
Senn, S.%
\end{APACrefauthors}%
\unskip\
\newblock
\APACrefYearMonthDay{2001}{}{}.
\newblock
{\BBOQ}\APACrefatitle {Two Cheers for {P}-Values?} {Two cheers for
  {P}-values?}{\BBCQ}
\newblock
\APACjournalVolNumPages{Journal of Epidemiology and
  Biostatistics}{6}{}{193--204}.
\PrintBackRefs{\CurrentBib}

\bibitem [\protect \citeauthoryear {%
Shafer%
}{%
Shafer%
}{%
{\protect \APACyear {1982}}%
}]{%
Shafer1982}
\APACinsertmetastar {%
Shafer1982}%
\begin{APACrefauthors}%
Shafer, G.%
\end{APACrefauthors}%
\unskip\
\newblock
\APACrefYearMonthDay{1982}{}{}.
\newblock
{\BBOQ}\APACrefatitle {{L}indley's Paradox} {{L}indley's paradox}.{\BBCQ}
\newblock
\APACjournalVolNumPages{Journal of the American Statistical
  Association}{77}{}{325--351}.
\PrintBackRefs{\CurrentBib}

\bibitem [\protect \citeauthoryear {%
Spanos%
}{%
Spanos%
}{%
{\protect \APACyear {2013}}%
}]{%
spanos2013should}
\APACinsertmetastar {%
spanos2013should}%
\begin{APACrefauthors}%
Spanos, A.%
\end{APACrefauthors}%
\unskip\
\newblock
\APACrefYearMonthDay{2013}{}{}.
\newblock
{\BBOQ}\APACrefatitle {Who should be afraid of the {Jeffreys-Lindley} paradox?}
  {Who should be afraid of the {Jeffreys-Lindley} paradox?}{\BBCQ}
\newblock
\APACjournalVolNumPages{Philosophy of Science}{80}{1}{73--93}.
\PrintBackRefs{\CurrentBib}

\bibitem [\protect \citeauthoryear {%
Sprenger%
}{%
Sprenger%
}{%
{\protect \APACyear {2013}}%
}]{%
sprenger2013testing}
\APACinsertmetastar {%
sprenger2013testing}%
\begin{APACrefauthors}%
Sprenger, J.%
\end{APACrefauthors}%
\unskip\
\newblock
\APACrefYearMonthDay{2013}{}{}.
\newblock
{\BBOQ}\APACrefatitle {Testing a precise null hypothesis: The case of
  {L}indley’s paradox} {Testing a precise null hypothesis: The case of
  {L}indley’s paradox}.{\BBCQ}
\newblock
\APACjournalVolNumPages{Philosophy of Science}{80}{5}{733--744}.
\PrintBackRefs{\CurrentBib}

\bibitem [\protect \citeauthoryear {%
Szab\'{o}%
\ \BBA {} {van der Vaart}%
}{%
Szab\'{o}%
\ \BBA {} {van der Vaart}%
}{%
{\protect \APACyear {2019}}%
}]{%
SzaboVaart2019}
\APACinsertmetastar {%
SzaboVaart2019}%
\begin{APACrefauthors}%
Szab\'{o}, B.%
\BCBT {}\ \BBA {} {van der Vaart}, A.%
\end{APACrefauthors}%
\unskip\
\newblock
\APACrefYearMonthDay{2019}{}{}.
\newblock
\APACrefbtitle {Bayesian Statistics [Lecture notes].} {Bayesian statistics
  [lecture notes].}
\newblock
\APACaddressPublisher{}{Leiden University}.
\PrintBackRefs{\CurrentBib}

\bibitem [\protect \citeauthoryear {%
Vehtari%
, Gelman%
\BCBL {}\ \BBA {} Gabry%
}{%
Vehtari%
\ \protect \BOthers {.}}{%
{\protect \APACyear {2017}}%
}]{%
VehtariEtAl2017}
\APACinsertmetastar {%
VehtariEtAl2017}%
\begin{APACrefauthors}%
Vehtari, A.%
, Gelman, A.%
\BCBL {}\ \BBA {} Gabry, J.%
\end{APACrefauthors}%
\unskip\
\newblock
\APACrefYearMonthDay{2017}{}{}.
\newblock
{\BBOQ}\APACrefatitle {Practical {B}ayesian Model Evaluation Using
  Leave--One--Out Cross--Validation and {WAIC}} {Practical {B}ayesian model
  evaluation using leave--one--out cross--validation and {WAIC}}.{\BBCQ}
\newblock
\APACjournalVolNumPages{Statistics and Computing}{27}{}{1413--1432}.
\PrintBackRefs{\CurrentBib}

\bibitem [\protect \citeauthoryear {%
Verdinelli%
\ \BBA {} Wasserman%
}{%
Verdinelli%
\ \BBA {} Wasserman%
}{%
{\protect \APACyear {1995}}%
}]{%
VerdinelliWasserman1995}
\APACinsertmetastar {%
VerdinelliWasserman1995}%
\begin{APACrefauthors}%
Verdinelli, I.%
\BCBT {}\ \BBA {} Wasserman, L.%
\end{APACrefauthors}%
\unskip\
\newblock
\APACrefYearMonthDay{1995}{}{}.
\newblock
{\BBOQ}\APACrefatitle {Computing {B}ayes Factors Using a Generalization of the
  {S}avage--{D}ickey Density Ratio} {Computing {B}ayes factors using a
  generalization of the {S}avage--{D}ickey density ratio}.{\BBCQ}
\newblock
\APACjournalVolNumPages{Journal of the American Statistical
  Association}{90}{}{614--618}.
\PrintBackRefs{\CurrentBib}

\bibitem [\protect \citeauthoryear {%
Villa%
\ \BBA {} Walker%
}{%
Villa%
\ \BBA {} Walker%
}{%
{\protect \APACyear {2017}}%
}]{%
VillaWalker2017}
\APACinsertmetastar {%
VillaWalker2017}%
\begin{APACrefauthors}%
Villa, C.%
\BCBT {}\ \BBA {} Walker, S.%
\end{APACrefauthors}%
\unskip\
\newblock
\APACrefYearMonthDay{2017}{}{}.
\newblock
{\BBOQ}\APACrefatitle {On the Mathematics of the {Jeffreys--Lindley} Paradox}
  {On the mathematics of the {Jeffreys--Lindley} paradox}.{\BBCQ}
\newblock
\APACjournalVolNumPages{Communications in Statistics -- Theory and
  Methods}{46}{}{12290--12298}.
\PrintBackRefs{\CurrentBib}

\bibitem [\protect \citeauthoryear {%
Wagenmakers%
}{%
Wagenmakers%
}{%
{\protect \APACyear {2007}}%
}]{%
Wagenmakers2007}
\APACinsertmetastar {%
Wagenmakers2007}%
\begin{APACrefauthors}%
Wagenmakers, E\BHBI J.%
\end{APACrefauthors}%
\unskip\
\newblock
\APACrefYearMonthDay{2007}{}{}.
\newblock
{\BBOQ}\APACrefatitle {A Practical Solution to the Pervasive Problems of $p$
  Values} {A practical solution to the pervasive problems of $p$
  values}.{\BBCQ}
\newblock
\APACjournalVolNumPages{Psychonomic Bulletin \& Review}{14}{}{779--804}.
\PrintBackRefs{\CurrentBib}

\bibitem [\protect \citeauthoryear {%
Wagenmakers%
, Gronau%
, Dablander%
\BCBL {}\ \BBA {} Etz%
}{%
Wagenmakers%
\ \protect \BOthers {.}}{%
{\protect \APACyear {{\protect \BIP {}}}}%
}]{%
WagenmakersEtAlinpressSupportInterval}
\APACinsertmetastar {%
WagenmakersEtAlinpressSupportInterval}%
\begin{APACrefauthors}%
Wagenmakers, E\BHBI J.%
, Gronau, Q\BPBI F.%
, Dablander, F.%
\BCBL {}\ \BBA {} Etz, A.%
\end{APACrefauthors}%
\unskip\
\newblock
\APACrefYearMonthDay{{\protect \BIP {}}}{}{}.
\newblock
{\BBOQ}\APACrefatitle {The Support Interval} {The support interval}.{\BBCQ}
\newblock
\APACjournalVolNumPages{Erkenntnis}{}{}{}.
\newblock
\begin{APACrefURL} \url{https://psyarxiv.com/zwnxb/} \end{APACrefURL}
\PrintBackRefs{\CurrentBib}

\bibitem [\protect \citeauthoryear {%
Wagenmakers%
\ \protect \BOthers {.}}{%
Wagenmakers%
\ \protect \BOthers {.}}{%
{\protect \APACyear {2017}}%
}]{%
WagenmakersEtAlScrutinyBook2017}
\APACinsertmetastar {%
WagenmakersEtAlScrutinyBook2017}%
\begin{APACrefauthors}%
Wagenmakers, E\BHBI J.%
, Verhagen, A\BPBI J.%
, Ly, A.%
, Matzke, D.%
, Steingroever, H.%
, Rouder, J\BPBI N.%
\BCBL {}\ \BBA {} Morey, R\BPBI D.%
\end{APACrefauthors}%
\unskip\
\newblock
\APACrefYearMonthDay{2017}{}{}.
\newblock
{\BBOQ}\APACrefatitle {The Need for {B}ayesian Hypothesis Testing in
  Psychological Science} {The need for {B}ayesian hypothesis testing in
  psychological science}.{\BBCQ}
\newblock
\BIn{} S\BPBI O.~Lilienfeld\ \BBA {} I.~Waldman\ (\BEDS), \APACrefbtitle
  {Psychological Science Under Scrutiny: {R}ecent Challenges and Proposed
  Solutions} {Psychological science under scrutiny: {R}ecent challenges and
  proposed solutions}\ (\BPGS\ 123--138).
\newblock
\APACaddressPublisher{}{John Wiley and Sons}.
\PrintBackRefs{\CurrentBib}

\bibitem [\protect \citeauthoryear {%
Wasserstein%
\ \BBA {} Lazar%
}{%
Wasserstein%
\ \BBA {} Lazar%
}{%
{\protect \APACyear {2016}}%
}]{%
wasserstein2016asa}
\APACinsertmetastar {%
wasserstein2016asa}%
\begin{APACrefauthors}%
Wasserstein, R\BPBI L.%
\BCBT {}\ \BBA {} Lazar, N\BPBI A.%
\end{APACrefauthors}%
\unskip\
\newblock
\APACrefYearMonthDay{2016}{}{}.
\newblock
{\BBOQ}\APACrefatitle {The {ASA}'s statement on p-values: Context, process, and
  purpose} {The {ASA}'s statement on p-values: Context, process, and
  purpose}.{\BBCQ}
\newblock
\APACjournalVolNumPages{The American Statistician}{70}{2}{129--133}.
\PrintBackRefs{\CurrentBib}

\bibitem [\protect \citeauthoryear {%
Wetzels%
, Grasman%
\BCBL {}\ \BBA {} Wagenmakers%
}{%
Wetzels%
\ \protect \BOthers {.}}{%
{\protect \APACyear {2010}}%
}]{%
WetzelsEtAl2010Borel}
\APACinsertmetastar {%
WetzelsEtAl2010Borel}%
\begin{APACrefauthors}%
Wetzels, R.%
, Grasman, R\BPBI P\BPBI P\BPBI P.%
\BCBL {}\ \BBA {} Wagenmakers, E\BHBI J.%
\end{APACrefauthors}%
\unskip\
\newblock
\APACrefYearMonthDay{2010}{}{}.
\newblock
{\BBOQ}\APACrefatitle {An Encompassing Prior Generalization of the
  {S}avage--{D}ickey Density Ratio Test} {An encompassing prior generalization
  of the {S}avage--{D}ickey density ratio test}.{\BBCQ}
\newblock
\APACjournalVolNumPages{Computational Statistics \& Data
  Analysis}{54}{}{2094--2102}.
\PrintBackRefs{\CurrentBib}

\bibitem [\protect \citeauthoryear {%
Wrinch%
\ \BBA {} Jeffreys%
}{%
Wrinch%
\ \BBA {} Jeffreys%
}{%
{\protect \APACyear {1919}}%
}]{%
WrinchJeffreys1919}
\APACinsertmetastar {%
WrinchJeffreys1919}%
\begin{APACrefauthors}%
Wrinch, D.%
\BCBT {}\ \BBA {} Jeffreys, H.%
\end{APACrefauthors}%
\unskip\
\newblock
\APACrefYearMonthDay{1919}{}{}.
\newblock
{\BBOQ}\APACrefatitle {On Some Aspects of the Theory of Probability} {On some
  aspects of the theory of probability}.{\BBCQ}
\newblock
\APACjournalVolNumPages{Philosophical Magazine}{38}{}{715--731}.
\PrintBackRefs{\CurrentBib}

\bibitem [\protect \citeauthoryear {%
Wrinch%
\ \BBA {} Jeffreys%
}{%
Wrinch%
\ \BBA {} Jeffreys%
}{%
{\protect \APACyear {1921}}%
}]{%
WrinchJeffreys1921}
\APACinsertmetastar {%
WrinchJeffreys1921}%
\begin{APACrefauthors}%
Wrinch, D.%
\BCBT {}\ \BBA {} Jeffreys, H.%
\end{APACrefauthors}%
\unskip\
\newblock
\APACrefYearMonthDay{1921}{}{}.
\newblock
{\BBOQ}\APACrefatitle {On Certain Fundamental Principles of Scientific Inquiry}
  {On certain fundamental principles of scientific inquiry}.{\BBCQ}
\newblock
\APACjournalVolNumPages{Philosophical Magazine}{42}{}{369--390}.
\PrintBackRefs{\CurrentBib}

\bibitem [\protect \citeauthoryear {%
Wrinch%
\ \BBA {} Jeffreys%
}{%
Wrinch%
\ \BBA {} Jeffreys%
}{%
{\protect \APACyear {1923}}%
}]{%
WrinchJeffreys1923}
\APACinsertmetastar {%
WrinchJeffreys1923}%
\begin{APACrefauthors}%
Wrinch, D.%
\BCBT {}\ \BBA {} Jeffreys, H.%
\end{APACrefauthors}%
\unskip\
\newblock
\APACrefYearMonthDay{1923}{}{}.
\newblock
{\BBOQ}\APACrefatitle {On Certain Fundamental Principles of Scientific Inquiry}
  {On certain fundamental principles of scientific inquiry}.{\BBCQ}
\newblock
\APACjournalVolNumPages{Philosophical Magazine}{45}{}{368--374}.
\PrintBackRefs{\CurrentBib}

\bibitem [\protect \citeauthoryear {%
Yin%
\ \BBA {} Shi%
}{%
Yin%
\ \BBA {} Shi%
}{%
{\protect \APACyear {2020}}%
}]{%
yin2020demystify}
\APACinsertmetastar {%
yin2020demystify}%
\begin{APACrefauthors}%
Yin, G.%
\BCBT {}\ \BBA {} Shi, H.%
\end{APACrefauthors}%
\unskip\
\newblock
\APACrefYearMonthDay{2020}{}{}.
\newblock
{\BBOQ}\APACrefatitle {Demystify {L}indley's Paradox by Interpreting P-value as
  Posterior Probability} {Demystify {L}indley's paradox by interpreting p-value
  as posterior probability}.{\BBCQ}
\newblock
\APACjournalVolNumPages{arXiv preprint arXiv:2002.10883}{}{}{}.
\PrintBackRefs{\CurrentBib}

\bibitem [\protect \citeauthoryear {%
Zellner%
}{%
Zellner%
}{%
{\protect \APACyear {{1971/1996}}}%
}]{%
Zellner19711996}
\APACinsertmetastar {%
Zellner19711996}%
\begin{APACrefauthors}%
Zellner, A.%
\end{APACrefauthors}%
\unskip\
\newblock
\APACrefYear{{1971/1996}}.
\newblock
\APACrefbtitle {An Introduction to {B}ayesian Inference in Econometrics} {An
  introduction to {B}ayesian inference in econometrics}.
\newblock
\APACaddressPublisher{New York}{Wiley}.
\PrintBackRefs{\CurrentBib}

\end{thebibliography}
\newpage

\section{Appendix: Jeffreys Discusses the Paradox Post 1957}
\label{App:Post1957}

As far as the paradox-related material in Jeffreys's books is concerned, the 1961 third edition of \emph{Theory of Probability} does not add anything to the 1948 second edition (cf. \citealp[pp. 221-222, p. 399]{Jeffreys1948} to \citealp[p. 248, p. 435]{Jeffreys1961}), which itself did not add much to the 1939 first edition \citep[p. 194, pp. 359-360]{Jeffreys1939}. Likewise, the 1973 third edition of \emph{Scientific Inference} repeats the short relevant fragment from the 1957 second edition provided in the main text (cf. \citealp[pp. 74-75]{Jeffreys1973} to \citealp[pp. 71-72]{Jeffreys1957SI}). 

Jeffreys does touch on the paradox in three papers published after 1957. First, in the 1974 article \emph{Fisher and inverse probability}, Jeffreys hints at the paradox when he writes:
\begin{quotation}
``I think that astronomers had found much earlier that discrepancies up to twice the standard error usually disappeared when more information became available, but those over three times usually persisted. In fact, \emph{with ordinary numbers of observations}, say 10 to 500, these rough rules are usually not far from the 95 per cent and 99 per cent rules or from the more detailed ones that I derive in my \emph{Theory of Probability}. (\citealp[p. 2]{Jeffreys1974}; first italics added for emphasis)
\end{quotation}

Later, the 1977 article \emph{Probability theory in geophysics} contains a relevant fragment that is highly similar to \citet[p. 349]{Jeffreys1957} cited above:
\begin{quotation}
``The theory leads to rules of significance for changes in laws, involving the introduction of new parameters in laws. They are usually approximately of the form
\begin{equation*}
    K = \frac{P(q \mid \theta p)}{P(q^{'} \mid \theta p)} \doteqdot (An)^{\frac{1}{2}} \exp \left( - \frac{a^2}{2s^2_a} \right).
\end{equation*}
Here $q$ is the hypothesis that the new parameter $\alpha$ is zero, that is, that the previous law needs no alteration; $q'$ the hypothesis that $\alpha$ is needed, having a value to be estimated from the observations; $a$ and $s_a$ are the estimate of $\alpha$ and its standard error as given by the method of least squares; $n$ is the number of observations; and $A$ is a constant, usually not far from 1. If $a<s_a$, the factor $n^{\frac{1}{2}}$ makes $K>1$ and the old law is supported; but with ordinary numbers of observations, if $a>2s_a$ or $3s_a$, $K<1$ and the new law is supported. To apply a test of this sort it is of course of the first importance that the number of observations shall be stated. This is in fact not often done by physicists, but thanks mainly to the work of Fisher (with whom I do not always agree) biologists usually do it, but with different rules. I once remarked to Fisher that in nearly all practical applications we should agree, and that when we differed we should both be doubtful.'' \citep[p. 89]{Jeffreys1977}
\end{quotation}

Finally, in 1980 Jeffreys published the chapter \emph{Some general points in probability theory} in the book \emph{Bayesian analysis in econometrics and statistics: Essays in honor of Harold Jeffreys}. Jeffreys summarizes his contributions and concludes as follows: 
\begin{quotation}
``Many complications have been dealt with. The usual form, if $y$ is used for the observational data, is approximately
\begin{equation*}
K = \frac{P(\mathcal{H}_0 \mid y)}{P(\mathcal{H}_1 \mid y)} = A n^{1/2} \, \text{exp}\left\{ -\frac{(a-\alpha_0)^2}{2s_a^2}\right\},   \end{equation*}
where $A$ is of order 1, $n$ the number of observations, $a$ and $s_a$ the estimates by maximum likelihood of the new parameter and its standard error. If $a < s_a$ and $n$ is large we get strong confirmation that no change in $\alpha$ is needed; if $a-\alpha_0$ is several times $s_a$ there is strong support for a change. For $n$ from about 10 to 500 the usual result is that $K=1$ when $(a-\alpha_0)/s_a$ is about 2, $10^{-1/2}$ when it is about 2.7, $10^{-1}$ about 3.2, and $10^-2$ about 4. These are not far from the rough rule long known to astronomers, i.e., that differences up to twice the standard error usually disappear when more or better observations become available, and that those of three or more time usually persist. They are also not far from the $0.05$, $0.01$ and so on limits for the usual $P$. I have always considered the arguments for the use of $P$ absurd. They amount to saying that a hypothesis that may or may not be true is rejected because a greater departure from the trial value was improbable; that is, that it has not predicted something that has not happened. As an argument astronomer's experience is far better. $P$ has a definite place when we already know what parameters are relevant, and we want to know their amounts; this is what I call a problem of estimation. A problem of significance is one where we are considering a change in the form of the law itself.'' \citep[p. 453]{Jeffreys1980}
\end{quotation}

\end{document}